\documentclass[aps,prl,twocolumn,reprint,floatfix,superscriptaddress]{revtex4-2}
\usepackage{graphicx}
\usepackage{amsmath,amssymb,mathrsfs,amsthm}
\usepackage[pdftex, colorlinks=true, linkcolor=blue, citecolor=blue, urlcolor=blue]{hyperref}
\usepackage{bbold}
\usepackage{times}
\usepackage{mathtools}
\usepackage{braket}
\usepackage{comment}

\usepackage{lipsum}
\usepackage{svg}
\usepackage[normalem]{ulem}
\usepackage{units}
\usepackage{xcolor}

\usepackage{float}

\begin{document}

\title{Broadband Dipole Absorption in Dispersive Photonic Time Crystals}

\author{Thomas F.\ Allard}
\author{Jaime E.\ Sustaeta-Osuna}
\author{Francisco J.\ Garc\'{i}a-Vidal}
\author{Paloma A.\ Huidobro}
\affiliation{Departamento de F\'{i}sica Te\'{o}rica de la Materia Condensada and Condensed Matter Physics Center (IFIMAC), Universidad Aut\'{o}noma de Madrid, E-28049 Madrid, Spain}
\affiliation{Condensed Matter Physics Center (IFIMAC), Universidad Aut\'{o}noma de Madrid, E-28049 Madrid, Spain}

\begin{abstract}
Photonic media modulated periodically in time, termed photonic time crystals (PTCs), have attracted considerable attention for their ability to open momentum bandgaps hosting amplifying modes.
These momentum gaps, however, generally appear only at the system's parametric resonance condition which constrain many features derived from amplification to a narrow frequency band.
Moreover, they are accompanied by exceptional points (EPs) and may drive the system into an instability, which render their analysis more intricate.
Here, we show that a careful consideration of dispersion and absorption can overcome these issues.
By investigating the dissipated power of a point-dipole embedded in a dispersive and absorptive PTC, we unveil that temporal modulation enables the conversion of dipole emission into dipole absorption within a broadband frequency window free of EPs.
We demonstrate that this effect is general, emerging in both the stable and unstable regimes, and occurs from weak modulation strength to low modulations frequencies that could be achieved for various material platforms.
\end{abstract}

\maketitle

\emph{Introduction}--\!
A sudden change in a material refractive index -- a time interface -- induces frequency conversion accompanied by both forward and backward propagating waves, a phenomenon termed time refraction and reflection \cite{Mendonça2002,Bacot2016}.
While early studies predicted these effects decades ago \cite{Morgenthaler1958,Felsen1970}, recent experimental advances have led to a surge in interest in the photonics of time-varying media, aiming at new possibilities of wave manipulation \cite{Alam2016,Shaltout2019,Bohn2021,Galiffi2022,Yin2022}.
Unique phenomena without counterpart in static systems have since been proposed \cite{Yu2009,Sounas2017,Deck2019,Pacheco-Pena2020,Zhou2020,Li2021,Pacheco-Pena2021,Prudencio2023,Vazquez2023,Hayran2024,Eswaran2025} and time reflection has been observed in transmission line metamaterials \cite{Moussa2023,Wang2023}.
Moreover, ultrafast modulation of the refractive index in transparent conductive oxides (TCOs) \cite{Reshef2019,Un2023} paves the way for time-varying media in the optical regime \cite{Vezzoli2018,Tirole2023,Lustig2023,Tirole2024}.

Within this burgeoning field, materials whose electric permittivity is modulated periodically in time \cite{Zurita2009} -- termed photonic time crystals (PTCs) -- have attracted much attention \cite{Asgari2024}.
Indeed, periodic modulation induces interference between time-refracted and -reflected waves, which open momentum bandgaps that host amplifying and decaying modes, allowing energy transfer between the temporal modulation and electromagnetic waves propagating through the material.

Although many effects can be drawn from simplified models of PTCs, the inclusion of realistic aspects such as material dispersion and losses becomes essential to accurately model experiments \cite{Solis2021,Mirmoosa2022,Horsley2023,Sloan2024,Koutserimpas2024,Asgari2024}.
Frequency dispersion also complexifies the concept of momentum gap, as it can induce extended gaps \cite{Wang2025}, as well as dispersive ones \cite{Feng2024,Ozlu2025,Wang2025APL}, namely, broadband \emph{frequency} windows of potentially amplifying modes.
The latter stand in sharp contrast with momentum gaps of nondispersive PTCs that, while encompassing many momenta, are constrained to the parametric resonance (PR) condition, i.e., to half of the modulation frequency.

A key question in the field of PTCs is their interaction with emitters.
Notably, what is their impact on dipole radiation and its quantum counterpart, spontaneous emission?
Enhanced charge radiations in PTCs have been predicted \cite{Dikopoltsev2022,Li2023}, and the authors of Ref.~\cite{Lyubarov2022} reported amplified emission from a dipolar emitter embedded in a nondispersive, lossless PTC, along with a modification of the spontaneous emission decay near the momentum gap frequency, depending on the initial modulation profile \cite{Lyubarov2024}.
On the other hand, the authors of Ref.~\cite{Park2025} recently proposed a classical non-Hermitian formalism considering losses.
They demonstrated that the spontaneous emission modification, proportional to the dipole's dissipated power, is accompanied by spontaneous excitation, which manifests classically as a negative dissipated power.
This negative power is intrinsic to the gain available in PTCs and can be interpreted as dipole \emph{absorption}.

Importantly, the above theoretical studies revealed three \textit{apriori} major drawbacks of PTCs.
First, the emission modification is constrained around the PR condition, making it very narrow band, a phenomenon recently observed experimentally using transmission-line metamaterials \cite{Lee_arxiv2025}.
Second, this particular frequency is associated to divergencies \cite{Park2025,Lyubarov2024}, and overlaps with exceptional points (EPs) \cite{Wang2018,Kazemi2019}. These non-Hermitian degeneracies, while offering potential interest for sensing \cite{Wang2024}, increase the spontaneous emission rate independently of amplification \cite{Pick2017}, which obscures the effect of the temporal modulation itself.
Third, amplifying modes of momentum gaps drive the system into an instability, greatly complicating the analysis.

In this Letter, we overcome these limitations by carefully considering dispersion and losses in our theoretical framework.
By investigating the dissipated power of a dipole embedded in a dispersive and absorptive PTC, we unveil new regimes of modulation where dispersive momentum gaps as well as negative Floquet replicas allow the conversion of emission into absorption in broadband frequency ranges.
Additionally, we show that inevitable losses eliminate EPs from dispersive momentum gaps, allowing us to disentangle the impacts of these points from those of modulation-induced gain.
We demonstrate that these effects are general, occurring from small modulation strengths to low modulation frequencies, for different material platforms, and in both the stable and unstable regimes of a PTC.

\emph{Dispersive and absorptive PTC}--\!
\begin{figure}[tb]
 \includegraphics[width=0.9\linewidth]{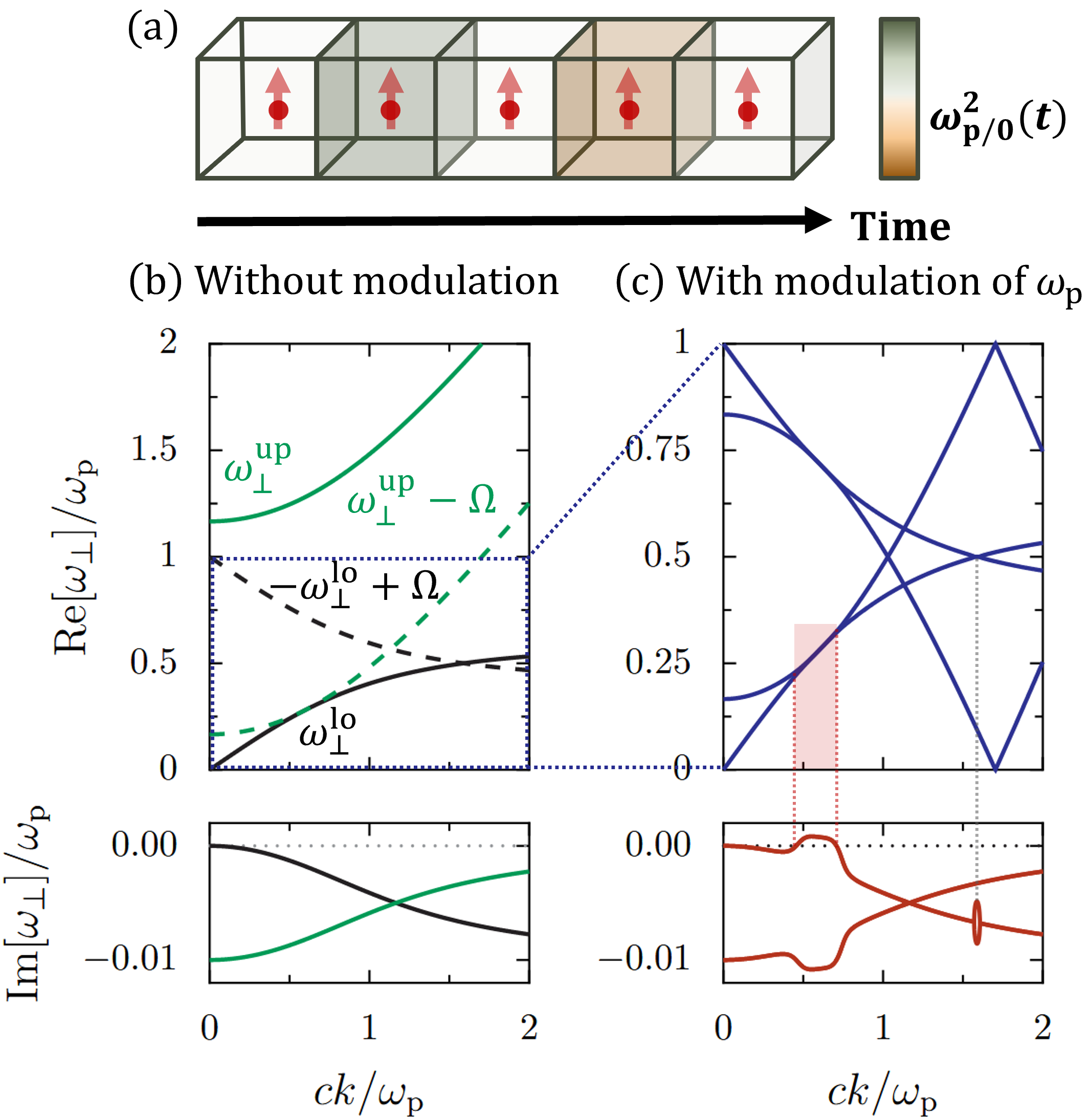}
 \caption{(a) Sketch of the system under consideration.
 A three-dimensional Drude-Lorentz medium, with either its plasma or resonance frequency modulated sinusoidally in time, contains an oscillating dipole.
 (b)-(c) Formation of dispersive momentum gaps.
 (b) Complex bandstructure of the static medium ($\alpha=0$).
 Dashed lines represent Floquet replicas of the original bands, shifted by a modulation frequency $\Omega=\omega_{\mathrm{p}}$.
 (c) Complex bandstructure of a medium with a modulated plasma frequency ($\alpha=0.05$) in the first FBZ.
 The red area highlights the modulation-induced gain region.
 In both panels, $\omega_0=0.6\omega_{\mathrm{p}}$ and, as in the remaining of this paper, $\gamma=0.02\omega_{\mathrm{p}}$.}
\label{fig:Sketch and dispersion}
\end{figure}
The system under consideration, sketched in Fig.~\ref{fig:Sketch and dispersion}(a), is modeled by the combination of Maxwell's equations $\nabla \times \mathbf{E} = -\mu_0 \partial_t \mathbf{H}$ and $\nabla \times \mathbf{H} = \partial_t \mathbf{D} + \mathbf{J}$ with the parametric Drude-Lorentz model
\begin{equation}
    \partial^2_t \mathbf{P} + \gamma \partial_t \mathbf{P} + \omega_0^2 \mathbf{P} = \epsilon_0 \omega_{\mathrm{p}}^2 \mathbf{E}
    \label{eq:Lorentz}
\end{equation}
that describes the dynamics of the three-dimensional polarization density $\mathbf{P}$ inside a general dispersive medium.
Here, $\gamma$, $\omega_{\mathrm{p}}$ and $\omega_0$ are the material's inverse relaxation time, plasma, and resonance frequency.

Importantly, our work aims at capturing the essential physics of various time-modulated platforms. To this end, we consider a periodic modulation in time of either the plasma or the resonance frequency as $\omega_{\mathrm{p}/0}^2(t) = \omega_{\mathrm{p}/0}^2[1+\alpha\sin(\Omega t)]$, with $\Omega$ and $\alpha$ the modulation's frequency and strength, respectively.
A time-dependent plasma frequency typically models ultrafast optical pumping of TCOs in the epsilon-near-zero (ENZ) regime \cite{Reshef2019}, where order-unity refractive index changes have been realized in both thin films \cite{Zhou2020,Tirole2023} and ENZ-based resonant metasurfaces \cite{Vezzoli2018,Pang2021}.
Recent promising experiments also demonstrated order-unity and sub-cycle periodic modulation of the conduction band electrons effective mass in a surface plasmon cavity metamaterial operating at terahertz frequencies \cite{Guo_arxiv2025}.
On the other hand, a time-dependent resonant frequency can be emulated by spatially structuring a modulated medium \cite{Wang2025}, by modulating the optical phonon frequency of polar insulators \cite{Cartella2018,Sloan2021}, or by varying the capacitance in state-of-the-art transmission-line metamaterials \cite{Park2022,Moussa2023}, for which the interaction between a dipole and a PTC has been recently measured \cite{Lee_arxiv2025}.

To obtain the fields in the medium, we follow Ref.~\cite{Raman2010} and combine Maxwell's equations with Eq.~\eqref{eq:Lorentz} into a Schrödinger-like matrix problem using the auxiliary field $\mathbf{\dot P} = \partial_t \mathbf{P}$.
Following Ref.~\cite{Park2025}, we solve this problem using a Floquet formalism, detailed in the Supplemental Material (SM) \cite{SupplementalMaterial}. \nocite{Barnett1992,Horsley2014,Thouin2025,Sun2025}
It is noteworthy that the complexity of our model accounts for the role of both longitudinal and transverse fields.
Recent works have shown that longitudinal modes can also undergo amplification in PTCs \cite{Zhang2024,Feinberg2025}.
Using an hydrodynamic nonlocal model, we demonstrate in the SM \cite{SupplementalMaterial} that longitudinal modes may indeed also modify a dipole's emission in a broadband manner. In the main text we focus however on transverse modes, as they match experiments using transmission-line metamaterials \cite{Park2022,Lee_arxiv2025}.

\emph{EP-free dispersive momentum gaps}--\!
Without modulation, the Drude-Lorentz model features transverse modes with a two-band complex bandstructure.
The upper and lower bands, denoted as $\omega_\perp^{\textrm{up}}$ and $\omega_\perp^{\textrm{lo}}$, are represented in Fig.~\ref{fig:Sketch and dispersion}(b).
The temporal periodicity in the modulation frequency $\Omega$ causes the folding of the bandstructure in the first Floquet Brillouin Zone (FBZ) $\omega \in [0,\Omega]$, inducing positive and negative Floquet replicas $\pm\omega_\perp^{\textrm{up}/\textrm{lo}} + n\Omega$, with $n \in \mathbb{Z}$.
In the modulated system, these replicas may interact with each other, through either an avoided crossing or their merging, which occurs at EPs \cite{Wang2018,Kazemi2019}.
Such a degeneracy of two replicas is what leads to momentum gaps, regions hosting a pair of eigenfrequencies with, respectively, increased and decreased imaginary parts with respect to the nonmodulated value. Depending on a competition between modulation strength and material losses, these regions can host amplified eigenmodes with Im$[\omega_\perp]>0$ \cite{Asgari2024}.

In nondispersive PTCs, these gaps are single-frequency so that we term them in the following as ``flat''.
They arise only at multiples of the PR condition, where an eigenmode $\omega$ is degenerate with one of its own negative replicas $-\omega+n\Omega$.
The two-band nature of a Drude-Lorentz PTC, however, enables new possibilities, such as the merging of the lower band with a replica of the upper band.
From their shape difference, this can lead to a dispersive, broadband in frequency, momentum gap \cite{Feng2024,Ozlu2025,Wang2025APL}.
We note that although we here leverage dispersion to achieve this phenomenon, it can also been found through space-time modulations \cite{Chamanara2018} or anisotropy \cite{Galiffi2022NatCommun,Wang2024,Dong2025}.

The complex bandstructure of a medium with modulated plasma frequency is presented in Fig.~\ref{fig:Sketch and dispersion}(c) and illustrates that mechanism.
Here, the modulation frequency $\Omega$ is chosen so that the lower band $\omega_\perp^{\textrm{lo}}$ couples with a downshifted replica of the upper band $\omega_\perp^{\textrm{up}}-\Omega$. By considering a modulation strength $\alpha$ large enough to counteract the material losses $\gamma$, this leads to a broadband gain region where the eigenfrequency's imaginary part is positive (see the red area), which drives the PTC into its unstable regime.
Importantly, dispersive momentum gaps originating from the coupling between two bands with a different imaginary dispersion, the inclusion of losses induces the imaginary component of the band to split in two within the gap.
This, critically, eliminates the EPs associated with dispersive momentum gaps.
In contrast, a flat momentum gap which preserves the EPs is still present at $\Omega/2$, but losses prevent it from allowing gain [see Fig.~\ref{fig:Sketch and dispersion}(c)].
We discuss the impact of losses on EPs through a calculation of the phase rigidity \cite{Wiersig2023} in the SM \cite{SupplementalMaterial}.

Interestingly, these EP-free momentum gaps occur in a wide range of parameters when modulating the plasma frequency.
Indeed, a simple analytical calculation shows that the condition $\omega_\perp^{\textrm{lo}} = \omega_\perp^{\textrm{up}} - n\Omega$ is satisfied as long as $n\Omega \geq \omega_{\mathrm{p}}$ \cite{SupplementalMaterial}.
While a small modulation strength only enables the first-order replica $n=1$ to interact, increasing $\alpha$ allows for higher-order replicas to contribute, lowering the required modulation frequency.
Notably, when $n\Omega=\omega_{\mathrm{p}}$ the two bands are degenerate at $ck=\omega_0$, allowing a crossing in the dispersive part of $\omega_\perp^{\textrm{lo}}$, hence maximizing the gain bandwidth.

\emph{Broadband dipole absorption}--\!
To embed an harmonic dipole oscillating at a frequency $\omega$ in the PTC, we set the source current density as $\mathbf{J}(\mathbf{r},t) = -i\omega\mathbf{p}(t)\delta(\mathbf{r})$, with $\mathbf{p}(t)=|\mathbf{p}|e^{-i\omega t}\hat{p}$ being the dipole moment.
The transverse field contribution of the power dissipated by the source is $P^{\perp}_{\omega}(t)= (\omega/2)\mathrm{Im}[\mathbf{p}^{*}(t)\cdot\mathbf{E}^{\perp}(\mathbf{0},t)]$.
Its time- and orientation-average can be rewritten as
\begin{equation}
     \bar{P}^{\perp}_\omega = \frac{\pi \omega^2 |\mathbf{p}|^2}{12\epsilon_0}  \int_{\mathbb{R}^3} d^3\mathbf{k}\, \tilde{\rho}^{\perp}_{k,\omega}.
     \label{eq:Dissipated power}
\end{equation}
Here, $\tilde{\rho}^{\perp}_{k,\omega}$ is the transverse part of the photonic momentum-resolved local density of states (LDOS), a quantity recently measured in a transmission-line-based PTC \cite{Lee_arxiv2025}, and which we compute through the Floquet Green dyadic of the fields \cite{SupplementalMaterial}.
As shown in Ref.~\cite{Park2025}, the gain-mechanism induced by the temporal modulation reveals itself through regions of reciprocal space with negative LDOS.
To analyze our results, we thus separate the positive and negative contributions of the LDOS to the dissipated power, and define $\bar{P}_\omega^{\perp,\textrm{loss (gain)}}$ as the integral over reciprocal space where the LDOS is positive (negative) \cite{Ren2024,Park2025}.
In that way, the total power $\bar{P}_\omega^{\perp,\mathrm{total}} = \bar{P}_\omega^{\perp,\mathrm{loss}} - \bar{P}_\omega^{\perp,\mathrm{gain}}$.
Importantly, such a calculation of dissipated power, as done in previous studies \cite{Lyubarov2022,Lyubarov2024,Park2025}, only reflects the effect of the modulation \emph{on the dipole itself}.
It does not account for the fields within the medium growing exponentially over time if the PTC operates in its unstable regime, with eigenfrequencies of positive imaginary part.
In that case, the quantity \eqref{eq:Dissipated power} is dominant only during the early-time dynamics of the system, when $t \ll 1/\textrm{max}(\mathrm{Im}[\omega_\perp])$.

\begin{figure}[t!]
 \includegraphics[width=0.95\columnwidth]{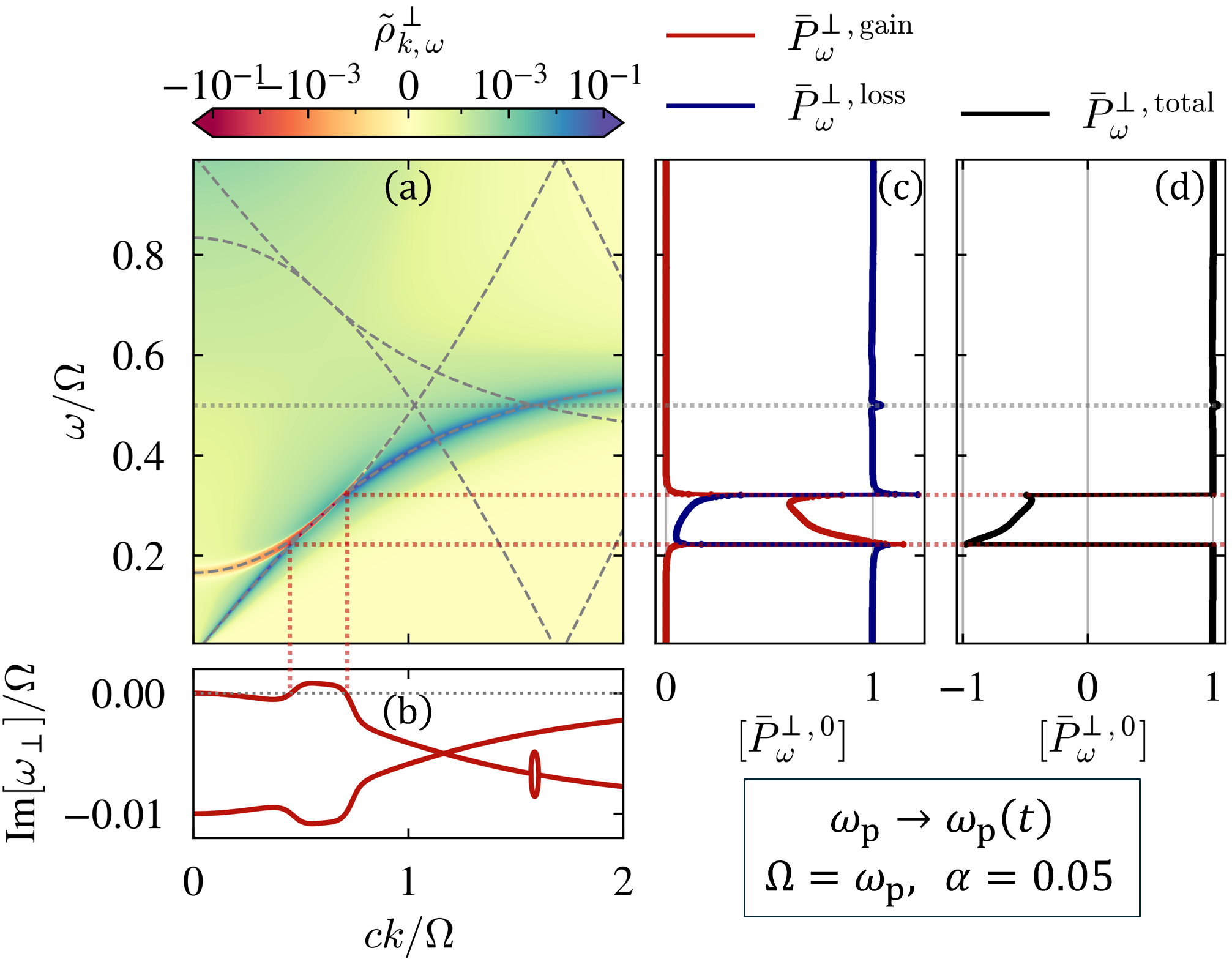}
 \caption{Broadband dipole absorption through weak modulation of the plasma frequency.
 (a) Momentum-resolved LDOS. The gray dashed lines show the real Floquet bandstructure Re$[\omega_\perp]$.
 (b) Imaginary Floquet bandstructure.
 (c) Positive $\bar{P}_\omega^{\perp,\mathrm{loss}}$ and negative $\bar{P}_\omega^{\perp,\mathrm{gain}}$ parts of the power dissipated by a point dipole with frequency $\omega$, in units of the nonmodulated value $\bar{P}_\omega^{\perp,0}$.
 (d) Total dissipated power $\bar{P}_\omega^{\perp,\mathrm{total}} = \bar{P}_\omega^{\perp,\mathrm{loss}} - \bar{P}_\omega^{\perp,\mathrm{gain}}$.}
\label{fig:weak modulation}
\end{figure}

The momentum-resolved LDOS, or spectral function, is presented alongside the real Floquet bandstructure in Fig.~\ref{fig:weak modulation}(a) in the scenario of weak yet fast modulation of the plasma frequency we exemplified in Fig.~\ref{fig:Sketch and dispersion}(c).
The LDOS is, as expected, positively peaked (blueish) along the original lower band $\omega_\perp^\textrm{lo}$.
Remarkably, however, a negative peak (reddish) is also present along the first downshifted replica of the upper band, $\omega_\perp^{\textrm{up}} - \Omega$.
Within the EP-free dispersive momentum gap generated by the coupling between these two bands, the negative spectral function arising from the downshifted replica dominates and supersedes the original positive peak.
This mechanism induces, as visible in Fig.~\ref{fig:weak modulation}(c), a large broadband gain contribution to the dissipated power (red line), together with an inhibition of the loss contribution (blue line).
Interestingly, both contributions present sharp peaks precisely at the boundaries of the gain bandwidth, where Im$[\omega_\perp]$$=0$
This is similar to what is found for flat momentum gaps, for which a divergence occurs at the EP's frequency \cite{Lyubarov2024} precisely due to momenta corresponding to real eigenfrequencies where the LDOS exhibits poles \cite{Park2025}.
The absence of EPs in our context confirms that these divergences and the associated peaks in dissipated power are not linked to exceptional-point physics \footnote{While these poles lead to an ill-defined dissipated power, we only approach them in all of our computations to keep convergent results.}.

The total dissipated power, shown in Fig.~\ref{fig:weak modulation}(d), hence exhibits negative values up to $1$ times what is found without modulation, in a bandwidth of about $0.1\Omega$.
We interpret this negative dissipated power as the absorption -- instead of emission -- of energy by the dipole, the temporal modulation transferring energy to the source, converting it into a sink.
Regarding the late-time dynamics, the broadband in frequency nature of the momentum gap let us predict a spectrally broad amplification of noise by the unstable PTC, in contrast with the narrow-band lasing induced by flat momentum gaps \cite{Lee_arxiv2025}.
Away from the dispersive momentum gap, the dissipated power is unchanged from that of a nonmodulated system, except for a slight increase at $\omega=\Omega/2$.
Indeed, the flat momentum gap, although not allowing any gain, induces an enhancement of dipole emission solely due to the EPs at its edges \cite{Pick2017}.

\begin{figure}[t!]
 \includegraphics[width=0.95\columnwidth]{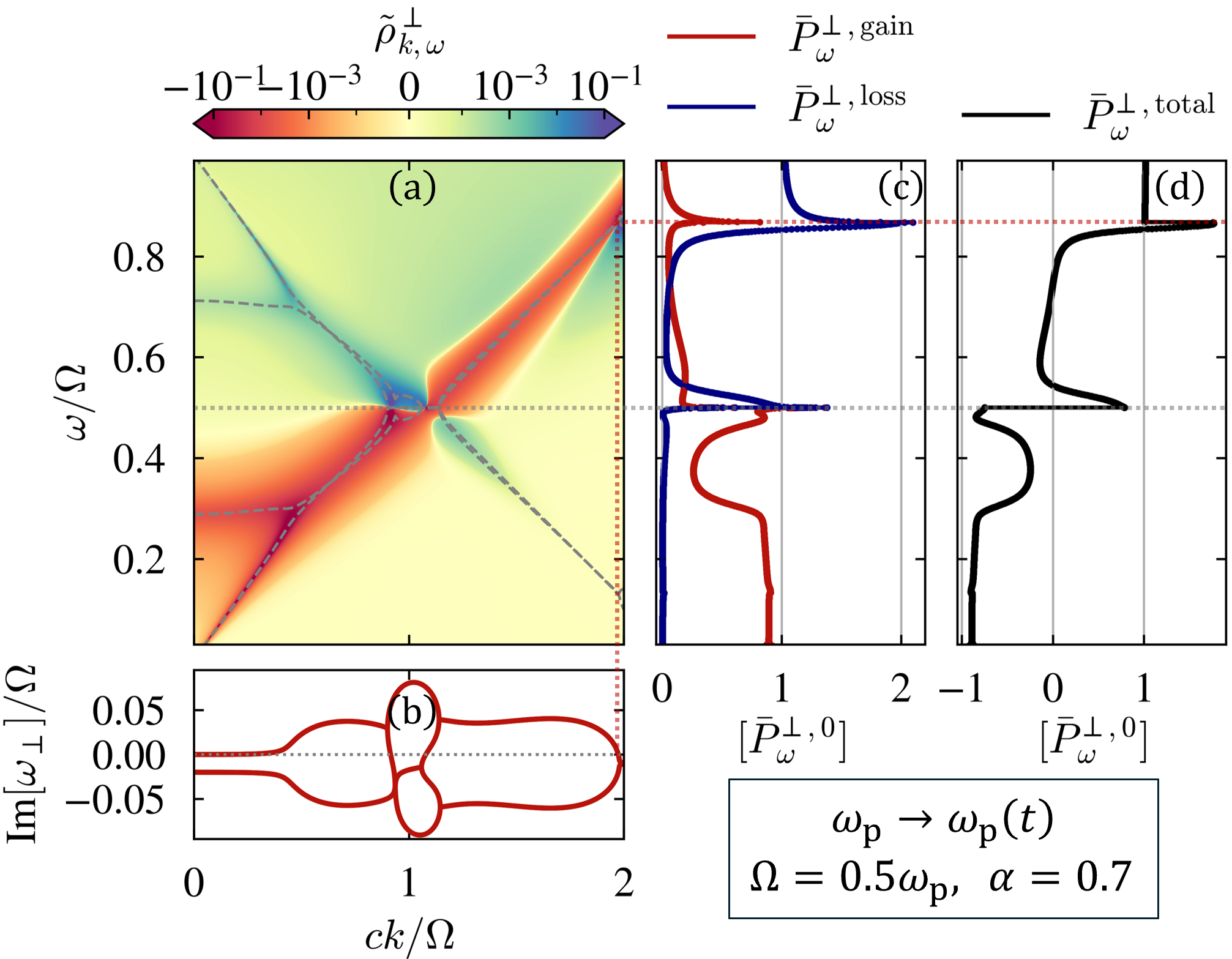}
 \caption{Broadband absorption and inhibition of dissipated power. Same quantities as in Fig.~\ref{fig:weak modulation}, but considering a stronger yet lower modulation of the plasma frequency.
 }
\label{fig:strong modulation}
\end{figure}

We now turn our attention to a scenario of plasma frequency modulation strong enough to allow several Floquet replicas to interact with each other, adding complexity to both the bandstructure and LDOS.
For the second-order replica $\omega_\perp^{\textrm{up}}-2\Omega$ to play the same role as the first-order one in the previous case, we consider a modulation with frequency $\Omega=\omega_\mathrm{p}/2$ and strength $\alpha=0.7$.
The same quantities as discussed previously are shown for that case in Fig.~\ref{fig:strong modulation}. Here, the interaction between $\omega_\perp^{\textrm{up}}-2\Omega$ and $\omega_\perp^{\textrm{lo}}$ induces a dispersive momentum gap occupying the majority of the first FBZ.
Modulating at $\Omega=\omega_{\mathrm{p}}/2$ also induces a flat momentum gap between $\omega_\perp^{\textrm{lo}}$ and  $-\omega_\perp^{\textrm{lo}} + \Omega$ in the same region as the dispersive one, leading
to an interplay that results in a larger value of positive imaginary part for the concerned eigenmodes [see Fig.~\ref{fig:strong modulation}(b)]

As visible in Fig.~\ref{fig:strong modulation}(a), the downshifted replica is again associated to negative peaks of the LDOS and converts the radiation of the lower band $\omega_\perp^{\textrm{lo}}$ into absorption in a wide bandwidth, suppressing the emission contribution of the dissipated power $\bar{P}^{\perp,\textrm{loss}}_\omega$[see Fig.~\ref{fig:strong modulation}(c)].
Interestingly, the splitting of the imaginary bands in two induced by losses allows for slightly positive imaginary parts at small momenta, further broadening the gain bandwidth.
The negative LDOS is however narrower and smaller at large wavenumbers, leading to a peculiar structure which reflects in the total dissipated power as two regimes of dipole frequencies [see Fig.~\ref{fig:strong modulation}(d)].
Below the PR condition ($\omega<\Omega/2$), the modulation-induced gain dominates and the dipole's emission is converted into absorption, with an almost unit ratio for small frequencies.
Above it ($\omega>\Omega/2$), however, both positive and negative LDOS values are small as compared to the nonmodulated case, resulting in a broadband suppression of dissipated power.
In that sense, the modulation-induced gain counteracts the dipole's regular emission and generates an effective frequency bandgap for the dipole, in which dissipated power is inhibited.
Two sharp peaks at $\omega=\Omega/2$ and $\omega\simeq0.87\Omega$ are also visible in the dissipated power, originating from poles of the LDOS where Im$[\omega_\perp]=0$ [see Fig.~\ref{fig:strong modulation}(b) near $ck\simeq\Omega$ and $ck\simeq2\Omega$]. At $\Omega/2$, a pair of EPs is also present, complicating the interpretation.
Finally, we stress that while we focused here on an unstable PTC, a modulated plasma frequency also allows significant modifications of a dipole's emission without relying on eigenmodes with positive imaginary part, as exemplified in the SM \cite{SupplementalMaterial}.

\begin{figure}[tb]
\includegraphics[width=0.958\columnwidth]{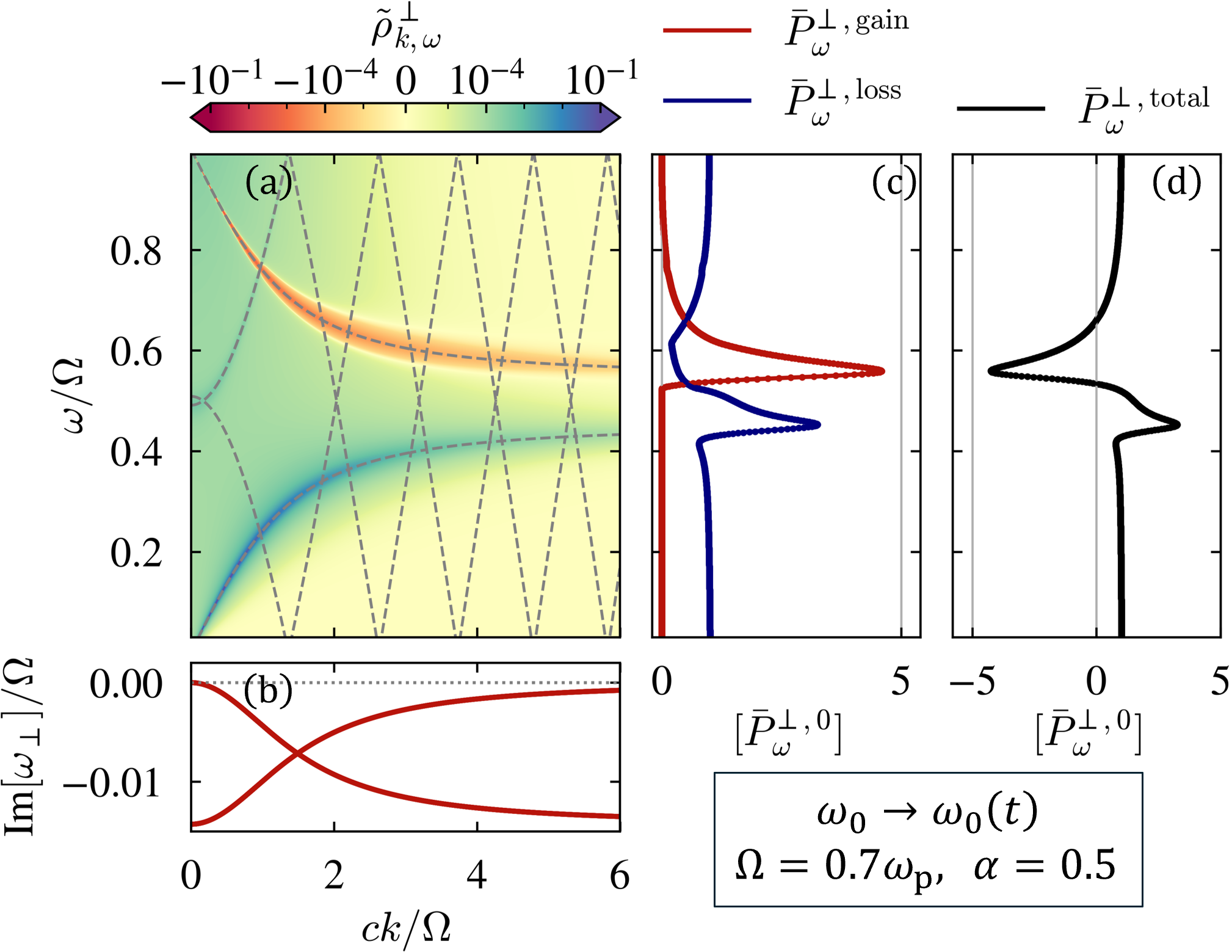}
 \caption{Drastic enhancement of emission and absorption in a stable PTC. Same quantities as in Figs.~\ref{fig:weak modulation} and \ref{fig:strong modulation}, but considering a modulation of the resonance frequency around the value $\omega_0=0.3\omega_\mathrm{p}$.}
\label{fig:resonance frequency}
\end{figure}

\emph{Modulated resonance frequency}--\!
As discussed in the presentation of our model,
PTCs have been realized experimentally using different forms of temporal modulation.
This motivates us to also examine a medium whose resonance frequency is modulated in time, for which we present the momentum-resolved LDOS and complex bandstructure in Fig.~\ref{fig:resonance frequency}(a)-(b).

Considering a resonance frequency $\omega_0=0.3\omega_\mathrm{p}$ modulated with strength $\alpha=0.5$ and frequency $\Omega=0.7\omega_\mathrm{p}>2\omega_0$, we place ourselves in a regime where the lower band and its own negative replica do not interact (see SM for detail \cite{SupplementalMaterial}).
In this way, no momentum gap is present, thereby keeping the PTC stable. Hence, the dissipated power \eqref{eq:Dissipated power} accurately represents the steady-state quantity, as no exponentially growing transient is present.
Critically, however, two distinct processes still allow a drastic and broadband modification of the dissipated power in that scenario.
First, the lower band, associated to positive LDOS, is slightly blue-shifted as compared to the nonmodulated case.
This induces a large LDOS in an almost flat band where it is nearly zero without modulation, producing an increase of the loss contribution to dissipated power up to $3$ times the static value [see Fig.~\ref{fig:resonance frequency}(c)].
Second, and in stark contrast with the plasma frequency modulation, here replicas of negative frequencies, such as $-\omega^\mathrm{lo}_\perp + \Omega$, are associated to negative LDOS. This enables the input of gain within all the frequency bandgap between the original lower and upper bands, and especially around its asymptotic close to $-\omega_0+\Omega$.
Interestingly, such a gain mechanism driven by negative replicas without momentum gaps also emerges in a plasmonic time crystal slab \cite{SustaetaOsuna_arxiv2025}.
Together, these processes leverage dispersion to enable either the enhancement of total emission, or the broadband absorption of a dipole, depending on its frequency being below or above the PR condition [see Fig.~\ref{fig:resonance frequency}(d)].
As this phenomenon does not occur when modulating the plasma frequency, it makes the modulation of a material's resonance a promising way of modifying dipole emission.

\emph{Conclusions}--\!
In this work, we provided a comprehensive study of the power dissipated by a point dipole embedded in a photonic time crystal.
We leveraged dispersion and absorption to unveil that temporal modulation enables a broadband in frequency conversion of dipole emission into dipole absorption that occurs across a wide range of modulation parameters and in both cases of modulated plasma and resonance frequencies.
In addition, for modulated resonance frequency, we demonstrated the possibility of both a broadband negative dissipated power and a large enhancement of emission while operating in the PTC's stable regime.
In our considered scenarios, these phenomena appear independently of exceptional points, allowing us to disentangle the effects associated to the latter from those of gain.
From the classical-quantum correspondence between dipole radiation and spontaneous emission \cite{Milonni1984}, our work, although being classical, paves the way for the investigation of quantum effects in PTCs \cite{Mendonça2000,Ganfornina2024,Mirmoosa2025,Sustaeta-Osuna2025,Bae2025}.
In particular, our results suggest that temporal modulation may force a multi-level emitter either to remain in its excited state or to climb its energy ladder through spontaneous excitation \cite{Park2025,Bae2025}, depending on its frequency, opening exciting perspectives in the field of time-varying media.

\nocite{Zenodo}
\emph{Acknowledgments}--\!
We acknowledge funding from the European Union through the ERC grant TIMELIGHT under GA 101115792, and from the Spanish Ministry for Science, Innovation, and Universities -- Agencia Estatal de Investigaci\'{o}n (AEI) through Grants No.~PID2022-141036NA-I00, PID2021-125894NB-I00, No.~CEX2018-000805-M (through the Mar\'{i}a de Maeztu program for Units of Excellence in Research and Developments) and Grant No.~RYC2021-031568-I (Ram\'{o}n y Cajal program). JES also acknowledges support from the CAM Consejer\'{i}a de Educaci\'{o}n, Ciencia y Universidades, Viceconsejer\'{i}a de Universidades, Investigaci\'{o}n y Ciencia, Direcci\'{o}n General de Investigaci\'{o}n e Innovaci\'{o}n Tecnol\'{o}gica (CAM FPI grant Ref.~A281).

\bibliography{Bibliography}

\begin{thebibliography}{76}%
\makeatletter
\providecommand \@ifxundefined [1]{%
 \@ifx{#1\undefined}
}%
\providecommand \@ifnum [1]{%
 \ifnum #1\expandafter \@firstoftwo
 \else \expandafter \@secondoftwo
 \fi
}%
\providecommand \@ifx [1]{%
 \ifx #1\expandafter \@firstoftwo
 \else \expandafter \@secondoftwo
 \fi
}%
\providecommand \natexlab [1]{#1}%
\providecommand \enquote  [1]{``#1''}%
\providecommand \bibnamefont  [1]{#1}%
\providecommand \bibfnamefont [1]{#1}%
\providecommand \citenamefont [1]{#1}%
\providecommand \href@noop [0]{\@secondoftwo}%
\providecommand \href [0]{\begingroup \@sanitize@url \@href}%
\providecommand \@href[1]{\@@startlink{#1}\@@href}%
\providecommand \@@href[1]{\endgroup#1\@@endlink}%
\providecommand \@sanitize@url [0]{\catcode `\\12\catcode `\$12\catcode `\&12\catcode `\#12\catcode `\^12\catcode `\_12\catcode `\%12\relax}%
\providecommand \@@startlink[1]{}%
\providecommand \@@endlink[0]{}%
\providecommand \url  [0]{\begingroup\@sanitize@url \@url }%
\providecommand \@url [1]{\endgroup\@href {#1}{\urlprefix }}%
\providecommand \urlprefix  [0]{URL }%
\providecommand \Eprint [0]{\href }%
\providecommand \doibase [0]{https://doi.org/}%
\providecommand \selectlanguage [0]{\@gobble}%
\providecommand \bibinfo  [0]{\@secondoftwo}%
\providecommand \bibfield  [0]{\@secondoftwo}%
\providecommand \translation [1]{[#1]}%
\providecommand \BibitemOpen [0]{}%
\providecommand \bibitemStop [0]{}%
\providecommand \bibitemNoStop [0]{.\EOS\space}%
\providecommand \EOS [0]{\spacefactor3000\relax}%
\providecommand \BibitemShut  [1]{\csname bibitem#1\endcsname}%
\let\auto@bib@innerbib\@empty
\bibitem [{\citenamefont {Mendonça}\ and\ \citenamefont {Shukla}(2002)}]{Mendonça2002}%
  \BibitemOpen
  \bibfield  {author} {\bibinfo {author} {\bibfnamefont {J.~T.}\ \bibnamefont {Mendonça}}\ and\ \bibinfo {author} {\bibfnamefont {P.~K.}\ \bibnamefont {Shukla}},\ }\bibfield  {title} {\bibinfo {title} {Time refraction and time reflection: Two basic concepts},\ }\href {https://doi.org/10.1238/Physica.Regular.065a00160} {\bibfield  {journal} {\bibinfo  {journal} {Phys. Scr.}\ }\textbf {\bibinfo {volume} {65}},\ \bibinfo {pages} {160} (\bibinfo {year} {2002})}\BibitemShut {NoStop}%
\bibitem [{\citenamefont {Bacot}\ \emph {et~al.}(2016)\citenamefont {Bacot}, \citenamefont {Labousse}, \citenamefont {Eddi}, \citenamefont {Fink},\ and\ \citenamefont {Fort}}]{Bacot2016}%
  \BibitemOpen
  \bibfield  {author} {\bibinfo {author} {\bibfnamefont {V.}~\bibnamefont {Bacot}}, \bibinfo {author} {\bibfnamefont {M.}~\bibnamefont {Labousse}}, \bibinfo {author} {\bibfnamefont {A.}~\bibnamefont {Eddi}}, \bibinfo {author} {\bibfnamefont {M.}~\bibnamefont {Fink}},\ and\ \bibinfo {author} {\bibfnamefont {E.}~\bibnamefont {Fort}},\ }\bibfield  {title} {\bibinfo {title} {Time reversal and holography with spacetime transformations},\ }\href {https://doi.org/10.1038/nphys3810} {\bibfield  {journal} {\bibinfo  {journal} {Nat. Phys.}\ }\textbf {\bibinfo {volume} {12}},\ \bibinfo {pages} {972} (\bibinfo {year} {2016})}\BibitemShut {NoStop}%
\bibitem [{\citenamefont {Morgenthaler}(1958)}]{Morgenthaler1958}%
  \BibitemOpen
  \bibfield  {author} {\bibinfo {author} {\bibfnamefont {F.}~\bibnamefont {Morgenthaler}},\ }\bibfield  {title} {\bibinfo {title} {Velocity modulation of electromagnetic waves},\ }\href {https://doi.org/10.1109/TMTT.1958.1124533} {\bibfield  {journal} {\bibinfo  {journal} {IEEE Trans. Microw. Theory Tech.}\ }\textbf {\bibinfo {volume} {6}},\ \bibinfo {pages} {167} (\bibinfo {year} {1958})}\BibitemShut {NoStop}%
\bibitem [{\citenamefont {Felsen}\ and\ \citenamefont {Whitman}(1970)}]{Felsen1970}%
  \BibitemOpen
  \bibfield  {author} {\bibinfo {author} {\bibfnamefont {L.}~\bibnamefont {Felsen}}\ and\ \bibinfo {author} {\bibfnamefont {G.}~\bibnamefont {Whitman}},\ }\bibfield  {title} {\bibinfo {title} {Wave propagation in time-varying media},\ }\href {https://doi.org/10.1109/TAP.1970.1139657} {\bibfield  {journal} {\bibinfo  {journal} {IEEE Trans. Antennas Propag.}\ }\textbf {\bibinfo {volume} {18}},\ \bibinfo {pages} {242} (\bibinfo {year} {1970})}\BibitemShut {NoStop}%
\bibitem [{\citenamefont {Alam}\ \emph {et~al.}(2016)\citenamefont {Alam}, \citenamefont {Leon},\ and\ \citenamefont {Boyd}}]{Alam2016}%
  \BibitemOpen
  \bibfield  {author} {\bibinfo {author} {\bibfnamefont {M.~Z.}\ \bibnamefont {Alam}}, \bibinfo {author} {\bibfnamefont {I.~D.}\ \bibnamefont {Leon}},\ and\ \bibinfo {author} {\bibfnamefont {R.~W.}\ \bibnamefont {Boyd}},\ }\bibfield  {title} {\bibinfo {title} {Large optical nonlinearity of indium tin oxide in its epsilon-near-zero region},\ }\href {https://doi.org/10.1126/science.aae0330} {\bibfield  {journal} {\bibinfo  {journal} {Science}\ }\textbf {\bibinfo {volume} {352}},\ \bibinfo {pages} {795} (\bibinfo {year} {2016})}\BibitemShut {NoStop}%
\bibitem [{\citenamefont {Shaltout}\ \emph {et~al.}(2019)\citenamefont {Shaltout}, \citenamefont {Shalaev},\ and\ \citenamefont {Brongersma}}]{Shaltout2019}%
  \BibitemOpen
  \bibfield  {author} {\bibinfo {author} {\bibfnamefont {A.~M.}\ \bibnamefont {Shaltout}}, \bibinfo {author} {\bibfnamefont {V.~M.}\ \bibnamefont {Shalaev}},\ and\ \bibinfo {author} {\bibfnamefont {M.~L.}\ \bibnamefont {Brongersma}},\ }\bibfield  {title} {\bibinfo {title} {Spatiotemporal light control with active metasurfaces},\ }\href {https://doi.org/10.1126/science.aat3100} {\bibfield  {journal} {\bibinfo  {journal} {Science}\ }\textbf {\bibinfo {volume} {364}},\ \bibinfo {pages} {eaat3100} (\bibinfo {year} {2019})}\BibitemShut {NoStop}%
\bibitem [{\citenamefont {Bohn}\ \emph {et~al.}(2021)\citenamefont {Bohn}, \citenamefont {Luk}, \citenamefont {Tollerton}, \citenamefont {Hutchings}, \citenamefont {Brener}, \citenamefont {Horsley}, \citenamefont {Barnes},\ and\ \citenamefont {Hendry}}]{Bohn2021}%
  \BibitemOpen
  \bibfield  {author} {\bibinfo {author} {\bibfnamefont {J.}~\bibnamefont {Bohn}}, \bibinfo {author} {\bibfnamefont {T.~S.}\ \bibnamefont {Luk}}, \bibinfo {author} {\bibfnamefont {C.}~\bibnamefont {Tollerton}}, \bibinfo {author} {\bibfnamefont {S.~W.}\ \bibnamefont {Hutchings}}, \bibinfo {author} {\bibfnamefont {I.}~\bibnamefont {Brener}}, \bibinfo {author} {\bibfnamefont {S.}~\bibnamefont {Horsley}}, \bibinfo {author} {\bibfnamefont {W.~L.}\ \bibnamefont {Barnes}},\ and\ \bibinfo {author} {\bibfnamefont {E.}~\bibnamefont {Hendry}},\ }\bibfield  {title} {\bibinfo {title} {All-optical switching of an epsilon-near-zero plasmon resonance in indium tin oxide},\ }\href {https://doi.org/10.1038/s41467-021-21332-y} {\bibfield  {journal} {\bibinfo  {journal} {Nat. Commun.}\ }\textbf {\bibinfo {volume} {12}},\ \bibinfo {pages} {1017} (\bibinfo {year} {2021})}\BibitemShut {NoStop}%
\bibitem [{\citenamefont {Galiffi}\ \emph {et~al.}(2022{\natexlab{a}})\citenamefont {Galiffi}, \citenamefont {Tirole}, \citenamefont {Yin}, \citenamefont {Li}, \citenamefont {Vezzoli}, \citenamefont {Huidobro}, \citenamefont {Silveirinha}, \citenamefont {Sapienza}, \citenamefont {Al{\`u}},\ and\ \citenamefont {Pendry}}]{Galiffi2022}%
  \BibitemOpen
  \bibfield  {author} {\bibinfo {author} {\bibfnamefont {E.}~\bibnamefont {Galiffi}}, \bibinfo {author} {\bibfnamefont {R.}~\bibnamefont {Tirole}}, \bibinfo {author} {\bibfnamefont {S.}~\bibnamefont {Yin}}, \bibinfo {author} {\bibfnamefont {H.}~\bibnamefont {Li}}, \bibinfo {author} {\bibfnamefont {S.}~\bibnamefont {Vezzoli}}, \bibinfo {author} {\bibfnamefont {P.~A.}\ \bibnamefont {Huidobro}}, \bibinfo {author} {\bibfnamefont {M.~G.}\ \bibnamefont {Silveirinha}}, \bibinfo {author} {\bibfnamefont {R.}~\bibnamefont {Sapienza}}, \bibinfo {author} {\bibfnamefont {A.}~\bibnamefont {Al{\`u}}},\ and\ \bibinfo {author} {\bibfnamefont {J.~B.}\ \bibnamefont {Pendry}},\ }\bibfield  {title} {\bibinfo {title} {{Photonics of time-varying media}},\ }\href {https://doi.org/10.1117/1.AP.4.1.014002} {\bibfield  {journal} {\bibinfo  {journal} {Adv. Photonics}\ }\textbf {\bibinfo {volume} {4}},\ \bibinfo {pages} {014002} (\bibinfo {year} {2022}{\natexlab{a}})}\BibitemShut {NoStop}%
\bibitem [{\citenamefont {Yin}\ \emph {et~al.}(2022)\citenamefont {Yin}, \citenamefont {Galiffi},\ and\ \citenamefont {Al{\`u}}}]{Yin2022}%
  \BibitemOpen
  \bibfield  {author} {\bibinfo {author} {\bibfnamefont {S.}~\bibnamefont {Yin}}, \bibinfo {author} {\bibfnamefont {E.}~\bibnamefont {Galiffi}},\ and\ \bibinfo {author} {\bibfnamefont {A.}~\bibnamefont {Al{\`u}}},\ }\bibfield  {title} {\bibinfo {title} {Floquet metamaterials},\ }\href {https://doi.org/10.1186/s43593-022-00015-1} {\bibfield  {journal} {\bibinfo  {journal} {eLight}\ }\textbf {\bibinfo {volume} {2}},\ \bibinfo {pages} {8} (\bibinfo {year} {2022})}\BibitemShut {NoStop}%
\bibitem [{\citenamefont {Yu}\ and\ \citenamefont {Fan}(2009)}]{Yu2009}%
  \BibitemOpen
  \bibfield  {author} {\bibinfo {author} {\bibfnamefont {Z.}~\bibnamefont {Yu}}\ and\ \bibinfo {author} {\bibfnamefont {S.}~\bibnamefont {Fan}},\ }\bibfield  {title} {\bibinfo {title} {Complete optical isolation created by indirect interband photonic transitions},\ }\href {https://doi.org/10.1038/nphoton.2008.273} {\bibfield  {journal} {\bibinfo  {journal} {Nat. Photonics}\ }\textbf {\bibinfo {volume} {3}},\ \bibinfo {pages} {91} (\bibinfo {year} {2009})}\BibitemShut {NoStop}%
\bibitem [{\citenamefont {Sounas}\ and\ \citenamefont {Al{\`u}}(2017)}]{Sounas2017}%
  \BibitemOpen
  \bibfield  {author} {\bibinfo {author} {\bibfnamefont {D.~L.}\ \bibnamefont {Sounas}}\ and\ \bibinfo {author} {\bibfnamefont {A.}~\bibnamefont {Al{\`u}}},\ }\bibfield  {title} {\bibinfo {title} {Non-reciprocal photonics based on time modulation},\ }\href {https://doi.org/10.1038/s41566-017-0051-x} {\bibfield  {journal} {\bibinfo  {journal} {Nat. Photonics}\ }\textbf {\bibinfo {volume} {11}},\ \bibinfo {pages} {774} (\bibinfo {year} {2017})}\BibitemShut {NoStop}%
\bibitem [{\citenamefont {Deck-L{\'e}ger}\ \emph {et~al.}(2019)\citenamefont {Deck-L{\'e}ger}, \citenamefont {Chamanara}, \citenamefont {Skorobogatiy}, \citenamefont {Silveirinha},\ and\ \citenamefont {Caloz}}]{Deck2019}%
  \BibitemOpen
  \bibfield  {author} {\bibinfo {author} {\bibfnamefont {Z.-L.}\ \bibnamefont {Deck-L{\'e}ger}}, \bibinfo {author} {\bibfnamefont {N.}~\bibnamefont {Chamanara}}, \bibinfo {author} {\bibfnamefont {M.}~\bibnamefont {Skorobogatiy}}, \bibinfo {author} {\bibfnamefont {M.~G.}\ \bibnamefont {Silveirinha}},\ and\ \bibinfo {author} {\bibfnamefont {C.}~\bibnamefont {Caloz}},\ }\bibfield  {title} {\bibinfo {title} {Uniform-velocity spacetime crystals},\ }\href {https://doi.org/https://doi.org/10.1117/1.AP.1.5.056002} {\bibfield  {journal} {\bibinfo  {journal} {Adv. Photonics}\ }\textbf {\bibinfo {volume} {1}},\ \bibinfo {pages} {056002} (\bibinfo {year} {2019})}\BibitemShut {NoStop}%
\bibitem [{\citenamefont {Pacheco-Pe\~na}\ and\ \citenamefont {Engheta}(2020)}]{Pacheco-Pena2020}%
  \BibitemOpen
  \bibfield  {author} {\bibinfo {author} {\bibfnamefont {V.}~\bibnamefont {Pacheco-Pe\~na}}\ and\ \bibinfo {author} {\bibfnamefont {N.}~\bibnamefont {Engheta}},\ }\bibfield  {title} {\bibinfo {title} {Antireflection temporal coatings},\ }\href {https://doi.org/10.1364/OPTICA.381175} {\bibfield  {journal} {\bibinfo  {journal} {Optica}\ }\textbf {\bibinfo {volume} {7}},\ \bibinfo {pages} {323} (\bibinfo {year} {2020})}\BibitemShut {NoStop}%
\bibitem [{\citenamefont {Zhou}\ \emph {et~al.}(2020)\citenamefont {Zhou}, \citenamefont {Alam}, \citenamefont {Karimi}, \citenamefont {Upham}, \citenamefont {Reshef}, \citenamefont {Liu}, \citenamefont {Willner},\ and\ \citenamefont {Boyd}}]{Zhou2020}%
  \BibitemOpen
  \bibfield  {author} {\bibinfo {author} {\bibfnamefont {Y.}~\bibnamefont {Zhou}}, \bibinfo {author} {\bibfnamefont {M.~Z.}\ \bibnamefont {Alam}}, \bibinfo {author} {\bibfnamefont {M.}~\bibnamefont {Karimi}}, \bibinfo {author} {\bibfnamefont {J.}~\bibnamefont {Upham}}, \bibinfo {author} {\bibfnamefont {O.}~\bibnamefont {Reshef}}, \bibinfo {author} {\bibfnamefont {C.}~\bibnamefont {Liu}}, \bibinfo {author} {\bibfnamefont {A.~E.}\ \bibnamefont {Willner}},\ and\ \bibinfo {author} {\bibfnamefont {R.~W.}\ \bibnamefont {Boyd}},\ }\bibfield  {title} {\bibinfo {title} {Broadband frequency translation through time refraction in an epsilon-near-zero material},\ }\href {https://doi.org/10.1038/s41467-020-15682-2} {\bibfield  {journal} {\bibinfo  {journal} {Nat. Commun.}\ }\textbf {\bibinfo {volume} {11}},\ \bibinfo {pages} {2180} (\bibinfo {year} {2020})}\BibitemShut {NoStop}%
\bibitem [{\citenamefont {Li}\ \emph {et~al.}(2021)\citenamefont {Li}, \citenamefont {Yin}, \citenamefont {Galiffi},\ and\ \citenamefont {Al\`u}}]{Li2021}%
  \BibitemOpen
  \bibfield  {author} {\bibinfo {author} {\bibfnamefont {H.}~\bibnamefont {Li}}, \bibinfo {author} {\bibfnamefont {S.}~\bibnamefont {Yin}}, \bibinfo {author} {\bibfnamefont {E.}~\bibnamefont {Galiffi}},\ and\ \bibinfo {author} {\bibfnamefont {A.}~\bibnamefont {Al\`u}},\ }\bibfield  {title} {\bibinfo {title} {{Temporal Parity-Time Symmetry for Extreme Energy Transformations}},\ }\href {https://doi.org/10.1103/PhysRevLett.127.153903} {\bibfield  {journal} {\bibinfo  {journal} {Phys. Rev. Lett.}\ }\textbf {\bibinfo {volume} {127}},\ \bibinfo {pages} {153903} (\bibinfo {year} {2021})}\BibitemShut {NoStop}%
\bibitem [{\citenamefont {Pacheco-Pe\~na}\ and\ \citenamefont {Engheta}(2021)}]{Pacheco-Pena2021}%
  \BibitemOpen
  \bibfield  {author} {\bibinfo {author} {\bibfnamefont {V.}~\bibnamefont {Pacheco-Pe\~na}}\ and\ \bibinfo {author} {\bibfnamefont {N.}~\bibnamefont {Engheta}},\ }\bibfield  {title} {\bibinfo {title} {Temporal equivalent of the brewster angle},\ }\href {https://doi.org/10.1103/PhysRevB.104.214308} {\bibfield  {journal} {\bibinfo  {journal} {Phys. Rev. B}\ }\textbf {\bibinfo {volume} {104}},\ \bibinfo {pages} {214308} (\bibinfo {year} {2021})}\BibitemShut {NoStop}%
\bibitem [{\citenamefont {Prud\^encio}\ and\ \citenamefont {Silveirinha}(2023)}]{Prudencio2023}%
  \BibitemOpen
  \bibfield  {author} {\bibinfo {author} {\bibfnamefont {F.~R.}\ \bibnamefont {Prud\^encio}}\ and\ \bibinfo {author} {\bibfnamefont {M.~G.}\ \bibnamefont {Silveirinha}},\ }\bibfield  {title} {\bibinfo {title} {{Synthetic Axion Response with Space-Time Crystals}},\ }\href {https://doi.org/10.1103/PhysRevApplied.19.024031} {\bibfield  {journal} {\bibinfo  {journal} {Phys. Rev. Appl.}\ }\textbf {\bibinfo {volume} {19}},\ \bibinfo {pages} {024031} (\bibinfo {year} {2023})}\BibitemShut {NoStop}%
\bibitem [{\citenamefont {V{\'a}zquez-Lozano}\ and\ \citenamefont {Liberal}(2023)}]{Vazquez2023}%
  \BibitemOpen
  \bibfield  {author} {\bibinfo {author} {\bibfnamefont {J.~E.}\ \bibnamefont {V{\'a}zquez-Lozano}}\ and\ \bibinfo {author} {\bibfnamefont {I.}~\bibnamefont {Liberal}},\ }\bibfield  {title} {\bibinfo {title} {Incandescent temporal metamaterials},\ }\href {https://doi.org/https://doi.org/10.1038/s41467-023-40281-2} {\bibfield  {journal} {\bibinfo  {journal} {Nat. Commun.}\ }\textbf {\bibinfo {volume} {14}},\ \bibinfo {pages} {4606} (\bibinfo {year} {2023})}\BibitemShut {NoStop}%
\bibitem [{\citenamefont {Hayran}\ and\ \citenamefont {Monticone}(2024)}]{Hayran2024}%
  \BibitemOpen
  \bibfield  {author} {\bibinfo {author} {\bibfnamefont {Z.}~\bibnamefont {Hayran}}\ and\ \bibinfo {author} {\bibfnamefont {F.}~\bibnamefont {Monticone}},\ }\bibfield  {title} {\bibinfo {title} {Beyond the rozanov bound on electromagnetic absorption via periodic temporal modulations},\ }\href {https://doi.org/10.1103/PhysRevApplied.21.044007} {\bibfield  {journal} {\bibinfo  {journal} {Phys. Rev. Appl.}\ }\textbf {\bibinfo {volume} {21}},\ \bibinfo {pages} {044007} (\bibinfo {year} {2024})}\BibitemShut {NoStop}%
\bibitem [{\citenamefont {Eswaran}\ \emph {et~al.}(2025)\citenamefont {Eswaran}, \citenamefont {Kopaei},\ and\ \citenamefont {Sacha}}]{Eswaran2025}%
  \BibitemOpen
  \bibfield  {author} {\bibinfo {author} {\bibfnamefont {K.~S.}\ \bibnamefont {Eswaran}}, \bibinfo {author} {\bibfnamefont {A.~E.}\ \bibnamefont {Kopaei}},\ and\ \bibinfo {author} {\bibfnamefont {K.}~\bibnamefont {Sacha}},\ }\bibfield  {title} {\bibinfo {title} {Anderson localization in photonic time crystals},\ }\href {https://doi.org/10.1103/PhysRevB.111.L180201} {\bibfield  {journal} {\bibinfo  {journal} {Phys. Rev. B}\ }\textbf {\bibinfo {volume} {111}},\ \bibinfo {pages} {L180201} (\bibinfo {year} {2025})}\BibitemShut {NoStop}%
\bibitem [{\citenamefont {Moussa}\ \emph {et~al.}(2023)\citenamefont {Moussa}, \citenamefont {Xu}, \citenamefont {Yin}, \citenamefont {Galiffi}, \citenamefont {Ra'di},\ and\ \citenamefont {Al{\`u}}}]{Moussa2023}%
  \BibitemOpen
  \bibfield  {author} {\bibinfo {author} {\bibfnamefont {H.}~\bibnamefont {Moussa}}, \bibinfo {author} {\bibfnamefont {G.}~\bibnamefont {Xu}}, \bibinfo {author} {\bibfnamefont {S.}~\bibnamefont {Yin}}, \bibinfo {author} {\bibfnamefont {E.}~\bibnamefont {Galiffi}}, \bibinfo {author} {\bibfnamefont {Y.}~\bibnamefont {Ra'di}},\ and\ \bibinfo {author} {\bibfnamefont {A.}~\bibnamefont {Al{\`u}}},\ }\bibfield  {title} {\bibinfo {title} {Observation of temporal reflection and broadband frequency translation at photonic time interfaces},\ }\href {https://doi.org/10.1038/s41567-023-01975-y} {\bibfield  {journal} {\bibinfo  {journal} {Nat. Phys.}\ }\textbf {\bibinfo {volume} {19}},\ \bibinfo {pages} {863} (\bibinfo {year} {2023})}\BibitemShut {NoStop}%
\bibitem [{\citenamefont {Wang}\ \emph {et~al.}(2023)\citenamefont {Wang}, \citenamefont {Mirmoosa}, \citenamefont {Asadchy}, \citenamefont {Rockstuhl}, \citenamefont {Fan},\ and\ \citenamefont {Tretyakov}}]{Wang2023}%
  \BibitemOpen
  \bibfield  {author} {\bibinfo {author} {\bibfnamefont {X.}~\bibnamefont {Wang}}, \bibinfo {author} {\bibfnamefont {M.~S.}\ \bibnamefont {Mirmoosa}}, \bibinfo {author} {\bibfnamefont {V.~S.}\ \bibnamefont {Asadchy}}, \bibinfo {author} {\bibfnamefont {C.}~\bibnamefont {Rockstuhl}}, \bibinfo {author} {\bibfnamefont {S.}~\bibnamefont {Fan}},\ and\ \bibinfo {author} {\bibfnamefont {S.~A.}\ \bibnamefont {Tretyakov}},\ }\bibfield  {title} {\bibinfo {title} {Metasurface-based realization of photonic time crystals},\ }\href {https://doi.org/10.1126/sciadv.adg7541} {\bibfield  {journal} {\bibinfo  {journal} {Sci. Adv.}\ }\textbf {\bibinfo {volume} {9}},\ \bibinfo {pages} {eadg7541} (\bibinfo {year} {2023})}\BibitemShut {NoStop}%
\bibitem [{\citenamefont {Reshef}\ \emph {et~al.}(2019)\citenamefont {Reshef}, \citenamefont {De~Leon}, \citenamefont {Alam},\ and\ \citenamefont {Boyd}}]{Reshef2019}%
  \BibitemOpen
  \bibfield  {author} {\bibinfo {author} {\bibfnamefont {O.}~\bibnamefont {Reshef}}, \bibinfo {author} {\bibfnamefont {I.}~\bibnamefont {De~Leon}}, \bibinfo {author} {\bibfnamefont {M.~Z.}\ \bibnamefont {Alam}},\ and\ \bibinfo {author} {\bibfnamefont {R.~W.}\ \bibnamefont {Boyd}},\ }\bibfield  {title} {\bibinfo {title} {Nonlinear optical effects in epsilon-near-zero media},\ }\href {https://doi.org/10.1038/s41578-019-0120-5} {\bibfield  {journal} {\bibinfo  {journal} {Nat. Rev. Mater.}\ }\textbf {\bibinfo {volume} {4}},\ \bibinfo {pages} {535} (\bibinfo {year} {2019})}\BibitemShut {NoStop}%
\bibitem [{\citenamefont {Un}\ \emph {et~al.}(2023)\citenamefont {Un}, \citenamefont {Sarkar},\ and\ \citenamefont {Sivan}}]{Un2023}%
  \BibitemOpen
  \bibfield  {author} {\bibinfo {author} {\bibfnamefont {I.-W.}\ \bibnamefont {Un}}, \bibinfo {author} {\bibfnamefont {S.}~\bibnamefont {Sarkar}},\ and\ \bibinfo {author} {\bibfnamefont {Y.}~\bibnamefont {Sivan}},\ }\bibfield  {title} {\bibinfo {title} {{Electronic-Based Model of the Optical Nonlinearity of Low-Electron-Density Drude Materials}},\ }\href {https://doi.org/10.1103/PhysRevApplied.19.044043} {\bibfield  {journal} {\bibinfo  {journal} {Phys. Rev. Appl.}\ }\textbf {\bibinfo {volume} {19}},\ \bibinfo {pages} {044043} (\bibinfo {year} {2023})}\BibitemShut {NoStop}%
\bibitem [{\citenamefont {Vezzoli}\ \emph {et~al.}(2018)\citenamefont {Vezzoli}, \citenamefont {Bruno}, \citenamefont {DeVault}, \citenamefont {Roger}, \citenamefont {Shalaev}, \citenamefont {Boltasseva}, \citenamefont {Ferrera}, \citenamefont {Clerici}, \citenamefont {Dubietis},\ and\ \citenamefont {Faccio}}]{Vezzoli2018}%
  \BibitemOpen
  \bibfield  {author} {\bibinfo {author} {\bibfnamefont {S.}~\bibnamefont {Vezzoli}}, \bibinfo {author} {\bibfnamefont {V.}~\bibnamefont {Bruno}}, \bibinfo {author} {\bibfnamefont {C.}~\bibnamefont {DeVault}}, \bibinfo {author} {\bibfnamefont {T.}~\bibnamefont {Roger}}, \bibinfo {author} {\bibfnamefont {V.~M.}\ \bibnamefont {Shalaev}}, \bibinfo {author} {\bibfnamefont {A.}~\bibnamefont {Boltasseva}}, \bibinfo {author} {\bibfnamefont {M.}~\bibnamefont {Ferrera}}, \bibinfo {author} {\bibfnamefont {M.}~\bibnamefont {Clerici}}, \bibinfo {author} {\bibfnamefont {A.}~\bibnamefont {Dubietis}},\ and\ \bibinfo {author} {\bibfnamefont {D.}~\bibnamefont {Faccio}},\ }\bibfield  {title} {\bibinfo {title} {{Optical Time Reversal from Time-Dependent Epsilon-Near-Zero Media}},\ }\href {https://doi.org/10.1103/PhysRevLett.120.043902} {\bibfield  {journal} {\bibinfo  {journal} {Phys. Rev. Lett.}\ }\textbf {\bibinfo {volume} {120}},\ \bibinfo {pages} {043902} (\bibinfo {year} {2018})}\BibitemShut {NoStop}%
\bibitem [{\citenamefont {Tirole}\ \emph {et~al.}(2023)\citenamefont {Tirole}, \citenamefont {Vezzoli}, \citenamefont {Galiffi}, \citenamefont {Robertson}, \citenamefont {Maurice}, \citenamefont {Tilmann}, \citenamefont {Maier}, \citenamefont {Pendry},\ and\ \citenamefont {Sapienza}}]{Tirole2023}%
  \BibitemOpen
  \bibfield  {author} {\bibinfo {author} {\bibfnamefont {R.}~\bibnamefont {Tirole}}, \bibinfo {author} {\bibfnamefont {S.}~\bibnamefont {Vezzoli}}, \bibinfo {author} {\bibfnamefont {E.}~\bibnamefont {Galiffi}}, \bibinfo {author} {\bibfnamefont {I.}~\bibnamefont {Robertson}}, \bibinfo {author} {\bibfnamefont {D.}~\bibnamefont {Maurice}}, \bibinfo {author} {\bibfnamefont {B.}~\bibnamefont {Tilmann}}, \bibinfo {author} {\bibfnamefont {S.~A.}\ \bibnamefont {Maier}}, \bibinfo {author} {\bibfnamefont {J.~B.}\ \bibnamefont {Pendry}},\ and\ \bibinfo {author} {\bibfnamefont {R.}~\bibnamefont {Sapienza}},\ }\bibfield  {title} {\bibinfo {title} {Double-slit time diffraction at optical frequencies},\ }\href {https://doi.org/10.1038/s41567-023-01993-w} {\bibfield  {journal} {\bibinfo  {journal} {Nat. Phys.}\ }\textbf {\bibinfo {volume} {19}},\ \bibinfo {pages} {999} (\bibinfo {year} {2023})}\BibitemShut {NoStop}%
\bibitem [{\citenamefont {Lustig}\ \emph {et~al.}(2023)\citenamefont {Lustig}, \citenamefont {Segal}, \citenamefont {Saha}, \citenamefont {Bordo}, \citenamefont {Chowdhury}, \citenamefont {Sharabi}, \citenamefont {Fleischer}, \citenamefont {Boltasseva}, \citenamefont {Cohen}, \citenamefont {Shalaev},\ and\ \citenamefont {Segev}}]{Lustig2023}%
  \BibitemOpen
  \bibfield  {author} {\bibinfo {author} {\bibfnamefont {E.}~\bibnamefont {Lustig}}, \bibinfo {author} {\bibfnamefont {O.}~\bibnamefont {Segal}}, \bibinfo {author} {\bibfnamefont {S.}~\bibnamefont {Saha}}, \bibinfo {author} {\bibfnamefont {E.}~\bibnamefont {Bordo}}, \bibinfo {author} {\bibfnamefont {S.~N.}\ \bibnamefont {Chowdhury}}, \bibinfo {author} {\bibfnamefont {Y.}~\bibnamefont {Sharabi}}, \bibinfo {author} {\bibfnamefont {A.}~\bibnamefont {Fleischer}}, \bibinfo {author} {\bibfnamefont {A.}~\bibnamefont {Boltasseva}}, \bibinfo {author} {\bibfnamefont {O.}~\bibnamefont {Cohen}}, \bibinfo {author} {\bibfnamefont {V.~M.}\ \bibnamefont {Shalaev}},\ and\ \bibinfo {author} {\bibfnamefont {M.}~\bibnamefont {Segev}},\ }\bibfield  {title} {\bibinfo {title} {Time-refraction optics with single cycle modulation},\ }\href {https://doi.org/doi:10.1515/nanoph-2023-0126} {\bibfield  {journal} {\bibinfo  {journal} {Nanophotonics}\ }\textbf {\bibinfo {volume} {12}},\ \bibinfo {pages} {2221} (\bibinfo {year}
  {2023})}\BibitemShut {NoStop}%
\bibitem [{\citenamefont {Tirole}\ \emph {et~al.}(2024)\citenamefont {Tirole}, \citenamefont {Vezzoli}, \citenamefont {Saxena}, \citenamefont {Yang}, \citenamefont {Raziman}, \citenamefont {Galiffi}, \citenamefont {Maier}, \citenamefont {Pendry},\ and\ \citenamefont {Sapienza}}]{Tirole2024}%
  \BibitemOpen
  \bibfield  {author} {\bibinfo {author} {\bibfnamefont {R.}~\bibnamefont {Tirole}}, \bibinfo {author} {\bibfnamefont {S.}~\bibnamefont {Vezzoli}}, \bibinfo {author} {\bibfnamefont {D.}~\bibnamefont {Saxena}}, \bibinfo {author} {\bibfnamefont {S.}~\bibnamefont {Yang}}, \bibinfo {author} {\bibfnamefont {T.~V.}\ \bibnamefont {Raziman}}, \bibinfo {author} {\bibfnamefont {E.}~\bibnamefont {Galiffi}}, \bibinfo {author} {\bibfnamefont {S.~A.}\ \bibnamefont {Maier}}, \bibinfo {author} {\bibfnamefont {J.~B.}\ \bibnamefont {Pendry}},\ and\ \bibinfo {author} {\bibfnamefont {R.}~\bibnamefont {Sapienza}},\ }\bibfield  {title} {\bibinfo {title} {Second harmonic generation at a time-varying interface},\ }\href {https://doi.org/10.1038/s41467-024-51588-z} {\bibfield  {journal} {\bibinfo  {journal} {Nat. Commun.}\ }\textbf {\bibinfo {volume} {15}},\ \bibinfo {pages} {7752} (\bibinfo {year} {2024})}\BibitemShut {NoStop}%
\bibitem [{\citenamefont {Zurita-S\'anchez}\ \emph {et~al.}(2009)\citenamefont {Zurita-S\'anchez}, \citenamefont {Halevi},\ and\ \citenamefont {Cervantes-Gonz\'alez}}]{Zurita2009}%
  \BibitemOpen
  \bibfield  {author} {\bibinfo {author} {\bibfnamefont {J.~R.}\ \bibnamefont {Zurita-S\'anchez}}, \bibinfo {author} {\bibfnamefont {P.}~\bibnamefont {Halevi}},\ and\ \bibinfo {author} {\bibfnamefont {J.~C.}\ \bibnamefont {Cervantes-Gonz\'alez}},\ }\bibfield  {title} {\bibinfo {title} {{Reflection and transmission of a wave incident on a slab with a time-periodic dielectric function $\epsilon(t)$}},\ }\href {https://doi.org/10.1103/PhysRevA.79.053821} {\bibfield  {journal} {\bibinfo  {journal} {Phys. Rev. A}\ }\textbf {\bibinfo {volume} {79}},\ \bibinfo {pages} {053821} (\bibinfo {year} {2009})}\BibitemShut {NoStop}%
\bibitem [{\citenamefont {Asgari}\ \emph {et~al.}(2024)\citenamefont {Asgari}, \citenamefont {Garg}, \citenamefont {Wang}, \citenamefont {Mirmoosa}, \citenamefont {Rockstuhl},\ and\ \citenamefont {Asadchy}}]{Asgari2024}%
  \BibitemOpen
  \bibfield  {author} {\bibinfo {author} {\bibfnamefont {M.~M.}\ \bibnamefont {Asgari}}, \bibinfo {author} {\bibfnamefont {P.}~\bibnamefont {Garg}}, \bibinfo {author} {\bibfnamefont {X.}~\bibnamefont {Wang}}, \bibinfo {author} {\bibfnamefont {M.~S.}\ \bibnamefont {Mirmoosa}}, \bibinfo {author} {\bibfnamefont {C.}~\bibnamefont {Rockstuhl}},\ and\ \bibinfo {author} {\bibfnamefont {V.}~\bibnamefont {Asadchy}},\ }\bibfield  {title} {\bibinfo {title} {Theory and applications of photonic time crystals: a tutorial},\ }\href {https://doi.org/10.1364/AOP.525163} {\bibfield  {journal} {\bibinfo  {journal} {Adv. Opt. Photon.}\ }\textbf {\bibinfo {volume} {16}},\ \bibinfo {pages} {958} (\bibinfo {year} {2024})}\BibitemShut {NoStop}%
\bibitem [{\citenamefont {Sol\'{\i}s}\ and\ \citenamefont {Engheta}(2021)}]{Solis2021}%
  \BibitemOpen
  \bibfield  {author} {\bibinfo {author} {\bibfnamefont {D.~M.}\ \bibnamefont {Sol\'{\i}s}}\ and\ \bibinfo {author} {\bibfnamefont {N.}~\bibnamefont {Engheta}},\ }\bibfield  {title} {\bibinfo {title} {{Functional analysis of the polarization response in linear time-varying media: A generalization of the Kramers-Kronig relations}},\ }\href {https://doi.org/10.1103/PhysRevB.103.144303} {\bibfield  {journal} {\bibinfo  {journal} {Phys. Rev. B}\ }\textbf {\bibinfo {volume} {103}},\ \bibinfo {pages} {144303} (\bibinfo {year} {2021})}\BibitemShut {NoStop}%
\bibitem [{\citenamefont {Mirmoosa}\ \emph {et~al.}(2022)\citenamefont {Mirmoosa}, \citenamefont {Koutserimpas}, \citenamefont {Ptitcyn}, \citenamefont {Tretyakov},\ and\ \citenamefont {Fleury}}]{Mirmoosa2022}%
  \BibitemOpen
  \bibfield  {author} {\bibinfo {author} {\bibfnamefont {M.~S.}\ \bibnamefont {Mirmoosa}}, \bibinfo {author} {\bibfnamefont {T.}~\bibnamefont {Koutserimpas}}, \bibinfo {author} {\bibfnamefont {G.}~\bibnamefont {Ptitcyn}}, \bibinfo {author} {\bibfnamefont {S.}~\bibnamefont {Tretyakov}},\ and\ \bibinfo {author} {\bibfnamefont {R.}~\bibnamefont {Fleury}},\ }\bibfield  {title} {\bibinfo {title} {Dipole polarizability of time-varying particles},\ }\href {https://doi.org/10.1088/1367-2630/ac6b4c} {\bibfield  {journal} {\bibinfo  {journal} {New J. Phys.}\ }\textbf {\bibinfo {volume} {24}},\ \bibinfo {pages} {063004} (\bibinfo {year} {2022})}\BibitemShut {NoStop}%
\bibitem [{\citenamefont {Horsley}\ \emph {et~al.}(2023)\citenamefont {Horsley}, \citenamefont {Galiffi},\ and\ \citenamefont {Wang}}]{Horsley2023}%
  \BibitemOpen
  \bibfield  {author} {\bibinfo {author} {\bibfnamefont {S.~A.~R.}\ \bibnamefont {Horsley}}, \bibinfo {author} {\bibfnamefont {E.}~\bibnamefont {Galiffi}},\ and\ \bibinfo {author} {\bibfnamefont {Y.-T.}\ \bibnamefont {Wang}},\ }\bibfield  {title} {\bibinfo {title} {{Eigenpulses of Dispersive Time-Varying Media}},\ }\href {https://doi.org/10.1103/PhysRevLett.130.203803} {\bibfield  {journal} {\bibinfo  {journal} {Phys. Rev. Lett.}\ }\textbf {\bibinfo {volume} {130}},\ \bibinfo {pages} {203803} (\bibinfo {year} {2023})}\BibitemShut {NoStop}%
\bibitem [{\citenamefont {Sloan}\ \emph {et~al.}(2024)\citenamefont {Sloan}, \citenamefont {Rivera}, \citenamefont {Joannopoulos},\ and\ \citenamefont {Soljacic}}]{Sloan2024}%
  \BibitemOpen
  \bibfield  {author} {\bibinfo {author} {\bibfnamefont {J.}~\bibnamefont {Sloan}}, \bibinfo {author} {\bibfnamefont {N.}~\bibnamefont {Rivera}}, \bibinfo {author} {\bibfnamefont {J.~D.}\ \bibnamefont {Joannopoulos}},\ and\ \bibinfo {author} {\bibfnamefont {M.}~\bibnamefont {Soljacic}},\ }\bibfield  {title} {\bibinfo {title} {Optical properties of dispersive time-dependent materials},\ }\href {https://doi.org/10.1021/acsphotonics.3c00773} {\bibfield  {journal} {\bibinfo  {journal} {ACS Photonics}\ }\textbf {\bibinfo {volume} {11}},\ \bibinfo {pages} {950} (\bibinfo {year} {2024})}\BibitemShut {NoStop}%
\bibitem [{\citenamefont {Koutserimpas}\ and\ \citenamefont {Monticone}(2024)}]{Koutserimpas2024}%
  \BibitemOpen
  \bibfield  {author} {\bibinfo {author} {\bibfnamefont {T.~T.}\ \bibnamefont {Koutserimpas}}\ and\ \bibinfo {author} {\bibfnamefont {F.}~\bibnamefont {Monticone}},\ }\bibfield  {title} {\bibinfo {title} {{Time-varying media, dispersion, and the principle of causality [Invited]}},\ }\href {https://doi.org/10.1364/OME.515957} {\bibfield  {journal} {\bibinfo  {journal} {Opt. Mater. Express}\ }\textbf {\bibinfo {volume} {14}},\ \bibinfo {pages} {1222} (\bibinfo {year} {2024})}\BibitemShut {NoStop}%
\bibitem [{\citenamefont {Wang}\ \emph {et~al.}(2025)\citenamefont {Wang}, \citenamefont {Garg}, \citenamefont {Mirmoosa}, \citenamefont {Lamprianidis}, \citenamefont {Rockstuhl},\ and\ \citenamefont {Asadchy}}]{Wang2025}%
  \BibitemOpen
  \bibfield  {author} {\bibinfo {author} {\bibfnamefont {X.}~\bibnamefont {Wang}}, \bibinfo {author} {\bibfnamefont {P.}~\bibnamefont {Garg}}, \bibinfo {author} {\bibfnamefont {M.~S.}\ \bibnamefont {Mirmoosa}}, \bibinfo {author} {\bibfnamefont {A.~G.}\ \bibnamefont {Lamprianidis}}, \bibinfo {author} {\bibfnamefont {C.}~\bibnamefont {Rockstuhl}},\ and\ \bibinfo {author} {\bibfnamefont {V.~S.}\ \bibnamefont {Asadchy}},\ }\bibfield  {title} {\bibinfo {title} {Expanding momentum bandgaps in photonic time crystals through resonances},\ }\href {https://doi.org/10.1038/s41566-024-01563-3} {\bibfield  {journal} {\bibinfo  {journal} {Nat. Photonics}\ }\textbf {\bibinfo {volume} {19}},\ \bibinfo {pages} {149} (\bibinfo {year} {2025})}\BibitemShut {NoStop}%
\bibitem [{\citenamefont {Feng}\ \emph {et~al.}(2024)\citenamefont {Feng}, \citenamefont {Wang},\ and\ \citenamefont {Wang}}]{Feng2024}%
  \BibitemOpen
  \bibfield  {author} {\bibinfo {author} {\bibfnamefont {F.}~\bibnamefont {Feng}}, \bibinfo {author} {\bibfnamefont {N.}~\bibnamefont {Wang}},\ and\ \bibinfo {author} {\bibfnamefont {G.~P.}\ \bibnamefont {Wang}},\ }\bibfield  {title} {\bibinfo {title} {Temporal transfer matrix method for lorentzian dispersive time-varying media},\ }\href {https://doi.org/10.1063/5.0187485} {\bibfield  {journal} {\bibinfo  {journal} {Appl. Phys. Lett.}\ }\textbf {\bibinfo {volume} {124}},\ \bibinfo {pages} {101701} (\bibinfo {year} {2024})}\BibitemShut {NoStop}%
\bibitem [{\citenamefont {Ozlu}\ \emph {et~al.}(2025)\citenamefont {Ozlu}, \citenamefont {Mkhitaryan}, \citenamefont {Fruhling}, \citenamefont {Boltasseva},\ and\ \citenamefont {Shalaev}}]{Ozlu2025}%
  \BibitemOpen
  \bibfield  {author} {\bibinfo {author} {\bibfnamefont {M.~G.}\ \bibnamefont {Ozlu}}, \bibinfo {author} {\bibfnamefont {V.}~\bibnamefont {Mkhitaryan}}, \bibinfo {author} {\bibfnamefont {C.~B.}\ \bibnamefont {Fruhling}}, \bibinfo {author} {\bibfnamefont {A.}~\bibnamefont {Boltasseva}},\ and\ \bibinfo {author} {\bibfnamefont {V.~M.}\ \bibnamefont {Shalaev}},\ }\bibfield  {title} {\bibinfo {title} {Floquet engineering of polaritonic amplification in dispersive photonic time crystals},\ }\href {https://doi.org/10.1103/PhysRevResearch.7.023214} {\bibfield  {journal} {\bibinfo  {journal} {Phys. Rev. Res.}\ }\textbf {\bibinfo {volume} {7}},\ \bibinfo {pages} {023214} (\bibinfo {year} {2025})}\BibitemShut {NoStop}%
\bibitem [{\citenamefont {Wang}\ and\ \citenamefont {Chen}(2025)}]{Wang2025APL}%
  \BibitemOpen
  \bibfield  {author} {\bibinfo {author} {\bibfnamefont {Y.-T.}\ \bibnamefont {Wang}}\ and\ \bibinfo {author} {\bibfnamefont {Y.-H.}\ \bibnamefont {Chen}},\ }\bibfield  {title} {\bibinfo {title} {Generating large complete momentum gaps in temporally modulated dispersive media},\ }\href {https://doi.org/10.1063/5.0249939} {\bibfield  {journal} {\bibinfo  {journal} {Appl. Phys. Lett.}\ }\textbf {\bibinfo {volume} {126}},\ \bibinfo {pages} {221701} (\bibinfo {year} {2025})}\BibitemShut {NoStop}%
\bibitem [{\citenamefont {Dikopoltsev}\ \emph {et~al.}(2022)\citenamefont {Dikopoltsev}, \citenamefont {Sharabi}, \citenamefont {Lyubarov}, \citenamefont {Lumer}, \citenamefont {Tsesses}, \citenamefont {Lustig}, \citenamefont {Kaminer},\ and\ \citenamefont {Segev}}]{Dikopoltsev2022}%
  \BibitemOpen
  \bibfield  {author} {\bibinfo {author} {\bibfnamefont {A.}~\bibnamefont {Dikopoltsev}}, \bibinfo {author} {\bibfnamefont {Y.}~\bibnamefont {Sharabi}}, \bibinfo {author} {\bibfnamefont {M.}~\bibnamefont {Lyubarov}}, \bibinfo {author} {\bibfnamefont {Y.}~\bibnamefont {Lumer}}, \bibinfo {author} {\bibfnamefont {S.}~\bibnamefont {Tsesses}}, \bibinfo {author} {\bibfnamefont {E.}~\bibnamefont {Lustig}}, \bibinfo {author} {\bibfnamefont {I.}~\bibnamefont {Kaminer}},\ and\ \bibinfo {author} {\bibfnamefont {M.}~\bibnamefont {Segev}},\ }\bibfield  {title} {\bibinfo {title} {Light emission by free electrons in photonic time-crystals},\ }\href {https://doi.org/10.1073/pnas.2119705119} {\bibfield  {journal} {\bibinfo  {journal} {Proc. Natl. Acad. Sci. U.S.A.}\ }\textbf {\bibinfo {volume} {119}},\ \bibinfo {pages} {e2119705119} (\bibinfo {year} {2022})}\BibitemShut {NoStop}%
\bibitem [{\citenamefont {Li}\ \emph {et~al.}(2023)\citenamefont {Li}, \citenamefont {Yin}, \citenamefont {He}, \citenamefont {Xu}, \citenamefont {Al\`u},\ and\ \citenamefont {Shapiro}}]{Li2023}%
  \BibitemOpen
  \bibfield  {author} {\bibinfo {author} {\bibfnamefont {H.}~\bibnamefont {Li}}, \bibinfo {author} {\bibfnamefont {S.}~\bibnamefont {Yin}}, \bibinfo {author} {\bibfnamefont {H.}~\bibnamefont {He}}, \bibinfo {author} {\bibfnamefont {J.}~\bibnamefont {Xu}}, \bibinfo {author} {\bibfnamefont {A.}~\bibnamefont {Al\`u}},\ and\ \bibinfo {author} {\bibfnamefont {B.}~\bibnamefont {Shapiro}},\ }\bibfield  {title} {\bibinfo {title} {{Stationary Charge Radiation in Anisotropic Photonic Time Crystals}},\ }\href {https://doi.org/10.1103/PhysRevLett.130.093803} {\bibfield  {journal} {\bibinfo  {journal} {Phys. Rev. Lett.}\ }\textbf {\bibinfo {volume} {130}},\ \bibinfo {pages} {093803} (\bibinfo {year} {2023})}\BibitemShut {NoStop}%
\bibitem [{\citenamefont {Lyubarov}\ \emph {et~al.}(2022)\citenamefont {Lyubarov}, \citenamefont {Lumer}, \citenamefont {Dikopoltsev}, \citenamefont {Lustig}, \citenamefont {Sharabi},\ and\ \citenamefont {Segev}}]{Lyubarov2022}%
  \BibitemOpen
  \bibfield  {author} {\bibinfo {author} {\bibfnamefont {M.}~\bibnamefont {Lyubarov}}, \bibinfo {author} {\bibfnamefont {Y.}~\bibnamefont {Lumer}}, \bibinfo {author} {\bibfnamefont {A.}~\bibnamefont {Dikopoltsev}}, \bibinfo {author} {\bibfnamefont {E.}~\bibnamefont {Lustig}}, \bibinfo {author} {\bibfnamefont {Y.}~\bibnamefont {Sharabi}},\ and\ \bibinfo {author} {\bibfnamefont {M.}~\bibnamefont {Segev}},\ }\bibfield  {title} {\bibinfo {title} {Amplified emission and lasing in photonic time crystals},\ }\href {https://doi.org/10.1126/science.abo3324} {\bibfield  {journal} {\bibinfo  {journal} {Science}\ }\textbf {\bibinfo {volume} {377}},\ \bibinfo {pages} {425} (\bibinfo {year} {2022})}\BibitemShut {NoStop}%
\bibitem [{\citenamefont {Lyubarov}\ \emph {et~al.}(2024)\citenamefont {Lyubarov}, \citenamefont {Dikopoltsev}, \citenamefont {Segal}, \citenamefont {Plotnik},\ and\ \citenamefont {Segev}}]{Lyubarov2024}%
  \BibitemOpen
  \bibfield  {author} {\bibinfo {author} {\bibfnamefont {M.}~\bibnamefont {Lyubarov}}, \bibinfo {author} {\bibfnamefont {A.}~\bibnamefont {Dikopoltsev}}, \bibinfo {author} {\bibfnamefont {O.}~\bibnamefont {Segal}}, \bibinfo {author} {\bibfnamefont {Y.}~\bibnamefont {Plotnik}},\ and\ \bibinfo {author} {\bibfnamefont {M.}~\bibnamefont {Segev}},\ }\bibfield  {title} {\bibinfo {title} {Controlling spontaneous emission through the preparation of a photonic time-crystal},\ }\href {https://doi.org/10.1364/OE.539636} {\bibfield  {journal} {\bibinfo  {journal} {Opt. Express}\ }\textbf {\bibinfo {volume} {32}},\ \bibinfo {pages} {39734} (\bibinfo {year} {2024})}\BibitemShut {NoStop}%
\bibitem [{\citenamefont {Park}\ \emph {et~al.}(2025)\citenamefont {Park}, \citenamefont {Lee}, \citenamefont {Zhang}, \citenamefont {Park}, \citenamefont {Ryu}, \citenamefont {Cho}, \citenamefont {Lee}, \citenamefont {Zhang}, \citenamefont {Park}, \citenamefont {Jeon}, \citenamefont {Shin}, \citenamefont {Chan},\ and\ \citenamefont {Min}}]{Park2025}%
  \BibitemOpen
  \bibfield  {author} {\bibinfo {author} {\bibfnamefont {J.}~\bibnamefont {Park}}, \bibinfo {author} {\bibfnamefont {K.}~\bibnamefont {Lee}}, \bibinfo {author} {\bibfnamefont {R.-Y.}\ \bibnamefont {Zhang}}, \bibinfo {author} {\bibfnamefont {H.-C.}\ \bibnamefont {Park}}, \bibinfo {author} {\bibfnamefont {J.-W.}\ \bibnamefont {Ryu}}, \bibinfo {author} {\bibfnamefont {G.~Y.}\ \bibnamefont {Cho}}, \bibinfo {author} {\bibfnamefont {M.~Y.}\ \bibnamefont {Lee}}, \bibinfo {author} {\bibfnamefont {Z.}~\bibnamefont {Zhang}}, \bibinfo {author} {\bibfnamefont {N.}~\bibnamefont {Park}}, \bibinfo {author} {\bibfnamefont {W.}~\bibnamefont {Jeon}}, \bibinfo {author} {\bibfnamefont {J.}~\bibnamefont {Shin}}, \bibinfo {author} {\bibfnamefont {C.~T.}\ \bibnamefont {Chan}},\ and\ \bibinfo {author} {\bibfnamefont {B.}~\bibnamefont {Min}},\ }\bibfield  {title} {\bibinfo {title} {Spontaneous emission decay and excitation in photonic time crystals},\ }\href {https://doi.org/10.1103/5v2w-yg7v} {\bibfield  {journal} {\bibinfo
  {journal} {Phys. Rev. Lett.}\ }\textbf {\bibinfo {volume} {135}},\ \bibinfo {pages} {133801} (\bibinfo {year} {2025})}\BibitemShut {NoStop}%
\bibitem [{\citenamefont {Lee}\ \emph {et~al.}()\citenamefont {Lee}, \citenamefont {Kyung}, \citenamefont {Kim}, \citenamefont {Park}, \citenamefont {Lee}, \citenamefont {Choi}, \citenamefont {Chan}, \citenamefont {Shin}, \citenamefont {Kim},\ and\ \citenamefont {Min}}]{Lee_arxiv2025}%
  \BibitemOpen
  \bibfield  {author} {\bibinfo {author} {\bibfnamefont {K.}~\bibnamefont {Lee}}, \bibinfo {author} {\bibfnamefont {M.}~\bibnamefont {Kyung}}, \bibinfo {author} {\bibfnamefont {Y.}~\bibnamefont {Kim}}, \bibinfo {author} {\bibfnamefont {J.}~\bibnamefont {Park}}, \bibinfo {author} {\bibfnamefont {H.}~\bibnamefont {Lee}}, \bibinfo {author} {\bibfnamefont {J.}~\bibnamefont {Choi}}, \bibinfo {author} {\bibfnamefont {C.~T.}\ \bibnamefont {Chan}}, \bibinfo {author} {\bibfnamefont {J.}~\bibnamefont {Shin}}, \bibinfo {author} {\bibfnamefont {K.~W.}\ \bibnamefont {Kim}},\ and\ \bibinfo {author} {\bibfnamefont {B.}~\bibnamefont {Min}},\ }\bibfield  {title} {\bibinfo {title} {Spontaneous emission and lasing in photonic time crystals},\ }\Eprint {https://arxiv.org/abs/2507.19916} {arXiv:2507.19916} \BibitemShut {NoStop}%
\bibitem [{\citenamefont {Wang}\ \emph {et~al.}(2018)\citenamefont {Wang}, \citenamefont {Zhang},\ and\ \citenamefont {Chan}}]{Wang2018}%
  \BibitemOpen
  \bibfield  {author} {\bibinfo {author} {\bibfnamefont {N.}~\bibnamefont {Wang}}, \bibinfo {author} {\bibfnamefont {Z.-Q.}\ \bibnamefont {Zhang}},\ and\ \bibinfo {author} {\bibfnamefont {C.~T.}\ \bibnamefont {Chan}},\ }\bibfield  {title} {\bibinfo {title} {Photonic floquet media with a complex time-periodic permittivity},\ }\href {https://doi.org/10.1103/PhysRevB.98.085142} {\bibfield  {journal} {\bibinfo  {journal} {Phys. Rev. B}\ }\textbf {\bibinfo {volume} {98}},\ \bibinfo {pages} {085142} (\bibinfo {year} {2018})}\BibitemShut {NoStop}%
\bibitem [{\citenamefont {Kazemi}\ \emph {et~al.}(2019)\citenamefont {Kazemi}, \citenamefont {Nada}, \citenamefont {Mealy}, \citenamefont {Abdelshafy},\ and\ \citenamefont {Capolino}}]{Kazemi2019}%
  \BibitemOpen
  \bibfield  {author} {\bibinfo {author} {\bibfnamefont {H.}~\bibnamefont {Kazemi}}, \bibinfo {author} {\bibfnamefont {M.~Y.}\ \bibnamefont {Nada}}, \bibinfo {author} {\bibfnamefont {T.}~\bibnamefont {Mealy}}, \bibinfo {author} {\bibfnamefont {A.~F.}\ \bibnamefont {Abdelshafy}},\ and\ \bibinfo {author} {\bibfnamefont {F.}~\bibnamefont {Capolino}},\ }\bibfield  {title} {\bibinfo {title} {Exceptional points of degeneracy induced by linear time-periodic variation},\ }\href {https://doi.org/10.1103/PhysRevApplied.11.014007} {\bibfield  {journal} {\bibinfo  {journal} {Phys. Rev. Appl.}\ }\textbf {\bibinfo {volume} {11}},\ \bibinfo {pages} {014007} (\bibinfo {year} {2019})}\BibitemShut {NoStop}%
\bibitem [{\citenamefont {Wang}\ \emph {et~al.}(2024)\citenamefont {Wang}, \citenamefont {Hong},\ and\ \citenamefont {Wang}}]{Wang2024}%
  \BibitemOpen
  \bibfield  {author} {\bibinfo {author} {\bibfnamefont {N.}~\bibnamefont {Wang}}, \bibinfo {author} {\bibfnamefont {B.}~\bibnamefont {Hong}},\ and\ \bibinfo {author} {\bibfnamefont {G.~P.}\ \bibnamefont {Wang}},\ }\bibfield  {title} {\bibinfo {title} {Higher-order exceptional points and enhanced polarization sensitivity in anisotropic photonic time-floquet crystals},\ }\href {https://doi.org/10.1364/OE.539505} {\bibfield  {journal} {\bibinfo  {journal} {Opt. Express}\ }\textbf {\bibinfo {volume} {32}},\ \bibinfo {pages} {40092} (\bibinfo {year} {2024})}\BibitemShut {NoStop}%
\bibitem [{\citenamefont {Pick}\ \emph {et~al.}(2017)\citenamefont {Pick}, \citenamefont {Zhen}, \citenamefont {Miller}, \citenamefont {Hsu}, \citenamefont {Hernandez}, \citenamefont {Rodriguez}, \citenamefont {Solja\v{c}i\'{c}},\ and\ \citenamefont {Johnson}}]{Pick2017}%
  \BibitemOpen
  \bibfield  {author} {\bibinfo {author} {\bibfnamefont {A.}~\bibnamefont {Pick}}, \bibinfo {author} {\bibfnamefont {B.}~\bibnamefont {Zhen}}, \bibinfo {author} {\bibfnamefont {O.~D.}\ \bibnamefont {Miller}}, \bibinfo {author} {\bibfnamefont {C.~W.}\ \bibnamefont {Hsu}}, \bibinfo {author} {\bibfnamefont {F.}~\bibnamefont {Hernandez}}, \bibinfo {author} {\bibfnamefont {A.~W.}\ \bibnamefont {Rodriguez}}, \bibinfo {author} {\bibfnamefont {M.}~\bibnamefont {Solja\v{c}i\'{c}}},\ and\ \bibinfo {author} {\bibfnamefont {S.~G.}\ \bibnamefont {Johnson}},\ }\bibfield  {title} {\bibinfo {title} {General theory of spontaneous emission near exceptional points},\ }\href {https://doi.org/10.1364/OE.25.012325} {\bibfield  {journal} {\bibinfo  {journal} {Opt. Express}\ }\textbf {\bibinfo {volume} {25}},\ \bibinfo {pages} {12325} (\bibinfo {year} {2017})}\BibitemShut {NoStop}%
\bibitem [{\citenamefont {Pang}\ \emph {et~al.}(2021)\citenamefont {Pang}, \citenamefont {Alam}, \citenamefont {Zhou}, \citenamefont {Liu}, \citenamefont {Reshef}, \citenamefont {Manukyan}, \citenamefont {Voegtle}, \citenamefont {Pennathur}, \citenamefont {Tseng}, \citenamefont {Su} \emph {et~al.}}]{Pang2021}%
  \BibitemOpen
  \bibfield  {author} {\bibinfo {author} {\bibfnamefont {K.}~\bibnamefont {Pang}}, \bibinfo {author} {\bibfnamefont {M.~Z.}\ \bibnamefont {Alam}}, \bibinfo {author} {\bibfnamefont {Y.}~\bibnamefont {Zhou}}, \bibinfo {author} {\bibfnamefont {C.}~\bibnamefont {Liu}}, \bibinfo {author} {\bibfnamefont {O.}~\bibnamefont {Reshef}}, \bibinfo {author} {\bibfnamefont {K.}~\bibnamefont {Manukyan}}, \bibinfo {author} {\bibfnamefont {M.}~\bibnamefont {Voegtle}}, \bibinfo {author} {\bibfnamefont {A.}~\bibnamefont {Pennathur}}, \bibinfo {author} {\bibfnamefont {C.}~\bibnamefont {Tseng}}, \bibinfo {author} {\bibfnamefont {X.}~\bibnamefont {Su}}, \emph {et~al.},\ }\bibfield  {title} {\bibinfo {title} {Adiabatic frequency conversion using a time-varying epsilon-near-zero metasurface},\ }\href {https://doi.org/10.1021/acs.nanolett.1c00550} {\bibfield  {journal} {\bibinfo  {journal} {Nano Lett.}\ }\textbf {\bibinfo {volume} {21}},\ \bibinfo {pages} {5907} (\bibinfo {year} {2021})}\BibitemShut {NoStop}%
\bibitem [{\citenamefont {Guo}\ \emph {et~al.}()\citenamefont {Guo}, \citenamefont {Sueiro}, \citenamefont {Andolina}, \citenamefont {Levchuk}, \citenamefont {Ponzoni}, \citenamefont {Grasset}, \citenamefont {Monthe}, \citenamefont {Aupiais}, \citenamefont {Daineka}, \citenamefont {Briatico}, \citenamefont {de~Oliveira}, \citenamefont {Ponomaryov}, \citenamefont {Arshad}, \citenamefont {Karimbana-Kandy}, \citenamefont {Prajapati}, \citenamefont {Ilyakov}, \citenamefont {Deinert}, \citenamefont {Perfetti}, \citenamefont {Schiro},\ and\ \citenamefont {Laplace}}]{Guo_arxiv2025}%
  \BibitemOpen
  \bibfield  {author} {\bibinfo {author} {\bibfnamefont {T.}~\bibnamefont {Guo}}, \bibinfo {author} {\bibfnamefont {J.}~\bibnamefont {Sueiro}}, \bibinfo {author} {\bibfnamefont {G.~M.}\ \bibnamefont {Andolina}}, \bibinfo {author} {\bibfnamefont {A.}~\bibnamefont {Levchuk}}, \bibinfo {author} {\bibfnamefont {S.}~\bibnamefont {Ponzoni}}, \bibinfo {author} {\bibfnamefont {R.}~\bibnamefont {Grasset}}, \bibinfo {author} {\bibfnamefont {D.}~\bibnamefont {Monthe}}, \bibinfo {author} {\bibfnamefont {I.}~\bibnamefont {Aupiais}}, \bibinfo {author} {\bibfnamefont {D.}~\bibnamefont {Daineka}}, \bibinfo {author} {\bibfnamefont {J.}~\bibnamefont {Briatico}}, \bibinfo {author} {\bibfnamefont {T.~V.}\ \bibnamefont {de~Oliveira}}, \bibinfo {author} {\bibfnamefont {A.}~\bibnamefont {Ponomaryov}}, \bibinfo {author} {\bibfnamefont {A.}~\bibnamefont {Arshad}}, \bibinfo {author} {\bibfnamefont {A.}~\bibnamefont {Karimbana-Kandy}}, \bibinfo {author} {\bibfnamefont {G.~L.}\ \bibnamefont {Prajapati}}, \bibinfo {author} {\bibfnamefont
  {I.}~\bibnamefont {Ilyakov}}, \bibinfo {author} {\bibfnamefont {J.-C.}\ \bibnamefont {Deinert}}, \bibinfo {author} {\bibfnamefont {L.}~\bibnamefont {Perfetti}}, \bibinfo {author} {\bibfnamefont {M.}~\bibnamefont {Schiro}},\ and\ \bibinfo {author} {\bibfnamefont {Y.}~\bibnamefont {Laplace}},\ }\bibfield  {title} {\bibinfo {title} {Plasmonic metamaterial time crystal},\ }\Eprint {https://arxiv.org/abs/2510.02845} {arXiv:2510.02845} \BibitemShut {NoStop}%
\bibitem [{\citenamefont {Cartella}\ \emph {et~al.}(2018)\citenamefont {Cartella}, \citenamefont {Nova}, \citenamefont {Fechner}, \citenamefont {Merlin},\ and\ \citenamefont {Cavalleri}}]{Cartella2018}%
  \BibitemOpen
  \bibfield  {author} {\bibinfo {author} {\bibfnamefont {A.}~\bibnamefont {Cartella}}, \bibinfo {author} {\bibfnamefont {T.~F.}\ \bibnamefont {Nova}}, \bibinfo {author} {\bibfnamefont {M.}~\bibnamefont {Fechner}}, \bibinfo {author} {\bibfnamefont {R.}~\bibnamefont {Merlin}},\ and\ \bibinfo {author} {\bibfnamefont {A.}~\bibnamefont {Cavalleri}},\ }\bibfield  {title} {\bibinfo {title} {Parametric amplification of optical phonons},\ }\href {https://doi.org/10.1073/pnas.1809725115} {\bibfield  {journal} {\bibinfo  {journal} {Proc. Natl. Acad. Sci. U.S.A.}\ }\textbf {\bibinfo {volume} {115}},\ \bibinfo {pages} {12148} (\bibinfo {year} {2018})}\BibitemShut {NoStop}%
\bibitem [{\citenamefont {Sloan}\ \emph {et~al.}(2021)\citenamefont {Sloan}, \citenamefont {Rivera}, \citenamefont {Joannopoulos},\ and\ \citenamefont {Solja\ifmmode \check{c}\else \v{c}\fi{}i\ifmmode~\acute{c}\else \'{c}\fi{}}}]{Sloan2021}%
  \BibitemOpen
  \bibfield  {author} {\bibinfo {author} {\bibfnamefont {J.}~\bibnamefont {Sloan}}, \bibinfo {author} {\bibfnamefont {N.}~\bibnamefont {Rivera}}, \bibinfo {author} {\bibfnamefont {J.~D.}\ \bibnamefont {Joannopoulos}},\ and\ \bibinfo {author} {\bibfnamefont {M.}~\bibnamefont {Solja\ifmmode \check{c}\else \v{c}\fi{}i\ifmmode~\acute{c}\else \'{c}\fi{}}},\ }\bibfield  {title} {\bibinfo {title} {{Casimir Light in Dispersive Nanophotonics}},\ }\href {https://doi.org/10.1103/PhysRevLett.127.053603} {\bibfield  {journal} {\bibinfo  {journal} {Phys. Rev. Lett.}\ }\textbf {\bibinfo {volume} {127}},\ \bibinfo {pages} {053603} (\bibinfo {year} {2021})}\BibitemShut {NoStop}%
\bibitem [{\citenamefont {Park}\ \emph {et~al.}(2022)\citenamefont {Park}, \citenamefont {Cho}, \citenamefont {Lee}, \citenamefont {Lee}, \citenamefont {Lee}, \citenamefont {Park}, \citenamefont {Ryu}, \citenamefont {Park}, \citenamefont {Jeon},\ and\ \citenamefont {Min}}]{Park2022}%
  \BibitemOpen
  \bibfield  {author} {\bibinfo {author} {\bibfnamefont {J.}~\bibnamefont {Park}}, \bibinfo {author} {\bibfnamefont {H.}~\bibnamefont {Cho}}, \bibinfo {author} {\bibfnamefont {S.}~\bibnamefont {Lee}}, \bibinfo {author} {\bibfnamefont {K.}~\bibnamefont {Lee}}, \bibinfo {author} {\bibfnamefont {K.}~\bibnamefont {Lee}}, \bibinfo {author} {\bibfnamefont {H.~C.}\ \bibnamefont {Park}}, \bibinfo {author} {\bibfnamefont {J.-W.}\ \bibnamefont {Ryu}}, \bibinfo {author} {\bibfnamefont {N.}~\bibnamefont {Park}}, \bibinfo {author} {\bibfnamefont {S.}~\bibnamefont {Jeon}},\ and\ \bibinfo {author} {\bibfnamefont {B.}~\bibnamefont {Min}},\ }\bibfield  {title} {\bibinfo {title} {Revealing non-hermitian band structure of photonic floquet media},\ }\href {https://doi.org/10.1126/sciadv.abo6220} {\bibfield  {journal} {\bibinfo  {journal} {Sci. Adv.}\ }\textbf {\bibinfo {volume} {8}},\ \bibinfo {pages} {eabo6220} (\bibinfo {year} {2022})}\BibitemShut {NoStop}%
\bibitem [{\citenamefont {Raman}\ and\ \citenamefont {Fan}(2010)}]{Raman2010}%
  \BibitemOpen
  \bibfield  {author} {\bibinfo {author} {\bibfnamefont {A.}~\bibnamefont {Raman}}\ and\ \bibinfo {author} {\bibfnamefont {S.}~\bibnamefont {Fan}},\ }\bibfield  {title} {\bibinfo {title} {{Photonic Band Structure of Dispersive Metamaterials Formulated as a Hermitian Eigenvalue Problem}},\ }\href {https://doi.org/10.1103/PhysRevLett.104.087401} {\bibfield  {journal} {\bibinfo  {journal} {Phys. Rev. Lett.}\ }\textbf {\bibinfo {volume} {104}},\ \bibinfo {pages} {087401} (\bibinfo {year} {2010})}\BibitemShut {NoStop}%
\bibitem [{Sup()}]{SupplementalMaterial}%
  \BibitemOpen
  \href@noop {} {\bibinfo {title} {{See Supplemental Material at [URL] for details on our formalism and complementary results, which includes Refs.~[57-60].}}}\BibitemShut {Stop}%
\bibitem [{\citenamefont {Barnett}\ \emph {et~al.}(1992)\citenamefont {Barnett}, \citenamefont {Huttner},\ and\ \citenamefont {Loudon}}]{Barnett1992}%
  \BibitemOpen
  \bibfield  {author} {\bibinfo {author} {\bibfnamefont {S.~M.}\ \bibnamefont {Barnett}}, \bibinfo {author} {\bibfnamefont {B.}~\bibnamefont {Huttner}},\ and\ \bibinfo {author} {\bibfnamefont {R.}~\bibnamefont {Loudon}},\ }\bibfield  {title} {\bibinfo {title} {Spontaneous emission in absorbing dielectric media},\ }\href {https://doi.org/10.1103/PhysRevLett.68.3698} {\bibfield  {journal} {\bibinfo  {journal} {Phys. Rev. Lett.}\ }\textbf {\bibinfo {volume} {68}},\ \bibinfo {pages} {3698} (\bibinfo {year} {1992})}\BibitemShut {NoStop}%
\bibitem [{\citenamefont {Horsley}\ and\ \citenamefont {Philbin}(2014)}]{Horsley2014}%
  \BibitemOpen
  \bibfield  {author} {\bibinfo {author} {\bibfnamefont {S.~A.~R.}\ \bibnamefont {Horsley}}\ and\ \bibinfo {author} {\bibfnamefont {T.~G.}\ \bibnamefont {Philbin}},\ }\bibfield  {title} {\bibinfo {title} {Canonical quantization of electromagnetism in spatially dispersive media},\ }\href {https://doi.org/10.1088/1367-2630/16/1/013030} {\bibfield  {journal} {\bibinfo  {journal} {New J. Phys.}\ }\textbf {\bibinfo {volume} {16}},\ \bibinfo {pages} {013030} (\bibinfo {year} {2014})}\BibitemShut {NoStop}%
\bibitem [{\citenamefont {Thouin}\ \emph {et~al.}(2025)\citenamefont {Thouin}, \citenamefont {Myers}, \citenamefont {Patri}, \citenamefont {Baloukas}, \citenamefont {Martinu}, \citenamefont {Fern{\'a}ndez-Dom{\'i}nguez},\ and\ \citenamefont {K{\'e}na-Cohen}}]{Thouin2025}%
  \BibitemOpen
  \bibfield  {author} {\bibinfo {author} {\bibfnamefont {F.}~\bibnamefont {Thouin}}, \bibinfo {author} {\bibfnamefont {D.~M.}\ \bibnamefont {Myers}}, \bibinfo {author} {\bibfnamefont {A.}~\bibnamefont {Patri}}, \bibinfo {author} {\bibfnamefont {B.}~\bibnamefont {Baloukas}}, \bibinfo {author} {\bibfnamefont {L.}~\bibnamefont {Martinu}}, \bibinfo {author} {\bibfnamefont {A.~I.}\ \bibnamefont {Fern{\'a}ndez-Dom{\'i}nguez}},\ and\ \bibinfo {author} {\bibfnamefont {S.}~\bibnamefont {K{\'e}na-Cohen}},\ }\bibfield  {title} {\bibinfo {title} {Field enhancement and nonlocal effects in epsilon-near-zero photonic gap antennas},\ }\href {https://doi.org/10.1021/acsnano.4c15531} {\bibfield  {journal} {\bibinfo  {journal} {ACS Nano}\ }\textbf {\bibinfo {volume} {19}},\ \bibinfo {pages} {7996} (\bibinfo {year} {2025})}\BibitemShut {NoStop}%
\bibitem [{\citenamefont {Sun}\ \emph {et~al.}(2025)\citenamefont {Sun}, \citenamefont {Fan},\ and\ \citenamefont {Hu}}]{Sun2025}%
  \BibitemOpen
  \bibfield  {author} {\bibinfo {author} {\bibfnamefont {Y.}~\bibnamefont {Sun}}, \bibinfo {author} {\bibfnamefont {S.}~\bibnamefont {Fan}},\ and\ \bibinfo {author} {\bibfnamefont {G.}~\bibnamefont {Hu}},\ }\bibfield  {title} {\bibinfo {title} {{Formulation of Dispersive and Dissipative Time-Varying Media as a Floquet Matrix Eigenproblem}},\ }\href {https://doi.org/10.1103/lcmw-qrrp} {\bibfield  {journal} {\bibinfo  {journal} {Phys. Rev. Lett.}\ }\textbf {\bibinfo {volume} {135}},\ \bibinfo {pages} {156903} (\bibinfo {year} {2025})}\BibitemShut {NoStop}%
\bibitem [{\citenamefont {Zhang}\ \emph {et~al.}(2024)\citenamefont {Zhang}, \citenamefont {Dong}, \citenamefont {Li}, \citenamefont {Xu},\ and\ \citenamefont {Shapiro}}]{Zhang2024}%
  \BibitemOpen
  \bibfield  {author} {\bibinfo {author} {\bibfnamefont {S.}~\bibnamefont {Zhang}}, \bibinfo {author} {\bibfnamefont {J.}~\bibnamefont {Dong}}, \bibinfo {author} {\bibfnamefont {H.}~\bibnamefont {Li}}, \bibinfo {author} {\bibfnamefont {J.}~\bibnamefont {Xu}},\ and\ \bibinfo {author} {\bibfnamefont {B.}~\bibnamefont {Shapiro}},\ }\bibfield  {title} {\bibinfo {title} {Longitudinal optical phonons in photonic time crystals containing a stationary charge},\ }\href {https://doi.org/10.1103/PhysRevB.110.L100306} {\bibfield  {journal} {\bibinfo  {journal} {Phys. Rev. B}\ }\textbf {\bibinfo {volume} {110}},\ \bibinfo {pages} {L100306} (\bibinfo {year} {2024})}\BibitemShut {NoStop}%
\bibitem [{\citenamefont {Feinberg}\ \emph {et~al.}(2025)\citenamefont {Feinberg}, \citenamefont {Fernandes}, \citenamefont {Shapiro},\ and\ \citenamefont {Silveirinha}}]{Feinberg2025}%
  \BibitemOpen
  \bibfield  {author} {\bibinfo {author} {\bibfnamefont {J.}~\bibnamefont {Feinberg}}, \bibinfo {author} {\bibfnamefont {D.~E.}\ \bibnamefont {Fernandes}}, \bibinfo {author} {\bibfnamefont {B.}~\bibnamefont {Shapiro}},\ and\ \bibinfo {author} {\bibfnamefont {M.~G.}\ \bibnamefont {Silveirinha}},\ }\bibfield  {title} {\bibinfo {title} {{Plasmonic Time Crystals}},\ }\href {https://doi.org/10.1103/PhysRevLett.134.183801} {\bibfield  {journal} {\bibinfo  {journal} {Phys. Rev. Lett.}\ }\textbf {\bibinfo {volume} {134}},\ \bibinfo {pages} {183801} (\bibinfo {year} {2025})}\BibitemShut {NoStop}%
\bibitem [{\citenamefont {Chamanara}\ \emph {et~al.}(2018)\citenamefont {Chamanara}, \citenamefont {Deck-L\'eger}, \citenamefont {Caloz},\ and\ \citenamefont {Kalluri}}]{Chamanara2018}%
  \BibitemOpen
  \bibfield  {author} {\bibinfo {author} {\bibfnamefont {N.}~\bibnamefont {Chamanara}}, \bibinfo {author} {\bibfnamefont {Z.-L.}\ \bibnamefont {Deck-L\'eger}}, \bibinfo {author} {\bibfnamefont {C.}~\bibnamefont {Caloz}},\ and\ \bibinfo {author} {\bibfnamefont {D.}~\bibnamefont {Kalluri}},\ }\bibfield  {title} {\bibinfo {title} {Unusual electromagnetic modes in space-time-modulated dispersion-engineered media},\ }\href {https://doi.org/10.1103/PhysRevA.97.063829} {\bibfield  {journal} {\bibinfo  {journal} {Phys. Rev. A}\ }\textbf {\bibinfo {volume} {97}},\ \bibinfo {pages} {063829} (\bibinfo {year} {2018})}\BibitemShut {NoStop}%
\bibitem [{\citenamefont {Galiffi}\ \emph {et~al.}(2022{\natexlab{b}})\citenamefont {Galiffi}, \citenamefont {Huidobro},\ and\ \citenamefont {Pendry}}]{Galiffi2022NatCommun}%
  \BibitemOpen
  \bibfield  {author} {\bibinfo {author} {\bibfnamefont {E.}~\bibnamefont {Galiffi}}, \bibinfo {author} {\bibfnamefont {P.~A.}\ \bibnamefont {Huidobro}},\ and\ \bibinfo {author} {\bibfnamefont {J.}~\bibnamefont {Pendry}},\ }\bibfield  {title} {\bibinfo {title} {{An Archimedes' screw for light}},\ }\href {https://doi.org/https://doi.org/10.1038/s41467-022-30079-z} {\bibfield  {journal} {\bibinfo  {journal} {Nat. Commun.}\ }\textbf {\bibinfo {volume} {13}},\ \bibinfo {pages} {2523} (\bibinfo {year} {2022}{\natexlab{b}})}\BibitemShut {NoStop}%
\bibitem [{\citenamefont {Dong}\ \emph {et~al.}(2025)\citenamefont {Dong}, \citenamefont {Zhang}, \citenamefont {He}, \citenamefont {Li},\ and\ \citenamefont {Xu}}]{Dong2025}%
  \BibitemOpen
  \bibfield  {author} {\bibinfo {author} {\bibfnamefont {J.}~\bibnamefont {Dong}}, \bibinfo {author} {\bibfnamefont {S.}~\bibnamefont {Zhang}}, \bibinfo {author} {\bibfnamefont {H.}~\bibnamefont {He}}, \bibinfo {author} {\bibfnamefont {H.}~\bibnamefont {Li}},\ and\ \bibinfo {author} {\bibfnamefont {J.}~\bibnamefont {Xu}},\ }\bibfield  {title} {\bibinfo {title} {{Nonuniform Wave Momentum Band Gap in Biaxial Anisotropic Photonic Time Crystals}},\ }\href {https://doi.org/10.1103/PhysRevLett.134.063801} {\bibfield  {journal} {\bibinfo  {journal} {Phys. Rev. Lett.}\ }\textbf {\bibinfo {volume} {134}},\ \bibinfo {pages} {063801} (\bibinfo {year} {2025})}\BibitemShut {NoStop}%
\bibitem [{\citenamefont {Wiersig}(2023)}]{Wiersig2023}%
  \BibitemOpen
  \bibfield  {author} {\bibinfo {author} {\bibfnamefont {J.}~\bibnamefont {Wiersig}},\ }\bibfield  {title} {\bibinfo {title} {Petermann factors and phase rigidities near exceptional points},\ }\href {https://doi.org/10.1103/PhysRevResearch.5.033042} {\bibfield  {journal} {\bibinfo  {journal} {Phys. Rev. Res.}\ }\textbf {\bibinfo {volume} {5}},\ \bibinfo {pages} {033042} (\bibinfo {year} {2023})}\BibitemShut {NoStop}%
\bibitem [{\citenamefont {Ren}\ \emph {et~al.}(2024)\citenamefont {Ren}, \citenamefont {Franke}, \citenamefont {VanDrunen},\ and\ \citenamefont {Hughes}}]{Ren2024}%
  \BibitemOpen
  \bibfield  {author} {\bibinfo {author} {\bibfnamefont {J.}~\bibnamefont {Ren}}, \bibinfo {author} {\bibfnamefont {S.}~\bibnamefont {Franke}}, \bibinfo {author} {\bibfnamefont {B.}~\bibnamefont {VanDrunen}},\ and\ \bibinfo {author} {\bibfnamefont {S.}~\bibnamefont {Hughes}},\ }\bibfield  {title} {\bibinfo {title} {Classical purcell factors and spontaneous emission decay rates in a linear gain medium},\ }\href {https://doi.org/10.1103/PhysRevA.109.013513} {\bibfield  {journal} {\bibinfo  {journal} {Phys. Rev. A}\ }\textbf {\bibinfo {volume} {109}},\ \bibinfo {pages} {013513} (\bibinfo {year} {2024})}\BibitemShut {NoStop}%
\bibitem [{Note1()}]{Note1}%
  \BibitemOpen
  \bibinfo {note} {While these poles lead to an ill-defined dissipated power, we only approach them in all of our computations to keep convergent results.}\BibitemShut {Stop}%
\bibitem [{\citenamefont {Sustaeta-Osuna}\ \emph {et~al.}(2026)\citenamefont {Sustaeta-Osuna}, \citenamefont {Allard}, \citenamefont {García-Vidal},\ and\ \citenamefont {Huidobro}}]{SustaetaOsuna_arxiv2025}%
  \BibitemOpen
  \bibfield  {author} {\bibinfo {author} {\bibfnamefont {J.~E.}\ \bibnamefont {Sustaeta-Osuna}}, \bibinfo {author} {\bibfnamefont {T.~F.}\ \bibnamefont {Allard}}, \bibinfo {author} {\bibfnamefont {F.~J.}\ \bibnamefont {García-Vidal}},\ and\ \bibinfo {author} {\bibfnamefont {P.~A.}\ \bibnamefont {Huidobro}},\ }\bibfield  {title} {\bibinfo {title} {Near-field gain and far-field control via a plasmonic time crystal slab},\ }\href {https://doi.org/10.1103/lfr1-dwlv} {\bibfield  {journal} {\bibinfo  {journal} {Phys. Rev. Lett., In Press}\ } (\bibinfo {year} {2026})}\BibitemShut {NoStop}%
\bibitem [{\citenamefont {Milonni}(1984)}]{Milonni1984}%
  \BibitemOpen
  \bibfield  {author} {\bibinfo {author} {\bibfnamefont {P.~W.}\ \bibnamefont {Milonni}},\ }\bibfield  {title} {\bibinfo {title} {{Why spontaneous emission?}},\ }\href {https://doi.org/10.1119/1.13886} {\bibfield  {journal} {\bibinfo  {journal} {Am. J. Phys.}\ }\textbf {\bibinfo {volume} {52}},\ \bibinfo {pages} {340} (\bibinfo {year} {1984})}\BibitemShut {NoStop}%
\bibitem [{\citenamefont {Mendon\ifmmode~\mbox{\c{c}}\else \c{c}\fi{}a}\ \emph {et~al.}(2000)\citenamefont {Mendon\ifmmode~\mbox{\c{c}}\else \c{c}\fi{}a}, \citenamefont {Guerreiro},\ and\ \citenamefont {Martins}}]{Mendonça2000}%
  \BibitemOpen
  \bibfield  {author} {\bibinfo {author} {\bibfnamefont {J.~T.}\ \bibnamefont {Mendon\ifmmode~\mbox{\c{c}}\else \c{c}\fi{}a}}, \bibinfo {author} {\bibfnamefont {A.}~\bibnamefont {Guerreiro}},\ and\ \bibinfo {author} {\bibfnamefont {A.~M.}\ \bibnamefont {Martins}},\ }\bibfield  {title} {\bibinfo {title} {Quantum theory of time refraction},\ }\href {https://doi.org/10.1103/PhysRevA.62.033805} {\bibfield  {journal} {\bibinfo  {journal} {Phys. Rev. A}\ }\textbf {\bibinfo {volume} {62}},\ \bibinfo {pages} {033805} (\bibinfo {year} {2000})}\BibitemShut {NoStop}%
\bibitem [{\citenamefont {Ganfornina-Andrades}\ \emph {et~al.}(2024)\citenamefont {Ganfornina-Andrades}, \citenamefont {V\'azquez-Lozano},\ and\ \citenamefont {Liberal}}]{Ganfornina2024}%
  \BibitemOpen
  \bibfield  {author} {\bibinfo {author} {\bibfnamefont {A.}~\bibnamefont {Ganfornina-Andrades}}, \bibinfo {author} {\bibfnamefont {J.~E.}\ \bibnamefont {V\'azquez-Lozano}},\ and\ \bibinfo {author} {\bibfnamefont {I.}~\bibnamefont {Liberal}},\ }\bibfield  {title} {\bibinfo {title} {Quantum vacuum amplification in time-varying media with arbitrary temporal profiles},\ }\href {https://doi.org/10.1103/PhysRevResearch.6.043320} {\bibfield  {journal} {\bibinfo  {journal} {Phys. Rev. Res.}\ }\textbf {\bibinfo {volume} {6}},\ \bibinfo {pages} {043320} (\bibinfo {year} {2024})}\BibitemShut {NoStop}%
\bibitem [{\citenamefont {Mirmoosa}\ \emph {et~al.}(2025)\citenamefont {Mirmoosa}, \citenamefont {Set\"al\"a},\ and\ \citenamefont {Norrman}}]{Mirmoosa2025}%
  \BibitemOpen
  \bibfield  {author} {\bibinfo {author} {\bibfnamefont {M.~S.}\ \bibnamefont {Mirmoosa}}, \bibinfo {author} {\bibfnamefont {T.}~\bibnamefont {Set\"al\"a}},\ and\ \bibinfo {author} {\bibfnamefont {A.}~\bibnamefont {Norrman}},\ }\bibfield  {title} {\bibinfo {title} {Quantum state engineering and photon statistics at electromagnetic time interfaces},\ }\href {https://doi.org/10.1103/PhysRevResearch.7.013120} {\bibfield  {journal} {\bibinfo  {journal} {Phys. Rev. Res.}\ }\textbf {\bibinfo {volume} {7}},\ \bibinfo {pages} {013120} (\bibinfo {year} {2025})}\BibitemShut {NoStop}%
\bibitem [{\citenamefont {Sustaeta-Osuna}\ \emph {et~al.}(2025)\citenamefont {Sustaeta-Osuna}, \citenamefont {Garc{\'i}a-Vidal},\ and\ \citenamefont {Huidobro}}]{Sustaeta-Osuna2025}%
  \BibitemOpen
  \bibfield  {author} {\bibinfo {author} {\bibfnamefont {J.~E.}\ \bibnamefont {Sustaeta-Osuna}}, \bibinfo {author} {\bibfnamefont {F.~J.}\ \bibnamefont {Garc{\'i}a-Vidal}},\ and\ \bibinfo {author} {\bibfnamefont {P.~A.}\ \bibnamefont {Huidobro}},\ }\bibfield  {title} {\bibinfo {title} {Quantum theory of photon pair creation in photonic time crystals},\ }\href {https://doi.org/10.1021/acsphotonics.4c02293} {\bibfield  {journal} {\bibinfo  {journal} {ACS Photonics}\ }\textbf {\bibinfo {volume} {12}},\ \bibinfo {pages} {1873} (\bibinfo {year} {2025})}\BibitemShut {NoStop}%
\bibitem [{\citenamefont {Bae}\ \emph {et~al.}(2025)\citenamefont {Bae}, \citenamefont {Lee}, \citenamefont {Min},\ and\ \citenamefont {Kim}}]{Bae2025}%
  \BibitemOpen
  \bibfield  {author} {\bibinfo {author} {\bibfnamefont {J.}~\bibnamefont {Bae}}, \bibinfo {author} {\bibfnamefont {K.}~\bibnamefont {Lee}}, \bibinfo {author} {\bibfnamefont {B.}~\bibnamefont {Min}},\ and\ \bibinfo {author} {\bibfnamefont {K.~W.}\ \bibnamefont {Kim}},\ }\bibfield  {title} {\bibinfo {title} {Quantum electrodynamics of photonic time crystals},\ }\href {https://doi.org/10.1038/s41467-025-67572-0} {\bibfield  {journal} {\bibinfo  {journal} {Nature Communications}\ }\textbf {\bibinfo {volume} {17}},\ \bibinfo {pages} {858} (\bibinfo {year} {2025})}\BibitemShut {NoStop}%
\bibitem [{\citenamefont {Allard}\ \emph {et~al.}()\citenamefont {Allard}, \citenamefont {Sustaeta-Osuna}, \citenamefont {García-Vidal},\ and\ \citenamefont {A.~Huidobro}}]{Zenodo}%
  \BibitemOpen
  \bibfield  {author} {\bibinfo {author} {\bibfnamefont {T.~F.}\ \bibnamefont {Allard}}, \bibinfo {author} {\bibfnamefont {J.~E.}\ \bibnamefont {Sustaeta-Osuna}}, \bibinfo {author} {\bibfnamefont {F.~J.}\ \bibnamefont {García-Vidal}},\ and\ \bibinfo {author} {\bibfnamefont {P.}~\bibnamefont {A.~Huidobro}},\ }\bibfield  {title} {\bibinfo {title} {{Dataset for ``Broadband Dipole Absorption in Dispersive Photonic Time Crystals"}},\ }\href {https://doi.org/10.5281/zenodo.18483040} {10.5281/zenodo.18483040}\BibitemShut {NoStop}%
\end{thebibliography}%

\end{document}


\beginsupplement

\title{Supplemental Material: Broadband Dipole Absorption in Dispersive Photonic Time Crystals}

\author{Thomas F.\ Allard}
\author{Jaime E.\ Sustaeta-Osuna}
\author{Francisco J.\ Garc\'{i}a-Vidal}
\author{Paloma A.\ Huidobro}
\affiliation{Departamento de F\'{i}sica Te\'{o}rica de la Materia Condensada and Condensed Matter Physics Center (IFIMAC), Universidad Aut\'{o}noma de Madrid, E-28049 Madrid, Spain}
\affiliation{Condensed Matter Physics Center (IFIMAC), Universidad Aut\'{o}noma de Madrid, E-28049 Madrid, Spain}

\maketitle

\tableofcontents

\section{Physical model}

\subsection{Setting the problem}

We consider a three-dimensional dispersive and lossy nonmagnetic unbounded medium whose electromagnetic fields are described by the usual Maxwell's equations
$\nabla \times \mathbf{E} = -\mu_0 \partial_t \mathbf{H}$
and
$\nabla \times \mathbf{H} = \partial_t \mathbf{D} + \mathbf{J}$,
together with the constitutive relations
$\mathbf{D} = \epsilon_0 \mathbf{E} + \mathbf{P}$
and
$\mathbf{H} = \frac{1}{\mu_0} \mathbf{B}$.
In this supplemental material, we assume the polarization density $\mathbf{P}$ to be described by a general parametric hydrodynamic Drude-Lorentz model, so that its dynamics follows the partial differential equation
\begin{equation}
    \partial^2_t \mathbf{P} + \gamma \partial_t \mathbf{P} + \omega_0^2 (t)\mathbf{P} + \beta^2 \nabla \rho = \epsilon_0 \omega_{\mathrm{p}}^2(t) \mathbf{E}.
    \label{eq:Lorentz}
\end{equation}
Here, $\gamma$, $\omega_0$ and $\omega_\mathrm{p}$ are, respectively, the inverse relaxation time, resonance and plasma frequencies of the material.
Moreover, $\beta$ is the hydrodynamic parameter that quantifies the nonlocal effects, and $\rho=-[\nabla\cdot\mathbf{P}]$ is the charge density.
In our model, such a nonlocality impacts only longitudinal modes, so that we did not discuss it in the main text, where our focus was on transverse modes.
Here, we keep this term as we will discuss the amplification of longitudinal modes in Sec.~\ref{sec:Longitudinal modes}.
In this supplemental material we consider the most general case of periodically varied in time plasma and resonance frequencies as $\omega_{\mathrm{p}/0}^2(t) = \omega_{\mathrm{p}/0}^2[1+\alpha_{\mathrm{p}/0}\sin(\Omega t)]$, with $\Omega$ the modulation frequency and $\alpha_{\mathrm{p}/0}$ the modulation strength.
We note that in the main text we have set either $\alpha_{\mathrm{p}}$ or $\alpha_0$ to zero, and removed the subindex in the remaining nonzero modulation strength $\alpha_{\mathrm{p}/0}\equiv\alpha$.

To solve this system of partial differential equations, we follow the work of Raman and Fan \cite{Raman2010} and reformulate it into a Schrödinger-like matrix problem using the auxiliary field $\mathbf{\dot P} = \partial_t \mathbf{P}$
\begin{equation}
    i \partial_t \begin{pmatrix}
                \mathbf{E} \\
                \mathbf{H} \\
                \mathbf{P} \\
                \mathbf{\dot P}
                \end{pmatrix}
    = \begin{pmatrix}
    \mathbb{0}_3 & \frac{i}{\epsilon_0}\nabla \times & \mathbb{0}_3 & -\frac{i}{\epsilon_0}\mathbb{1}_3 \\
    -\frac{i}{\mu_0}\nabla \times & \mathbb{0}_3 & \mathbb{0}_3 & \mathbb{0}_3 \\
    \mathbb{0}_3 & \mathbb{0}_3 & \mathbb{0}_3 & i\mathbb{1}_3 \\
    i\epsilon_0\omega_{\mathrm{p}}^2(t)\mathbb{1}_3 & \mathbb{0}_3 & -i\omega_0^2(t)\mathbb{1}_3 + i\beta^2\nabla\nabla\cdot & -i\gamma\mathbb{1}_3
    \end{pmatrix}
    \begin{pmatrix}
        \mathbf{E} \\
        \mathbf{H} \\
        \mathbf{P} \\
        \mathbf{\dot P}
    \end{pmatrix}
    +
    \begin{pmatrix}
    -\frac{i}{\epsilon_0}\mathbf{J} \\
    \mathbf{0} \\
    \mathbf{0} \\
    \mathbf{0}
    \end{pmatrix}.
    \label{eq:eigenproblem}
\end{equation}
Using the diagonal transformation \cite{Raman2010}
\begin{equation}
\mathcal{D} = \mathrm{diag}[ \sqrt{\epsilon_0}, \sqrt{\mu_0}, \frac{\omega_0}{\sqrt{\epsilon_0}\omega_{\mathrm{p}}}, \frac{1}{\sqrt{\epsilon_0}\omega_{\mathrm{p}}}]
\label{eq:diagonal transformation}
\end{equation}
the system of equations \eqref{eq:eigenproblem} can be symmetrized as
\begin{equation}
    i \partial_t \begin{pmatrix}
                {\sqrt{\epsilon_0} \mathbf{E}} \\
                {\sqrt{\mu_0}  \mathbf{H}} \\
                {\frac{\omega_0}{\sqrt{\epsilon_0}\omega_{\mathrm{p}}}  \mathbf{P}} \\
                {\frac{1}{\sqrt{\epsilon_0}\omega_{\mathrm{p}}} \dot{\mathbf{P}}}
                \end{pmatrix}
    = \begin{pmatrix}
    \mathbb{0}_3 & ic_0 \nabla \times & \mathbb{0}_3 & -i\omega_{\mathrm{p}}\mathbb{1}_3 \\
    -i c_0 \nabla \times & \mathbb{0}_3 & \mathbb{0}_3 & \mathbb{0}_3 \\
    \mathbb{0}_3 & \mathbb{0}_3 & \mathbb{0}_3 & i\omega_0\mathbb{1}_3 \\
    i\frac{\omega_{\mathrm{p}}^2(t)}{\omega_{\mathrm{p}}}\mathbb{1}_3 & \mathbb{0}_3 & -i\frac{\omega_0^2(t)}{\omega_0}\mathbb{1}_3 + i\frac{\beta^2}{\omega_0}\nabla \nabla \cdot & -i\gamma\mathbb{1}_3
    \end{pmatrix}
    \begin{pmatrix}
            {\sqrt{\epsilon_0} \mathbf{E}} \\
            {\sqrt{\mu_0}  \mathbf{H}} \\
            {\frac{\omega_0}{\sqrt{\epsilon_0}\omega_{\mathrm{p}}}  \mathbf{P}} \\
            {\frac{1}{\sqrt{\epsilon_0}\omega_{\mathrm{p}}} {\dot{\mathbf{P}}}}
    \end{pmatrix}
    +
    \begin{pmatrix}
    -\frac{i}{\sqrt{\epsilon_0}}\mathbf{{J}} \\
    \mathbf{0} \\
    \mathbf{0} \\
    \mathbf{0}
    \end{pmatrix} \;\; \Leftrightarrow \; \; i\partial_t \mathbf{\Psi} = \mathcal{H} \mathbf{\Psi} + \boldsymbol{\mathcal{J}},
    \label{eq:eigenproblem symmetrized}
\end{equation}
so that the matrix $\mathcal{H}$ is Hermitian in the case of an unmodulated and lossless media ($\alpha_{\mathrm{p}}=\alpha_0=\gamma=0$).
In the above equation, we note that $c_0=(\epsilon_0\mu_0)^{-1/2}$ is the velocity of light in vacuum.

Taking advantage of the isotropy and homogeneity of the three-dimensional medium under consideration, we define the spatial Fourier transformation of a given field $\mathbf{X}$ as
\begin{equation}
    \mathbf{X}(\mathbf{r},t) = \frac{1}{(2\pi)^3}\int_{\mathbb{R}^3} d^3\mathbf{k} \, \tilde{\mathbf{X}}_\mathbf{k}(t) e^{i\mathbf{k}\cdot\mathbf{r}},
\label{Fourier transformation}
\end{equation}
and rewrite the system of equation \eqref{eq:eigenproblem symmetrized} for each Fourier components of the fields
\begin{align}
    &i \partial_t \begin{pmatrix}
                \sqrt{\epsilon_0} \tilde{\mathbf{E}}_\mathbf{k}(t) \\
                \sqrt{\mu_0}  \tilde{\mathbf{H}}_\mathbf{k}(t) \\
                \frac{\omega_0}{\sqrt{\epsilon_0}\omega_{\mathrm{p}}}  \tilde{\mathbf{P}}_\mathbf{k}(t) \\
                \frac{1}{\sqrt{\epsilon_0}\omega_{\mathrm{p}}} \tilde{\dot{\mathbf{P}}}_\mathbf{k}(t)
                \end{pmatrix}
    = \begin{pmatrix}
    \mathbb{0}_3 & -c_0 \mathbf{k} \times & \mathbb{0}_3 & -i\omega_{\mathrm{p}}\mathbb{1}_3 \\
    c_0 \mathbf{k} \times & \mathbb{0}_3 & \mathbb{0}_3 & \mathbb{0}_3 \\
    \mathbb{0}_3 & \mathbb{0}_3 & \mathbb{0}_3 & i\omega_0\mathbb{1}_3 \\
    i\frac{\omega_{\mathrm{p}}^2(t)}{\omega_{\mathrm{p}}}\mathbb{1}_3 & \mathbb{0}_3 & -i\frac{\omega_0^2(t)}{\omega_0}\mathbb{1}_3 - i\frac{\beta^2}{\omega_0}\mathbf{k}\mathbf{k}\cdot & -i\gamma\mathbb{1}_3
    \end{pmatrix}
    \begin{pmatrix}
        \sqrt{\epsilon_0} \tilde{\mathbf{E}}_\mathbf{k}(t) \\
        \sqrt{\mu_0}  \tilde{\mathbf{H}}_\mathbf{k}(t) \\
        \frac{\omega_0}{\sqrt{\epsilon_0}\omega_{\mathrm{p}}}  \tilde{\mathbf{P}}_\mathbf{k}(t) \\
        \frac{1}{\sqrt{\epsilon_0}\omega_{\mathrm{p}}} \tilde{\dot{\mathbf{P}}}_\mathbf{k}(t)
    \end{pmatrix}
    +
    \begin{pmatrix}
    -\frac{i}{\sqrt{\epsilon_0}}\tilde{\mathbf{J}}_\mathbf{k}(t) \\
    \mathbf{0} \\
    \mathbf{0} \\
    \mathbf{0}
    \end{pmatrix}
    \\
    & \Leftrightarrow \; \; i\partial_t \tilde{\mathbf{\Psi}}_\mathbf{k}(t) = \tilde{\mathcal{H}}_\mathbf{k}(t) \tilde{\mathbf{\Psi}}_\mathbf{k}(t) + \tilde{\boldsymbol{\mathcal{J}}}_\mathbf{k}(t).
    \label{eq:eigenproblem fourier}
\end{align}

\subsection{Transverse and longitudinal components of the fields}

To separate the transverse and longitudinal components of the fields, we use the unitary transformation
\begin{equation}
    \mathbb{U}=
    \begin{pmatrix}
        \mathbf{u}_\parallel & \mathbf{u}_+ & \mathbf{u}_-
    \end{pmatrix}, \;\;\;\;\;
    \mathbf{u}_\parallel=\frac{\mathbf{k}}{k}, \;\;\;\;\;
    \mathbf{u}_{\pm}=\frac{1}{\sqrt{2}k\sqrt{k_x^2 + k_y^2}}\begin{pmatrix}
                                                            k_x k_z \pm i k k_y \\
                                                           k_y k_z \mp i k k_x \\
                                                            -k_x^2 - k_y^2
                                                        \end{pmatrix},
\label{eq:unitary transformation}
\end{equation}
which acts on a field $\mathbf{X}$ as
\begin{equation}
    \mathbb{U}^{-1} \mathbf{{X}} = 
    \begin{pmatrix}
        \mathbf{u}^*_\parallel \cdot \mathbf{{X}} \\
        \mathbf{u}^*_+ \cdot \mathbf{{X}} \\
        \mathbf{u}^*_- \cdot \mathbf{{X}}     
    \end{pmatrix} = 
    \begin{pmatrix}
        {{X}_\parallel} \\
        {{X}_+} \\
        {{X}_-}     
    \end{pmatrix} = \mathbf{{X}}',
\end{equation}
so that its Helmholtz decomposition $\mathbf{{X}} =  \mathbf{{X}_\parallel} +  \mathbf{{X}_\perp} = {{X}_\parallel}\mathbf{u}_\parallel + {{X}_+}\mathbf{u}_+ + {{X}_-}\mathbf{u}_-$.
This allows us to reduce the $12-$dimensional system of equations \eqref{eq:eigenproblem fourier} to three independent $4-$dimensional systems, one for each component $\sigma \in \{\parallel, +,-\}$ of the fields
\begin{equation}
    i\partial_t \tilde{\mathbf{\Psi}}_{\sigma,{k}}(t) = \tilde{\mathcal{H}}_{\sigma,{k}}(t) \tilde{\mathbf{\Psi}}_{\sigma,{k}}(t) + \tilde{\boldsymbol{\mathcal{J}}}_{\sigma,{k}}(t),
\label{eq:projected equation}
\end{equation}
with
\begin{equation}
     \tilde{\mathbf{\Psi}}_{\sigma,{k}}(t) =\begin{pmatrix}
                {\sqrt{\epsilon_0} \tilde{E}_{\sigma,{k}}(t)} \\
                {\sqrt{\mu_0}  \tilde{H}_{\sigma,{k}}(t)} \\
                {\frac{\omega_0}{\sqrt{\epsilon_0}\omega_{\mathrm{p}}}  \tilde{P}_{\sigma,{k}}(t)} \\
                {\frac{1}{\sqrt{\epsilon_0}\omega_{\mathrm{p}}} \tilde{\dot P}_{\sigma,{k}}(t)}
                \end{pmatrix}, \;\;\;\;\;
                 \tilde{\boldsymbol{\mathcal{J}}}_{\sigma,{k}}(t) = \begin{pmatrix}
    -\frac{i}{\sqrt{\epsilon_0}}{\tilde{J}_{\sigma,{k}}}(t) \\
    \mathbf{0} \\
    \mathbf{0} \\
    \mathbf{0}
    \end{pmatrix},
\end{equation}
and where the effective Hamiltonian matrices for the transverse and longitudinal matrices read
\begin{equation}
    \tilde{\mathcal{H}}_{+,{k}}(t) =
    \begin{pmatrix}
     0 & -ic_0 k & 0 & -i\omega_{\mathrm{p}} \\
     i c_0 k & 0 & 0 & 0 \\
     0 & 0 & 0 & i\omega_0 \\
     i\frac{\omega_{\mathrm{p}}^2(t)}{\omega_{\mathrm{p}}} & 0 & -i\frac{\omega_0^2(t)}{\omega_0} & -i\gamma
    \end{pmatrix}
    \;\;\;\;\; \textrm{and} \;\;\;\;\;
    \tilde{\mathcal{H}}_{\parallel,{k}}(t) =
    \begin{pmatrix}
     0 & 0 & 0 & -i\omega_{\mathrm{p}} \\
     0 & 0 & 0 & 0 \\
     0 & 0 & 0 & i\omega_0 \\
     i\frac{\omega_{\mathrm{p}}^2(t)}{\omega_{\mathrm{p}}} & 0 & -i\frac{\omega_0^2(t)}{\omega_0} - i\frac{\beta^2 k^2}{\omega_0} & -i\gamma
    \end{pmatrix}
\label{eq:H transverse and longitudinal}.
\end{equation}
We note that from isotropy and homogeneity, the projected reduced systems of equations \eqref{eq:projected equation} only depends on the wavenumber $k=|\mathbf{k}|$, and one has $\tilde{\mathcal{H}}_{-,k}(t)=\tilde{\mathcal{H}}_{+,-k}(t)$.

\subsection{Floquet resolution}

From now on we work on the projected $4-$dimensional subspaces corresponding to either the longitudinal or transverse components of the fields.
Following the formalism developed by Park \textit{et al.}~\cite{Park2025}, we use the time-periodicity of the effective Hamiltonian to expand it in terms of its Floquet modes
\begin{equation}
    \tilde{\mathcal{H}}_{\sigma,{k}}(t) = \sum_{m=-N_F}^{N_F} e^{-i m \Omega t}\tilde{\mathcal{H}}^{(m)}_{\sigma,k},
\end{equation}
where the first modes
\begin{equation}
    \tilde{\mathcal{H}}^{(m = \pm 1)}_{\sigma,{k}} =
    \begin{pmatrix}
     0 & 0 & 0 & 0 \\
     0 & 0 & 0 & 0 \\
     0 & 0 & 0 & 0 \\
     \mp \frac{\alpha_{\mathrm{p}}}{2}\omega_{\mathrm{p}} & 0 & \pm\frac{\alpha_0}{2}\omega_0 & 0
    \end{pmatrix},
\end{equation}
while all the higher-order modes $\tilde{\mathcal{H}}^{(|m|>1)}_{\sigma,{k}} = \mathbb{0}_4$.
Here, $N_F \rightarrow \infty$ is half the number of Floquet modes under consideration.
We note that in all our computations we considered $N_F=15$, i.e., $30$ Floquet replicas, and verified that it was sufficient for convergence.
We also assume an harmonic source current density so that
\begin{equation}
    \tilde{\boldsymbol{\mathcal{J}}}_{\sigma,{k}}(t) = \begin{pmatrix}
-\frac{i}{\sqrt{\epsilon_0}}{\tilde{J}_{\sigma,{k}}}e^{-i \omega t} \\
\mathbf{0} \\
\mathbf{0} \\
\mathbf{0}
\end{pmatrix} = \tilde{\boldsymbol{\mathcal{J}}}_{\sigma,{k}}e^{-i \omega t},
\end{equation}
and consider the Floquet decomposition of the fields
\begin{equation}
    \tilde{\mathbf{\Psi}}_{\sigma,k}(t) = e^{-i \omega t} \sum_{n=-N_F}^{N_F} e^{-i n \Omega t} \tilde{\boldsymbol{\phi}}^{(n)}_{\sigma,k},
    \;\;\;\;\; \textrm{with} \;\;\;\;\;
    \tilde{\boldsymbol{\phi}}^{(n)}_{\sigma,k} =\begin{pmatrix}
            {\sqrt{\epsilon_0} \tilde{E}^{(n)}_{\sigma,k}} \\
            {\sqrt{\mu_0}  \tilde{H}^{(n)}_{\sigma,k}} \\
            {\frac{\omega_0}{\sqrt{\epsilon_0}\omega_{\mathrm{p}}}  \tilde{P}^{(n)}_{\sigma,k}} \\
            {\frac{1}{\sqrt{\epsilon_0}\omega_{\mathrm{p}}} {\dot{\tilde{P}}}^{(n)}_{\sigma,k}}
            \end{pmatrix}.
\end{equation}
Note that with our current notations, a field $\mathbf{X}$ reads in real space
\begin{equation}
    \mathbf{X}(\mathbf{r},t) = \frac{1}{(2\pi)^3} \int_{\mathbb{R}^3} d^3 \mathbf{k} \,   e^{i\mathbf{k}\cdot\mathbf{r}} e^{-i \omega t} \sum_n e^{-i n \Omega t} \sum_\sigma \tilde{X}^{(n)}_{\sigma,k} \mathbf{u}_{\sigma,k}.
\end{equation}
Plugging the Floquet decomposition of the fields into Eq.~\eqref{eq:projected equation} leads to
\begin{equation}
    \sum_n (\omega + n \Omega) e^{-i n \Omega t} \tilde{\boldsymbol{\phi}}^{(n)}_{\sigma,k} = \sum_m \sum_n \tilde{\mathcal{H}}^{(n-m)}_{\sigma,k}  e^{-i n \Omega t} \tilde{\boldsymbol{\phi}}^{(m)}_{\sigma,k} + \tilde{\boldsymbol{\mathcal{J}}}_{\sigma,{k}},
\end{equation}
which can be rewritten as
\begin{equation}
    \omega \underline{\tilde{\phi}_{\sigma,k}} = \underline{\underline{\tilde{\mathcal{H}}_{\sigma,k}}} \underline{\tilde{\phi}_{\sigma,k}} + \underline{ \tilde{\boldsymbol{\mathcal{J}}}_{\sigma,{k}}},
\label{eq:Floquet eigenproblem}
\end{equation}
where $\underline{\tilde{\phi}_{\sigma,k}}$, $\underline{ \tilde{\boldsymbol{\mathcal{J}}}_{\sigma,{k}}}$ and $\underline{\underline{\tilde{\mathcal{H}}_{\sigma,k}}}$ are, respectively, two vectors and a matrix in the extended Floquet space of dimension $4(2N_F+1)$ that read
\begin{equation}
\underline{\tilde{\phi}_{\sigma,k}} =
\begin{pmatrix}
\vdots \\
\boldsymbol{\tilde{\phi}}^{(-1)}_{\sigma,k} \\
\boldsymbol{\tilde{\phi}}^{(0)}_{\sigma,k} \\
\boldsymbol{\tilde{\phi}}^{(+1)}_{\sigma,k} \\
\vdots
\end{pmatrix},
\;\;\;\;\;
\underline{ \tilde{\boldsymbol{\mathcal{J}}}_{\sigma,{k}}} =
\begin{pmatrix}
\vdots \\
\mathbf{0} \\
 \tilde{\boldsymbol{\mathcal{J}}}_{\sigma,{k}} \\
\mathbf{0} \\
\vdots
\end{pmatrix}
\;\;\;\;\; \textrm{and} \;\;\;\;\;
\underline{\underline{\tilde{\mathcal{H}}_{\sigma,k}}} =
\begin{pmatrix}
\ddots & \ddots & \mathbb{0}_4 &   &  \\
\ddots & \tilde{\mathcal{H}}^{(0)}_\sigma + \Omega \mathbb{1}_4  & \tilde{\mathcal{H}}^{(-1)}_\sigma & \mathbb{0}_4 &   \\
\mathbb{0}_4 & \tilde{\mathcal{H}}^{(+1)}_\sigma & \tilde{\mathcal{H}}^{(0)}_\sigma & \tilde{\mathcal{H}}^{(-1)}_\sigma & \mathbb{0}_4 \\
  & \mathbb{0}_4 & \tilde{\mathcal{H}}^{(+1)}_\sigma & \tilde{\mathcal{H}}^{(0)}_\sigma - \Omega \mathbb{1}_4 & \ddots \\
  &   & \mathbb{0}_4  & \ddots & \ddots
\end{pmatrix}.
\label{eq:floquet matrix}
\end{equation}
Importantly, the $4(2N_F + 1)$ eigenvalues of $\underline{\underline{\tilde{\mathcal{H}}_{\sigma,k}}}$ correspond to the complex Floquet frequencies $\omega^{(n)}_{\sigma,\pm}(k) = \pm \omega^{(0)}_{\sigma,\pm}(k) + n \Omega$.
To obtain the fields, we define from Eq.~\eqref{eq:Floquet eigenproblem} the Fourier space Floquet Green matrix for the component $\sigma$
\begin{equation}
    \underline{\underline{\tilde{\mathcal{G}}_{\sigma,\omega,k}}} = -\frac{c_0^2}{\omega} \left( \omega\underline{\underline{\mathbb{1}}} - \underline{\underline{\tilde{\mathcal{H}}_{\sigma,k}}}\right)^{-1} =   \begin{pmatrix}
   \ddots & \vdots & \vdots & \vdots & \reflectbox{$\ddots$} \\
    \cdots & \tilde{\mathcal{G}}^{(-1,-1)}_{\sigma,\omega,k} & \tilde{\mathcal{G}}^{(-1,0)}_{\sigma,\omega,k} & \tilde{\mathcal{G}}^{(-1,1)}_{\sigma,\omega,k} & \cdots \\
    \cdots & \tilde{\mathcal{G}}^{(0,-1)}_{\sigma,\omega,k} & \tilde{\mathcal{G}}^{(0,0)}_{\sigma,\omega,k} & \tilde{\mathcal{G}}^{(0,1)}_{\sigma,\omega,k} & \cdots \\
    \cdots & \tilde{\mathcal{G}}^{(1,-1)}_{\sigma,\omega,k} & \tilde{\mathcal{G}}^{(1,0)}_{\sigma,\omega,k} & \tilde{\mathcal{G}}^{(1,1)}_{\sigma,\omega,k} & \cdots \\
    \reflectbox{$\ddots$} & \vdots & \vdots & \vdots & \ddots
  \end{pmatrix}.
\label{eq:Floquet Green dyadic}
\end{equation}
In that way, the $n$th Floquet mode of the field vector
\begin{equation}
    \tilde{\boldsymbol{\phi}}^{(n)}_{\sigma,k} = -\frac{\omega}{c_0^2}\tilde{\mathcal{G}}^{(n,0)}_{\sigma,\omega,k}\tilde{\boldsymbol{\mathcal{J}}}_{\sigma,{k}},
\label{eq:Floquet fields}
\end{equation}
so that the $n$th Floquet mode of the $\sigma$ component of the electric field reads
\begin{equation}
    \tilde{E}^{(n)}_{\sigma,k} = i\omega\mu_0 \tilde{\mathcal{G}}^{E, (n,0)}_{\sigma,\omega,k} \tilde{J}_{\sigma,k},
\label{eq:Floquet electric field}
\end{equation}
where $\tilde{\mathcal{G}}^{E,(n,0)}_{\sigma,\omega,k}$ is the scalar element of the Fourier space Floquet Green matrix that selects the electric field only.
We note that during the reviewing process of our manuscript, Ref.~\cite{Sun2025}, which develops a similar Floquet formalism to dispersive and lossy PTCs as ours, has been published.

\subsection{Power dissipated by a point-dipole}

We consider a current source density $\mathbf{J}(\mathbf{r},t)$ corresponding to an harmonic point-dipole with frequency $\omega$ and dipole moment $\mathbf{p}(t)=\mathbf{p}e^{-i\omega t}$, so that
\begin{align}
    \mathbf{J}(\mathbf{r},t) &= \tilde{\mathbf{J}}_\mathbf{k}(t)\delta(\mathbf{r}) = -i\omega\mathbf{p}e^{-i\omega t}\delta(\mathbf{r}),
\end{align}
with $\tilde{\mathbf{J}}_\mathbf{k}(t) = \tilde{\mathbf{J}}(t)$ the (constant) Fourier components of the current density.
The power dissipated by such a point-dipole reads
\begin{equation}
   P_{\omega}(t)= \frac{\omega}{2}\mathrm{Im}[\mathbf{p}^{*}(t)\cdot\mathbf{E}(\mathbf{0},t)],
\end{equation}
while its average over a cycle of modulation is
\begin{equation}
    \bar{P}_{\omega} = \frac{\Omega}{2\pi}\int_{0}^{\frac{2\pi}{\Omega}}dt\,P_{\omega}(t).
\end{equation}
We rewrite this time-averaged dissipated power in terms of the Fourier components of the fields as
\begin{align}
    \bar{P}_{\omega} &= \frac{1}{(2\pi)^3}\int_{\mathbb{R}^3} d^3\mathbf{k}\, \sum_\sigma \frac{\omega }{2}\mathrm{Im}\left[p^{*}_{\sigma}\tilde{E}^{(0)}_{\sigma,k} \right] \\
    &=  \frac{1}{(2\pi)^3}\int_{\mathbb{R}^3} d^3\mathbf{k}\, \sum_\sigma \frac{\omega^3 \mu_0 |p_\sigma|^2}{2} \mathrm{Im}\left[ \tilde{\mathcal{G}}^{E,(0,0)}_{\sigma,\omega,k} \right].
\end{align}
where we used the Fourier space Floquet Green matrix \eqref{eq:Floquet Green dyadic}. Averaging over the dipole orientations and using that from isotropy and homogeneity $|p_\sigma|^2=|\mathbf{p}|^2/3$, we can rewrite the above dissipated power in its usual form as
\begin{equation}
    \bar{P}_{\omega} = \frac{\pi \omega^2 |\mathbf{p}|^2}{12\epsilon_0} \rho(\mathbf{0},\omega),
\end{equation}
where the photonic local density of states (LDOS)
\begin{align}
    \rho(\mathbf{0},\omega) = \frac{1}{(2\pi)^3}\int_{\mathbb{R}^3} d^3\mathbf{k}\, \frac{2\omega}{\pi c_0^2} \mathrm{Tr}\left( \mathrm{Im}\left[ \overleftrightarrow{\tilde{\mathcal{G}}}^{E,(0,0)}_{\omega,k} \right] \right)
\end{align}
is written in terms of the Green dyadic of the 0th Floquet mode of the electric field $\overleftrightarrow{\tilde{\mathcal{G}}}^{E,(0,0)}_{\omega,k} = \mathrm{diag}[\tilde{\mathcal{G}}^{E,(0,0)}_{\parallel,\omega,k},\tilde{\mathcal{G}}^{E,(0,0)}_{+,\omega,k},\tilde{\mathcal{G}}^{E,(0,0)}_{-,\omega,k}]$.
The contribution of the dissipated power originating from a component $\sigma$ of the fields can therefore be expressed as
\begin{equation}
    \bar{P}^{\sigma}_{\omega} = \frac{\pi \omega^2 |\mathbf{p}|^2}{12\epsilon_0} \int_{\mathbb{R}^3} d^3\mathbf{k}\, \tilde{\rho}^{\sigma}_{\omega,k},
\end{equation}
with the $\sigma$ component of the momentum-resolved LDOS
\begin{equation}
    \tilde{\rho}^{\sigma}_{\omega,k} = \frac{1}{(2\pi)^3}\frac{2\omega}{\pi c_0^2} \mathrm{Im}\left[ \tilde{\mathcal{G}}^{E,(0,0)}_{\sigma,\omega,k} \right].
\end{equation}

Finally, we separate the contributions of the dissipated power originating from the longitudinal and transverse fields as $\bar{P}_\omega = \bar{P}^{\parallel}_\omega +\bar{P}^{\perp}_\omega$, with
\begin{equation}
     \bar{P}^{\parallel/\perp}_\omega = \frac{\pi \omega^2 |\mathbf{p}|^2}{12\epsilon_0} \int_{\mathbb{R}^3} d^3\mathbf{k}\, \tilde{\rho}^{\parallel/\perp}_{\omega,k},
\end{equation}
where the longitudinal and transverse momentum-resolved LDOS read
\begin{equation}
    \tilde{\rho}^{\parallel}_{\omega,k} = \frac{1}{(2\pi)^3}\frac{2\omega}{\pi c_0^2} \mathrm{Im}\left[ \tilde{\mathcal{G}}^{E,(0,0)}_{\parallel,\omega,k} \right] \;\;\;\;\; \textrm{and} \;\;\;\;\; \tilde{\rho}^{\perp}_{\omega,k} = \frac{1}{(2\pi)^3}\frac{4\omega}{\pi c_0^2} \mathrm{Im}\left[ \tilde{\mathcal{G}}^{E,(0,0)}_{+,\omega,k} \right].
\end{equation}

\section{Analytical solutions for the static case}

In the case of a static ($\alpha_{\mathrm{p}/0}=0$) dispersive and lossy medium, we can recover analytically the usual values of power dissipated by a point dipole.

\subsection{Transverse fields}

In the static case, the $(0,0)$ element of the transverse effective Floquet Green dyadic read
\begin{equation}
    \mathcal{G}^{0,(0,0)}_{\perp,\omega,k} \equiv \mathcal{G}^{0,(0,0)}_{+,\omega,k} = -\frac{c_0^2}{\omega}\left( \omega \mathbb{1}_4 - \mathcal{H}_{\perp,k}^{(0)} \right)^{-1} = -\frac{c_0^2}{\omega} \begin{pmatrix}
     \omega & ic_0 k & 0 & i\omega_{\mathrm{p}} \\
     -i c_0 k & \omega & 0 & 0 \\
     0 & 0 & \omega & -i\omega_0 \\
     -i\omega_{\mathrm{p}} & 0 & i\omega_0 & \omega + i\gamma
    \end{pmatrix}^{-1}.
\end{equation}
A matrix inversion reveals that the $(0,0)$ element of the dyadic, which corresponds to the electric field, reads
\begin{equation}
    \mathcal{G}^{0,E,(0,0)}_{\perp,\omega,k} = \frac{c_0^2}{(c_0 k)^2 - \omega^2 \epsilon(\omega)},
\end{equation}
where the complex static Drude-Lorentz permittivity
\begin{align}
    \epsilon(\omega) = \epsilon'(\omega) + i\epsilon''(\omega) = 1 + \frac{\omega_{\mathrm{p}}^2}{\omega_0^2 - \omega^2 - i\omega\gamma}.
\label{eq:permittivity}
\end{align}
The average power dissipated by the transverse fields is therefore
\begin{align}
     \bar{P}^{0,\perp}_\omega &= \frac{\omega^3 |\mathbf{p}|^2}{3\epsilon_0 c_0^2} \frac{1}{(2\pi)^3}\int_{\mathbb{R}^3} d^3\mathbf{k}\, \mathrm{Im}\left[\mathcal{G}^{0,E,(0,0)}_{\perp,\omega,k} \right] \\
    &= \frac{2\omega^5|\mathbf{p}|^2}{3\epsilon(2\pi)^2}\epsilon''(\omega)\int_{0}^{\infty} dk\, \frac{k^2}{|\omega^2\epsilon(\omega) - (c_0 k)^2|^2}  \\
    &= \frac{\omega^4 |\mathbf{p}|^2}{12\pi\epsilon_0 c_0^3}\mathrm{Re}\left[\sqrt{\epsilon(\omega)}\right],
\label{eq:P perp static}
\end{align}
where we used that the imaginary part of the permittivity $\epsilon''(\omega)>0$ to evaluate the integral.
In Eq.~\eqref{eq:P perp static}, we recognize the average power dissipated by a point dipole in vacuum
\begin{equation}
    \bar{P}^{\textrm{vac}}_\omega = \frac{\omega^4 |\mathbf{p}|^2}{12\pi\epsilon_0 c_0^3},
\end{equation}
so that we recover the usual result of a dispersive and lossy media \cite{Barnett1992}
\begin{equation}
     \bar{P}^{\perp,0}_\omega = \bar{P}^{\textrm{vac}}_{\omega}  \mathrm{Re}\left[n(\omega)\right],
\end{equation}
where $n(\omega)$ is the complex refractive index. Finally, we note that the static transverse eigenfrequencies are found by diagonalizing the static effective Hamiltonian $\mathcal{H}^{(0)}_{\perp,k}$.
In the lossless case ($\gamma=0$) the two pairs of real eigenfrequencies correspond to the upper and lower polaritonic bands
\begin{equation}
   \pm \omega^{\mathrm{up}}_{\perp, \gamma=0}(k) = \pm \frac{1}{\sqrt{2}}\sqrt{ (c_0 k)^2 + \omega_0^2 + \omega_{\mathrm{p}}^2 + \sqrt{[(c_0 k)^2 + \omega_0^2 + \omega_{\mathrm{p}}^2 ]^2 - 4(c_0 k)^2\omega_0^2}},
\end{equation}
and
\begin{equation}
   \pm \omega^{\mathrm{lo}}_{\perp, \gamma=0}(k) = \pm \frac{1}{\sqrt{2}}\sqrt{ (c_0 k)^2 + \omega_0^2 + \omega_{\mathrm{p}}^2 - \sqrt{[(c_0 k)^2 + \omega_0^2 + \omega_{\mathrm{p}}^2 ]^2 - 4(c_0 k)^2\omega_0^2}}.
\end{equation}

This analytical solution of the lossless static bands allows us to easily predict the formation of dispersive momentum gaps.
After some manipulations, one shows that the wavenumber $k$ allowing the equation $\omega^{\mathrm{up}}_{\perp}(k) - n\Omega = \omega^{\mathrm{lo}}_{\perp}(k)$ to be satisfied has the two solutions $k_{\pm} = \omega_0/c \pm \sqrt{(n\Omega/c)^2-(\omega_{\mathrm{p}}/c)^2}$.
Real solutions therefore exist only if $n\Omega \geq \omega_{\mathrm{p}}$.
The two solutions are degenerate so that the bands touch where $n\Omega=\omega_{\mathrm{p}}$, at a wavenumber $k=\omega_0/c$,
and two crossings appear whenever $n\Omega > \omega_{\mathrm{p}}$.
Since the lower band $\omega^{\mathrm{lo}}_{\perp}$ is highly dispersive around $k=\omega_0/c$, a modulation frequency close to unit fractions of the plasma frequency allows to maximize the gain bandwidth of dispersive momentum gaps.
For larges value of $\Omega>\omega_{\mathrm{p}}$, however, the crossings will arise at large wavenumbers $k_\pm$, where the lower band $\omega_{\perp}^{\mathrm{lo}}$ is almost flat.
In these cases, the gain bandwidth related to the dispersive momentum gap, although being nonzero, is relatively small.

In our study, we use these nonmodulated bandstructures to infer our choice of dispersion and modulation parameters.
The value of $\omega_0$ has been fixed at $0.6\omega_\mathrm{p}$ when modulating the plasma frequency. A smaller ratio $\omega_0/\omega_\mathrm{p}$ enlarges the frequency gap $\Delta_{\gamma=0}=\omega_{\perp,\gamma=0}^{\mathrm{up}}(k=0) - \omega_{\perp,\gamma=0}^{\mathrm{lo}}(k\rightarrow\infty)=\omega_0(\sqrt{1+(\omega_\mathrm{p}/\omega_0)^2}-1)$ between the lower and upper bands.
This leads to the requirement of either a larger strength $\alpha$ or a higher frequency $\Omega$ to couple replicas of one band to another. Conversely, a larger ratio $\omega_0/\omega_\mathrm{p}$ facilitates such a coupling by reducing the frequency gap, and would thus lead to modulation-induced effects for smaller $\alpha$ and $\Omega$.

\begin{figure*}[h!]
 \includegraphics[width=0.48\columnwidth]{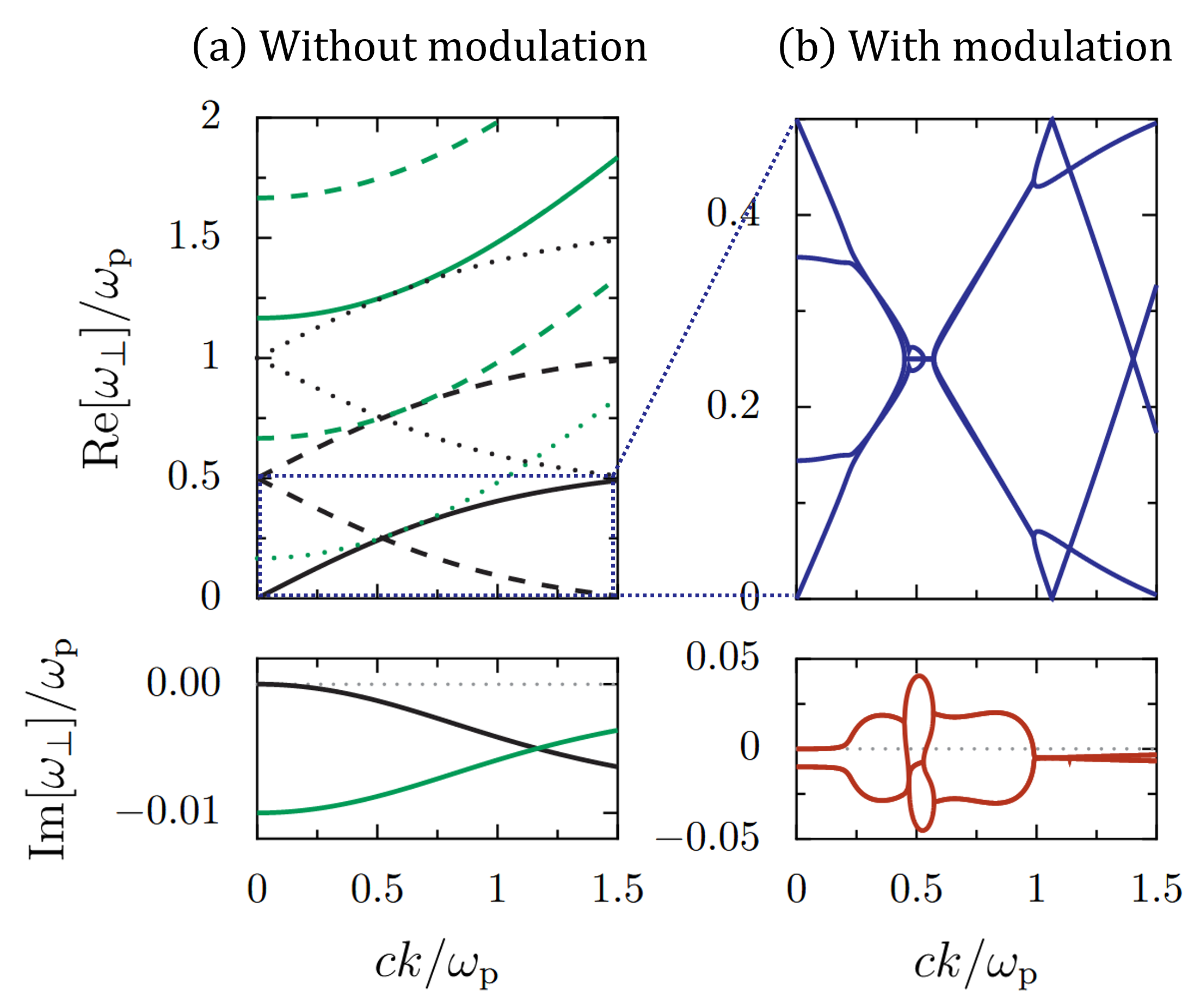}
 \hspace{0.39cm}
 \includegraphics[width=0.48\columnwidth]{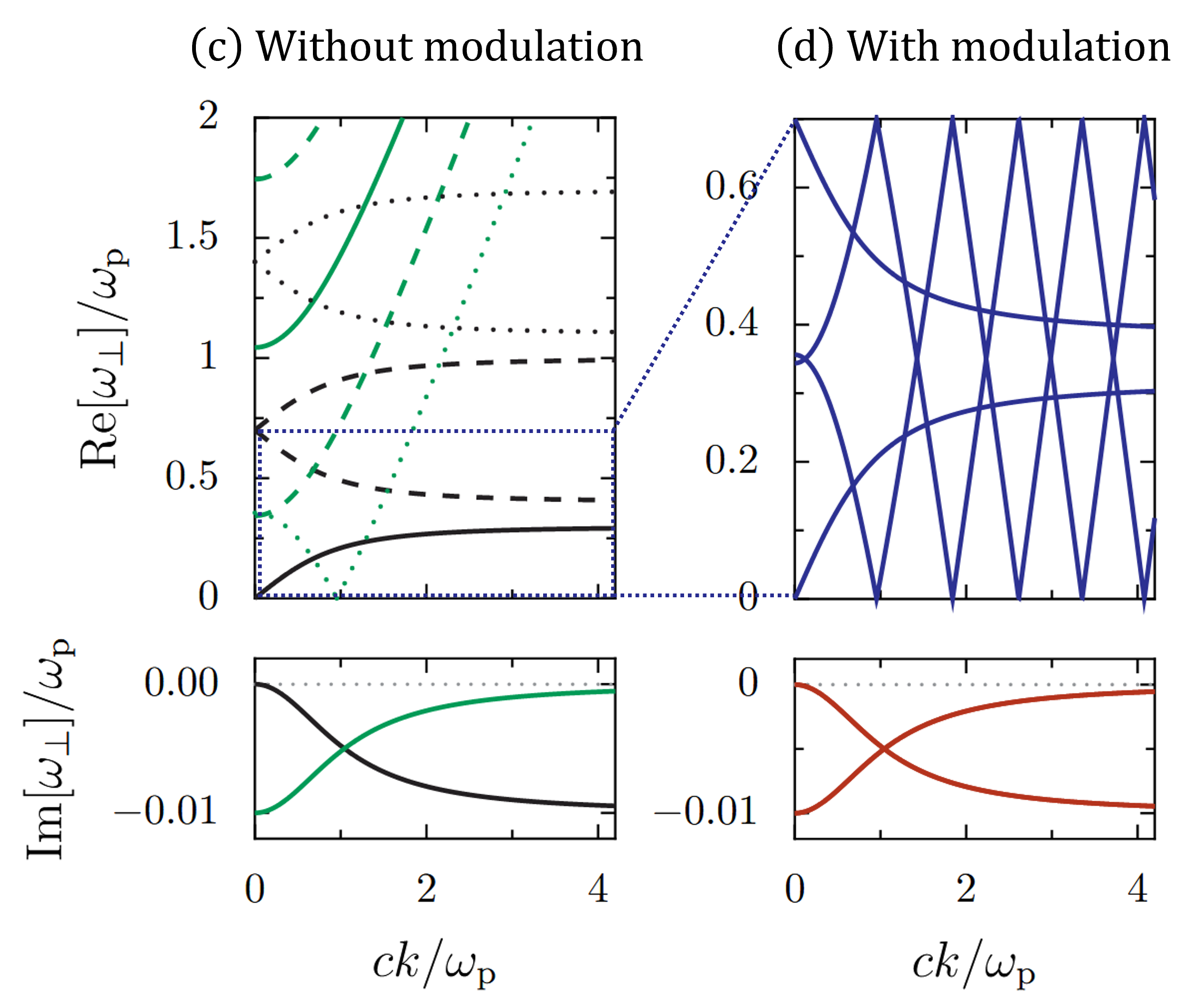}
 \centering
 \caption{(a)-(b) Complex bandstructure of the static (a) and modulated (b) medium, considering the parameters of Fig.~3 of the main text.
 (c)-(d) Complex bandstructure of the static (c) and modulated (d) medium, considering the parameters of Fig.~4 of the main text.
 In the static bandstructures, dashed lines are first-order replicas $\pm \omega_\perp^{\mathrm{up}/\mathrm{lo}} \pm \Omega$, while dotted lines are second-order ones $\pm \omega_\perp^{\mathrm{up}/\mathrm{lo}} \pm 2\Omega$.}
\label{fig:sketches}
\end{figure*}

All the modulation parameters are chosen from simple band coupling arguments.
As discussed in the main text and exemplified in Figs.~1(b)-(c), the modulation parameters of Fig.~2 have been chosen so that $\omega_\perp^\mathrm{up}-\Omega$ overlaps with $\omega_\perp^\mathrm{lo}$ in its dispersive region.
This arises for a modulation frequency $\Omega=\omega_p$.
In Fig.~\ref{fig:sketches}(a)-(b), we show how the modulation parameters of Fig.~3 of the main text have been chosen.
In order for the second-order replica $\omega_\perp^\mathrm{up}-2\Omega$ [green dotted line in Fig.~\ref{fig:sketches}(a)] to take the same role as the first-order one in Fig.~2, we consider a modulation frequency $\Omega=0.5\omega_\mathrm{p}$.
This also induces a coupling between the first order negative replica $-\omega_\perp^\mathrm{lo}+\Omega$ [black dashed line in Fig.~\ref{fig:sketches}(a)] and both $\omega_\perp^\mathrm{up}-2\Omega$ and $\omega_\perp^\mathrm{lo}$.
Such a three-band coupling leads the complicated bandstructure obtained in the modulated system [Fig.~\ref{fig:sketches}(b)].
The modulation strength $\alpha$ is then increased as much as required for that second-order replica to interact and so that almost all the eigenfrequencies belonging to the 1st FBZ feature a positive imaginary part. We note that smaller $\alpha$ would lead to the same qualitative effect, however inducing positive imaginary parts in a reduced real frequency window.

We present the same results in Fig.~\ref{fig:sketches}(c)-(d) for the parameters used in Fig.~4 of the main text.
Here we lowered the resonance frequency to $\omega_0=0.3\omega_\mathrm{p}$ in order to increase the frequency gap $\Delta$.
A modulation frequency $\Omega>2\omega_0$ allows the negative replica $\Omega - \omega_\perp^\mathrm{lo}$ to be present within this frequency gap.
The choice of a slightly larger $\Omega=0.7\omega_\mathrm{p}\simeq2.33\omega_0$ permits that such a replica does not interact with the upper or lower band, nor the lower band does interact with a negative replica of the upper band.
The modulation strength $\alpha=0.5$, on the other hand, is chosen to ensure a large negative local density of states along the negative replica, but kept small enough to not induce coupling with second-order replicas.

\subsection{Longitudinal fields}

The same exercise can be done for the longitudinal fields, where the corresponding effective Floquet Green dyadic read in the static case
\begin{equation}
    \mathcal{G}^{0,(0,0)}_{\parallel,\omega,k} = -\frac{c_0^2}{\omega}\left( \omega \mathbb{1}_4 - \mathcal{H}_{\parallel,k}^{(0)} \right)^{-1} = -\frac{c_0^2}{\omega}\begin{pmatrix}
     \omega & 0 & 0 & i\omega_{\mathrm{p}} \\
     0 & \omega & 0 & 0 \\
     0 & 0 & \omega & -i\omega_0 \\
     -i\omega_{\mathrm{p}} & 0 & i\omega_0 + i\frac{\beta^2}{\omega_0} & \omega + i\gamma
    \end{pmatrix}^{-1},
\end{equation}
so that its $(0,0)$ element corresponding to the electric field reads
\begin{equation}
    \mathcal{G}^{0,E,(0,0)}_{\parallel,\omega,k} = \frac{-c_0^2}{\omega^2 \epsilon(\omega,k)},
\end{equation}
where the non-local hydrodynamic Drude-Lorentz permittivity
\begin{equation}
     \epsilon(\omega,k) = 1 + \frac{\omega_{\mathrm{p}}^2}{\omega_0^2+\beta^2 k^2 - \omega^2 - i\omega\gamma}.
\label{eq:nonlocal permittivity}
\end{equation}
In the lossless and nondispersive case ($\gamma=\omega_{\mathrm{p}}=0$), we note that $\mathcal{G}^{0,E,(0,0)}_{\parallel,\omega,k}$ is real so that no longitudinal modes are present and the electromagnetic fields are purely transverse.
Importantly, in the dispersive and lossy but local case, where $\beta=0$, the longitudinal modes are independent of the wavevector $\mathbf{k}$ so that the associated power dissipated by a point-dipole diverges.
The consideration of nonlocality through an hydrodynamic model allows us to circumvent this issue, and the $k$ dependence of the nonlocal permittivity \eqref{eq:nonlocal permittivity} allows us to compute the average power dissipated by the longitudinal fields as
\begin{align}
     \bar{P}^{\parallel,0}_\omega  =\frac{\omega \omega_{\mathrm{p}}^3|\mathbf{p}|^2}{24\pi\epsilon_0\beta^3}\mathrm{Re}\left[ \sqrt{\frac{\epsilon(\omega)}{1-\epsilon(\omega)}} \right],
\label{eq:P longit}
\end{align}
where we note that $\epsilon(\omega)$ is here the \emph{local} Drude-Lorentz permittivity \eqref{eq:permittivity}.
We note that the result Eq.~\eqref{eq:P longit} is consistent with what has been found for the longitudinal contribution of the electric field intensity in a dispersive and nonlocal media \cite{Horsley2014}.
Finally, the longitudinal complex eigenfrequencies have real and imaginary parts
\begin{equation}
    \pm\mathrm{Re}[\omega_{\parallel}(k)] = \pm \sqrt{\omega_0^2 + \beta^2 k^2 + \omega_{\mathrm{p}}^2 - \left(\frac{\gamma}{2}\right)^2} \;\;\;\;\; \textrm{and} \;\;\;\;\; \mathrm{Im}[\omega_{\parallel}(k)] = -\frac{\gamma}{2}.
\end{equation}

\section{Effects of material losses on dispersive momentum gaps}

\subsection{Lifting of exceptional points}

As discussed in the main text, the inclusion of material losses eliminates exceptional points (EPs) associated to dispersive momentum gaps.
To further support that claim, we compute here the phase rigidity associated to the eigenstates of the non-Hermitian Floquet matrix describing the transverse modes in the PTC [see Eq.~\eqref{eq:floquet matrix}].
The phase rigidity is defined for a given Floquet band and a given momentum $k$ as \cite{Wiersig2023}
\begin{equation}
    r_\perp = \frac{|\braket{L|R}|}{\sqrt{\braket{L|L}\braket{R|R}}},
\label{eq:rigidity}
\end{equation}
where $\ket{L}$ and $\ket{R}$ are the left and right eigenstates of the Floquet matrix.
The phase rigidity characterizes the difference between the latter left and right eigenstates and, interestingly, vanishes at EPs \cite{Wiersig2023}.

\begin{figure}[h!]
 \includegraphics[width=0.3\columnwidth]{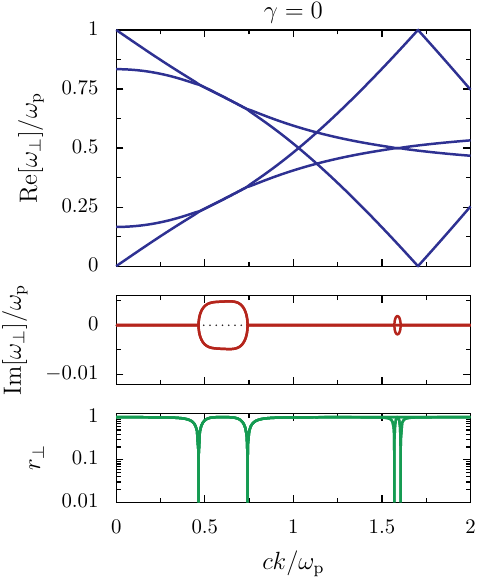}
 \hspace{2cm}
  \includegraphics[width=0.3\columnwidth]{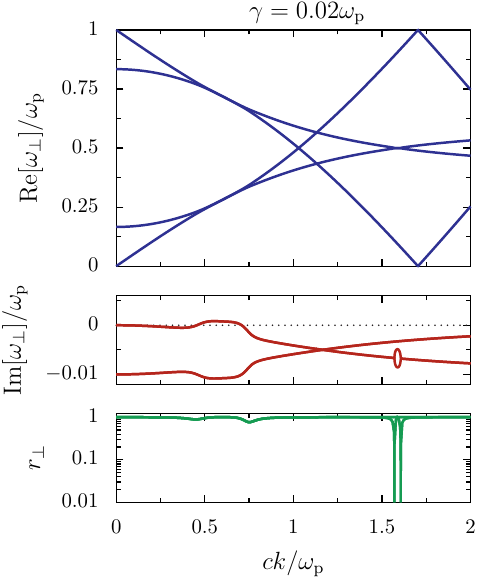}
    \caption{Complex bandstructure and rigidity [as defined in Eq.~\eqref{eq:rigidity}] of the transverse eigenmodes in the case of a lossless ($\gamma=0$, left panels) and lossy ($\gamma=0.02\omega_{\mathrm{p}}$, right panels) medium.
    The other considered parameters are the same as discussed in Figs.~1 and 2 of the main text, namely, $\omega_0=0.6\omega_{\mathrm{p}}$ as well as a weak yet high modulation of the plasma frequency with strength $\alpha=0.05$ and frequency $\Omega=\omega_{\mathrm{p}}$.}
\label{fig:rigidity weak}
\end{figure}

The complex bandstructure of the transverse eigenmodes in the case studied in Figs.~1 and 2 of the main text is shown in Fig.~\ref{fig:rigidity weak} along with the phase rigidity of each of the four bands in the first FBZ.
In the left panels, we consider the case of a lossless medium, fixing the inverse relaxation time to $\gamma=0$.
In that specific case, we observe that the dispersive momentum gap features EPs, as marked by the vanishing phase rigidity at the momenta corresponding to the edges of the gap.
The flat momentum gap at $\Omega/2$ presents a similar vanishing phase rigidity, confirming the presence of EPs at its edges as well.
We note that from the absence of material losses, the imaginary part of the eigenfrequencies is symmetric around zero.
In the right panels, we now present the case of a lossy medium, with $\gamma=0.02\omega_{\mathrm{p}}$ as considered in the main text.
Interestingly, we see that the phase rigidity is not anymore vanishing around the dispersive momentum gap, confirming that once material losses are included, dispersive momentum gaps are free of EPs.
The phase rigidity approaches however still zero at the edges of the flat momentum gap, signaling that the EPs associated to the latter gap are robust to losses.

Finally, we note that in a lossless system, EPs of dispersive and flat momentum gaps can be merged into an EP of order $4$, similarly as what has been done using anisotropy \cite{Wang2024}.
However, the elimination of EPs associated to dispersive momentum gaps signifies the impossibility of forming any higher-order EPs in a lossy PTC through the use of dispersion.

\subsection{Loss-induced amplification}

\begin{figure}[h!]
 \includegraphics[width=0.3\columnwidth]{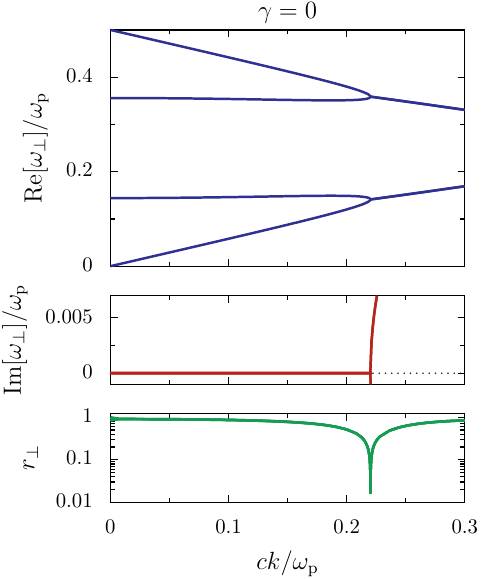}
 \hspace{2cm}
\includegraphics[width=0.3\columnwidth]{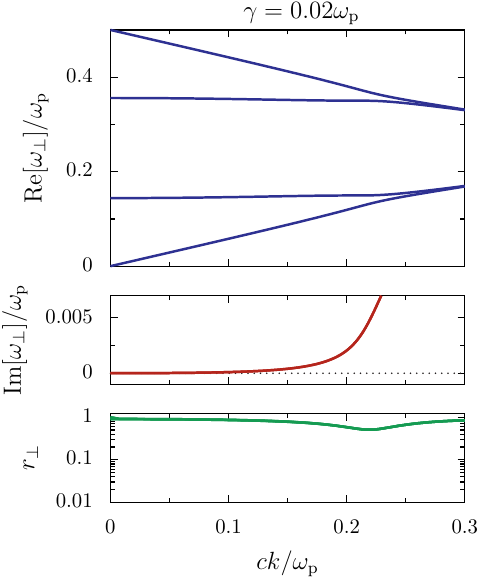}
    \caption{Same quantities as in Fig.~\ref{fig:rigidity weak}, but here in the case of the strong yet low modulation of the plasma frequency discussed in Fig.~3 of the main text.
    Only small momenta are shown in order to emphasize the loss-induced positive imaginary part of the transverse eigenfrequencies.}
\label{fig:rigidity strong}
\end{figure}

In the main text, when discussing the case of a strong yet low modulation of the plasma frequency, we unveiled that the lifting of EPs also induces a positive imaginary part in a larger region of momenta.
To detail that mechanism, we show in Fig.~\ref{fig:rigidity strong} the complex bandstructure along with the phase rigidity of the system, in both the cases of a lossless (left panels) and lossy (right panels) medium.
Only the region of small momenta where this effect appears is shown.
Interestingly, when considering a nonzero inverse relaxation time, we observe that the imaginary part of the eigenfrequencies acquires a positive value for momenta smaller than the edge of the dispersive momentum gap in the lossless case, here for $ck<0.22\omega_\textrm{p}$.
This is precisely what allows the broadening of the gain bandwidth at small momenta discussed in Fig.~3 of the main text.

\vspace{-1.15cm}

\section{Dipole absorption in the stable regime with a modulated plasma frequency}
\label{sec:Stable modulated plasma}
In the main text, we demonstrated that modulating the resonance frequency of a PTC enables both a broadband enhancement of dissipated power and a total negative dissipated power, while keeping the system in its stable regime.
In the case of modulated plasma frequency, however, we unveiled that negative dissipated power is obtained through dispersive momentum gaps. In that case, dipole absorption is present solely when the PTC is in its unstable regime, and the dipole frequencies corresponding to a total negative dissipated power are precisely the ones for which the PTC present eigenfrequencies of positive imaginary part.

\begin{figure}[h!]
 \includegraphics[width=0.57\columnwidth]{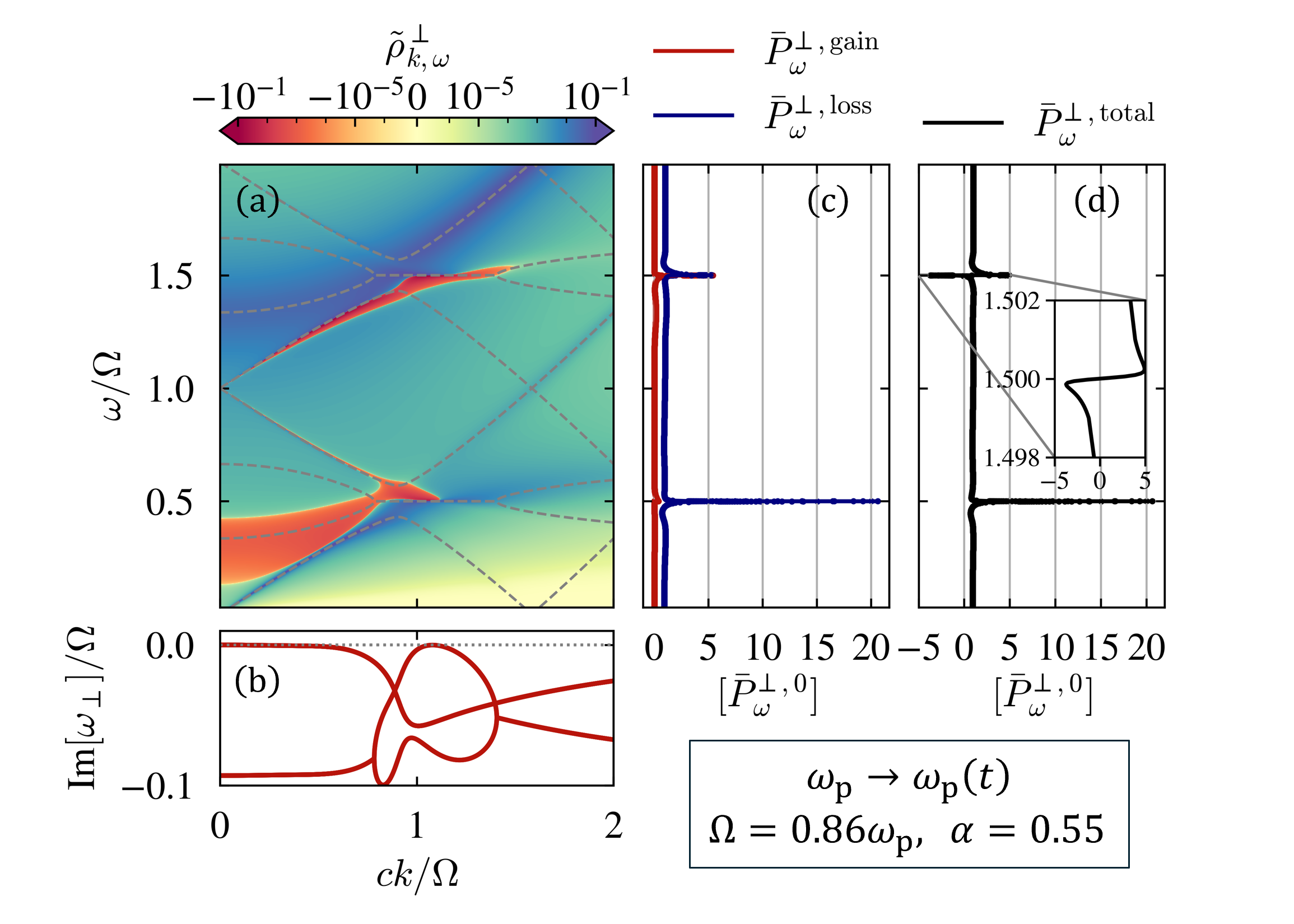}
 \centering
 \caption{Dipole absorption and enhanced dipole emission occurring in the stable regime of a PTC whose plasma frequency is modulated at a frequency $\Omega=0.86\omega_\textrm{p}$, with strength $\alpha=0.55$.
 Here the resonance frequency $\omega_0=0.6\omega_\textrm{p}$, and the damping parameter has been increased to $\gamma=0.16\omega_\textrm{p}$.
 These parameters were here chosen specifically so that the replica of the lower band $\omega_\perp^\mathrm{lo}+\Omega$, associated to a negative LDOS, interacts with a second-order momentum gap formed by the upper band $\omega_\perp^\mathrm{up}$ and its second-order negative replica $2\Omega-\omega_\perp^\mathrm{up}$. The damping parameter has been increased in order to compensate the required large modulation strength, so that the PTC remains stable.}
\label{fig:transverse modes stable}
\end{figure}

Here, we present in Fig.~\ref{fig:transverse modes stable} as a proof of principle an exemple of total negative dissipated power realized in a modulated plasma frequency stable PTC, where the eigenfrequencies imaginary part stay negative.
While, as explained in the caption, the modulation parameters need to be fine-tuned to achieve this effect, it demonstrates its theoretical possibility.

\section{Amplification of longitudinal modes}
\label{sec:Longitudinal modes}

\begin{figure}[h!]
 \includegraphics[width=0.8\columnwidth]{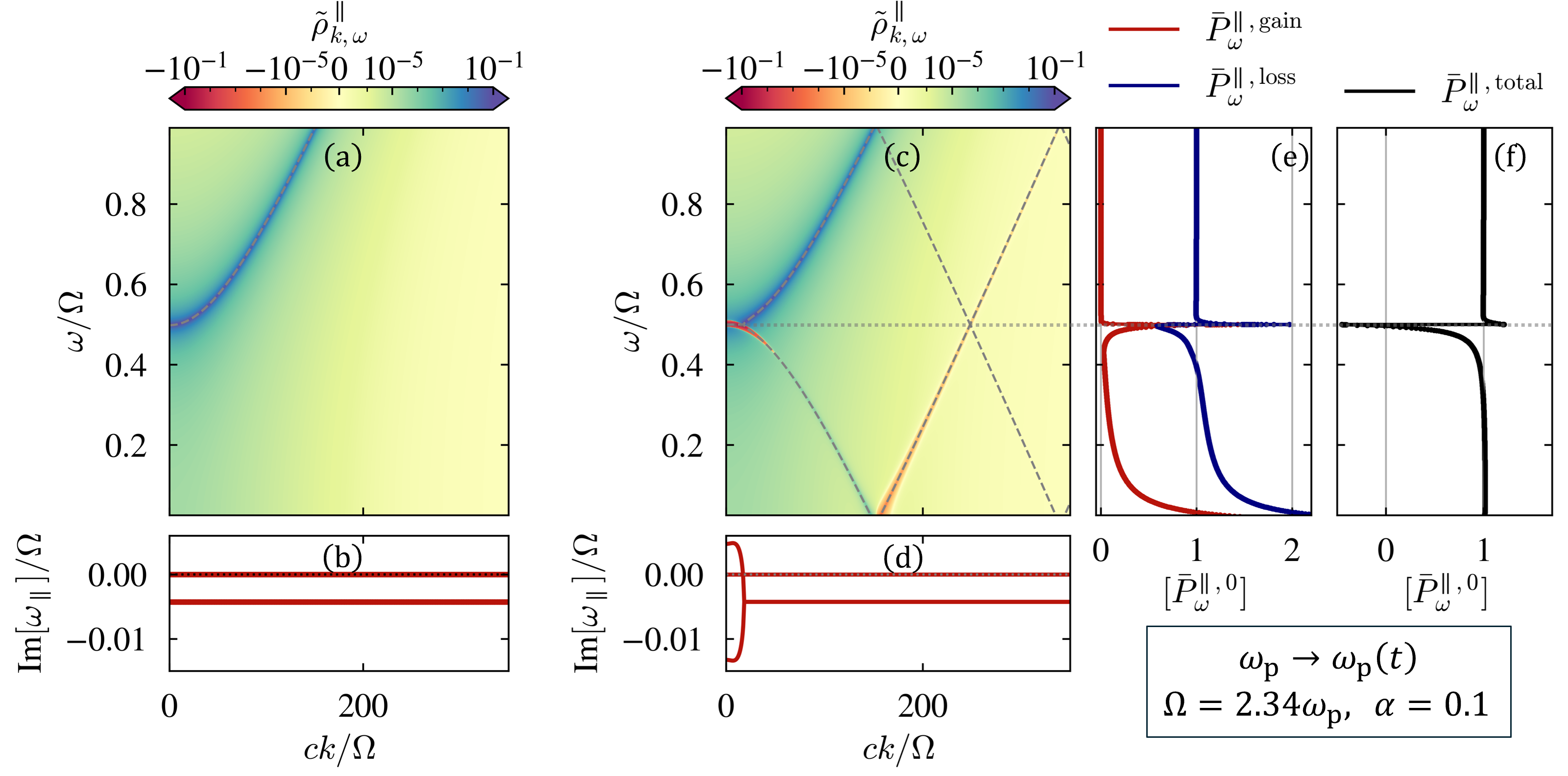}
    \caption{(a)-(b) Momentum-resolved LDOS and imaginary part of the longitudinal eigenfrequencies of a nonmodulated dispersive media.
    (c)-(d) Momentum-resolved LDOS and imaginary part of the longitudinal eigenfrequencies of a dispersive media whose plasma frequency is modulated periodically in time.
    (e) Gain $\bar{P}^{\parallel,\textrm{gain}}_\omega$ and loss $\bar{P}^{\parallel,\textrm{loss}}_\omega$ contributions to the power dissipated by a point dipole of frequency $\omega$, in units of the value obtained in a nonmodulated media $\bar{P}^{\parallel,0}_\omega$.
    (f) Total power dissipated $\bar{P}^{\parallel,\textrm{total}}_\omega$ = $\bar{P}^{\parallel,\textrm{loss}}_\omega$ - $\bar{P}^{\parallel,\textrm{gain}}_\omega$.
    In all panels, the resonance frequency $\omega_0=0.6\omega_{\mathrm{p}}$, the inverse relaxation time $\gamma=0.02\omega_{\mathrm{p}}$, the modulation frequency $\Omega=2.34\omega_{\mathrm{p}}$, and the non-local parameter $\beta=0.0057c$, a value typical for indium-tin-oxide (ITO) \cite{Thouin2025}. In panels (c)-(f), the modulation strength $\alpha=0.1$.
    }
\label{fig:longitudinal}
\end{figure}

While electromagnetic fields in vacuum are purely transverse, a dispersive or absorptive material hosts longitudinal modes that contribute to the density of states and hence modify the emission of an embedded emitter \cite{Barnett1992}.
Recent studies revealed that temporal modulation may amplify longitudinal modes \cite{Zhang2024,Feinberg2025}, however, only local models were considered.
Here, we take advantage of our nonlocal model to reveal the impact of amplified longitudinal modes on dipole emission.
Therefore, we show in Fig.~\ref{fig:longitudinal} the same quantities as discussed in the main text but considering the longitudinal parts of the fields only.
Figs.~\ref{fig:longitudinal}(a)-(b) presents the case of a nonmodulated medium, while a modulation of the plasma frequency is considered in Figs.~\ref{fig:longitudinal}(c)-(f).

By modulating the system at a frequency of more than twice the lowest longitudinal mode $\omega_{\parallel}(k=0)$, we produce a flat momentum gap at the PR condition.
This is exactly similar to the formation of momentum gaps in usual nondispersive PTCs, except that here the effective mass of the longitudinal modes imposes a requirement for a very fast modulation frequency to couple to its first negative replica $-\omega_{\parallel} + \Omega$.
Interestingly, however, the region of negative LDOS is very broad in momentum, encompassing wavenumbers up to $k=20\Omega/c$.
Nevertheless, this specificity does not alter the behavior of dipole emission beyond what is found for usual flat momentum gaps in nondispersive PTCs, the gain and loss contributions being both enhanced at the PR condition, resulting solely in a very narrow effect at $\omega=\Omega/2$.

Interestingly, we note that the downshifted replica $\omega_{\parallel} - \Omega$ is also associated to a negative LDOS, inducing a gain contribution of dissipated power for all dipole frequencies $\omega<\Omega/2$.
As discussed in the main text, a modulation of the resonance frequency would not induce such a negative LDOS for downshifted replicas.
However, as visible in the bottom left of Fig.~\ref{fig:longitudinal}(c), the negative replica $-\omega_{\parallel} + \Omega$ carries a positive LDOS in the same region of frequencies, enhancing the loss contribution of dissipated power, so that the total one is eventually mostly unaffected by the modulation.
This implies that the impact of a modulation of the resonance frequency on longitudinal modes would be qualitatively similar to what we observe here.
Therefore, the impact of momentum gaps of longitudinal modes on dipole emission in PTC is very similar to what has been observed for nondispersive transverse modes \cite{Park2025}, except that longitudinal modes allow naturally for amplification over a wide momentum range \cite{Feinberg2025}.

\begin{figure}[h!]
 \includegraphics[width=0.57\columnwidth]{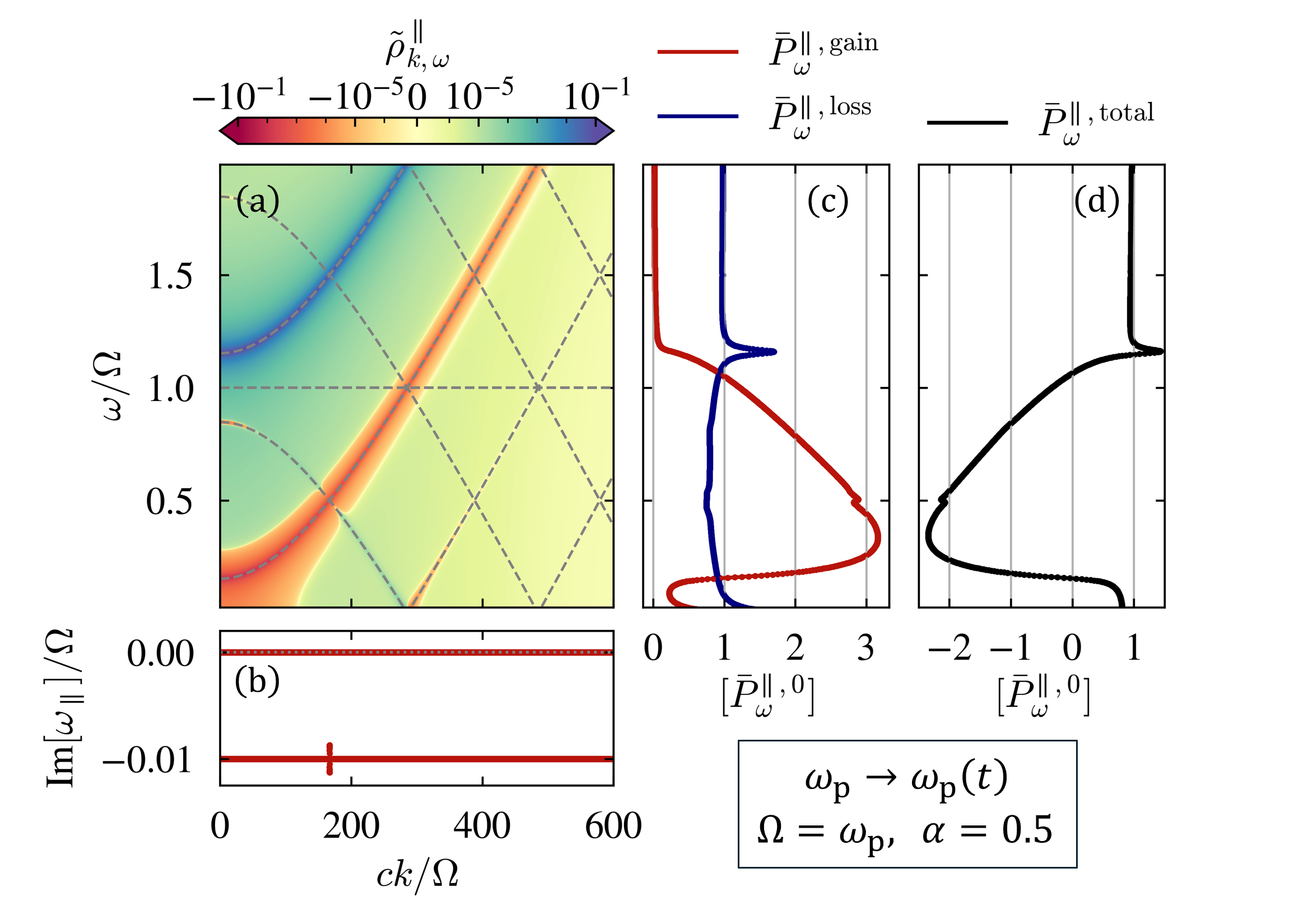}
 \centering
 \caption{Broadband absorption of dissipated power for longitudinal modes. The plasma frequency is modulated at a frequency $\Omega=\omega_\textrm{p}$, with strength $\alpha=0.5$.
 In contrast to the case of transverse modes with modulated plasma frequency discussed in the main text, here the gain arises from the fact that the chosen modulation frequency allows the downshifted replica $\omega_\parallel-\Omega$ to be in the frequency bandgap.
 This allows the appearance of broadband absorption in the stable regime, independently of momentum gaps and exceptional points.
 As in the main text, the resonance frequency $\omega_0=0.6\omega_\textrm{p}$, and the damping $\gamma=0.02\omega_\textrm{p}$.}
\label{fig:longitudinal modes stable}
\end{figure}

The fact that longitudinal modes present a large frequency gap ($\omega<\omega_{\parallel}$) also allows us to use the negative local density of states associated to downshifted replicas when modulating the plasma frequency to induce broadband negative dissipated power without resorting to momentum gaps.
This is exemplified in Fig.~\ref{fig:longitudinal modes stable}, where considering a larger strength $\alpha=0.5$ but a smaller modulation frequency $\Omega=\omega_\mathrm{p}<2\omega_{\parallel}$ prevents the opening of a large momentum gap but enables negative local density of states along $\omega_\parallel-\Omega$.
As a result, a broadband negative dissipated power is observed in most of the frequency gap, while keeping the PTC in its stable regime.

\bibliography{Bibliography}